\documentclass[structabstract]{aa}  
\usepackage{graphicx}
\usepackage{enumerate}
\usepackage{natbib}
\bibpunct{(}{)}{;}{a}{}{,} % to follow the A&A style
\usepackage{txfonts}
\usepackage{enumitem}
\usepackage{float}
\usepackage{array}

\newcolumntype{M}{>{$\vcenter\bgroup\hbox\bgroup}c<{\egroup\egroup$}}

\begin{document}

 \title{Constraining the structure of the transition disk HD~135344B (SAO~206462)
 by simultaneous modeling of multiwavelength gas~and~dust~observations
 \thanks{Based on PIONIER, CRIRES, and UVES observations collected at the VLTI and VLT (European Southern Observatory, Paranal, Chile)  
 with programs 087.C-0702(A,B,D),
 087.C-0458(C), 087.C-0703(B), 
 179.C-0151(A), 077.C-0521(A)}$^,$
 \thanks{{\it Herschel} is an ESA space observatory with science instruments provided by European-led Principal Investigator consortia and with important participation from NASA.}}
\titlerunning{Simultaneous multiwavelength modeling of gas~and~dust~observations of HD~135344B}
\authorrunning{A. Carmona et al. }
   \author{
  A. Carmona\inst{1},
  C. Pinte\inst{2,1},
  W.F. Thi\inst{1},
  M. Benisty\inst{1},
  F. M\'enard\inst{2,1},
  C. Grady\inst{3,4,5},  
  I. Kamp\inst{6},
  P. Woitke\inst{7},
  J. Olofsson\inst{8},
  A. Roberge\inst{5},
  S. Brittain\inst{9},
  G. Duch\^ene\inst{10,1},
  G. Meeus\inst{11},
  C. Martin-Za\"idi\inst{1},
  B. Dent\inst{12},
  J.B. Le Bouquin\inst{1},
  J.P. Berger\inst{13,1}
          }

  \institute{UJF-Grenoble 1 / CNRS-INSU, 
  Institut de Plan\'etologie et d'Astrophysique de Grenoble (IPAG) UMR 5274, Grenoble, F-38041, France,
                \email{andres.carmona@obs.ujf-grenoble.fr}
 \and
 UMI-FCA, CNRS / INSU France (UMI 3386), and Departamento de Astronom\'ia, Universidad de Chile, Santiago, Chile
 \and
 Eureka Scientific, 2452 Delmer, Suite 100, Oakland CA 96002, USA 
 \and
 ExoPlanets and Stellar Astrophysics Laboratory, Code 667, Goddard Space Flight Center, Greenbelt, MD 20771, USA
 \and
 Goddard Center for Astrobiology, Goddard Space Flight Center, Greenbelt, MD 20771, USA 
 \and
 Kapteyn Astronomical Institute, P.O. Box 800, 9700 AV Groningen, The Netherlands
 \and 
 SUPA, School of Physics and Astronomy, University of St Andrews, KY16 9SS, UK
 \and
 Max Planck Institut f\"ur Astronomie, K\"onigstuhl 17, D-69117 Heidelberg, Germany
  \and
 Department of Physics \& Astronomy, 118 Kinard Laboratory, Clemson University, Clemson, SC 29634, USA  
 \and
 Astronomy Department, University of California, Berkeley, CA 94720-3411, USA
  \and
 Departamento de F\'isica Te\'orica, Universidad Autonoma de Madrid, Campus Cantoblanco, Spain 
 \and
  Joint ALMA Observatory, Alonso de C\'ordova 3107, Vitacura 763-0355, Santiago, Chile 
  \and
  European Southern Observatory, Alonso de C\'ordova, 3107 Vitacura, Chile
 }

   \date{Received 25 August 2013; accepted 23 March 2014}

 \abstract{
Constraining the gas and dust disk structure of transition disks, particularly in the inner dust cavity, 
is a crucial step toward understanding the link between them and planet formation.
HD~135344B is an accreting (pre-) transition disk that displays the CO 4.7~$\mu$m emission extending tens of AU inside its 30 AU dust cavity.
 }
 {We constrain HD~135344B's disk structure from multi-instrument gas and dust observations.}
 {
We used the dust radiative transfer code MCFOST and the 
thermochemical code ProDiMo to derive the disk structure from
the simultaneous modeling of 
the spectral energy distribution (SED),
VLT/CRIRES CO~P(10)~4.75~$\mu$m,  
Herschel/PACS [\ion{O}{i}]~63~$\mu$m, 
Spitzer-IRS,  and JCMT~$^{12}$CO~J=3-2 spectra,
VLTI/PIONIER  H-band visibilities, 
and constraints from (sub-)mm continuum interferometry and near-IR imaging.
}
{
We found a disk model able to describe the current gas and dust observations simultaneously.
This disk has the following structure.
(1) To simultaneously reproduce the SED, the near-IR interferometry data, and the CO ro-vibrational emission,  
refractory grains (we suggest carbon) are present inside the silicate sublimation radius ($0.08<R<0.2$ AU).
(2) The dust cavity (R$<$30 AU) is filled with gas, the surface density of the gas inside the cavity must increase with radius to fit the CO ro-vibrational line profile, 
a small gap of a few AU in the gas distribution is compatible with current data, 
and a large gap of tens of AU in the gas does not appear likely.
(4) The gas-to-dust ratio inside the cavity is $>$ 100 to account for the 870 $\mu$m continuum upper limit and the CO P(10) line flux.
(5) The gas-to-dust ratio in the outer disk ($30<R<200$~AU) is $<$ 10 to simultaneously describe the [\ion{O}{i}]~63~$\mu$m line flux and the CO~P(10) line profile.
(6) In the outer disk, most of the gas and dust mass should be located in the midplane,
and a significant fraction of the dust should be in large grains.
}
{
Simultaneous modeling of the gas and dust is required to break the model degeneracies and constrain the disk structure.
An increasing gas surface density with radius in the inner cavity echoes the effect of a migrating jovian planet 
in the disk structure.
The low gas mass (a few~Jupiter~masses) throughout the HD~135344B disk supports the idea that it is an evolved disk 
that has already lost a large portion of its mass. 
}

   \maketitle
%
%________________________________________________________________

\section{Introduction}

Observations of young stars of different ages
reveal that protoplanetary disks evolve from optically thick, 
gas-rich disks to optically thin, 
gas-poor debris disks \citep[e.g, see review by][]{WilliamsCieza2011}.
The transition between these two classes of objects is believed to occur relatively fast (10$^5$ years) compared to the disk's life time 
(5 - 10 Myr) \citep[e.g.,][]{Cieza2007,Damjanov2007,CurrieAguilar2011}.
Infrared space observatories have
unveiled several young stars with spectral energy distributions (SED) 
characterized by IR-excess at $>$ 10 $\mu$m
and significantly reduced excesses at shorter wavelengths \citep[e.g.,][]{Strom1989, Calvet2005, Sicilia-Aguilar2006, 
Hernandez2006, Espaillat2007, Fang2009, Merin2010, Rebull2010, Cieza2010}.
These sources are commonly named ``transition disks", 
because they are believed to be the objects in the transition phase between a young star with an optically thick gas-rich  disk
and a star with an optically thin gas-poor debris disk.

Generally, the lack of strong emission in the near or mid-IR  SED is interpreted as evidence of a gap 
or cavity from a few up to tens of AU in the disk.
Follow-up imaging of several transition disks with  sub-mm interferometers have 
confirmed that transition disks indeed have a deficit of sub-mm continuum emission at a few mJy levels 
inside tens of AU \citep[e.g.,][]{Pietu2007,Brown2009,Hughes2009, Andrews2011, Isella2012}, 
thus providing further evidence for such inner disk cavities, at least on the large dust grains. 

Several scenarios have been discussed in the literature to explain the presence of a cavity and the transitional disk SED shape: 
grain growth \citep[e.g.,][]{DullemondDominik2005,Birnstiel2012},
migrating giant planets that opened a gap \citep[e.g.,][]{Varniere2006,Zhu2011,Dodson-RobinsonSalyk2011},
dust filtration by a giant planet \citep[e.g.,][]{Rice2006, Zhu2012,Pinilla2012}, 
disk dissipation due to a photoevaporative disk wind \citep[e.g.,][]{AlexanderArmitage2007,Owen2012}, and
dust free inner holes due to radiation pressure \citep[][]{ChiangMurray-Clay2007}.
One important step toward understanding the transitional disk phenomenon
is to understand how the gas and dust structures compare.
Do they follow each other in density structure?
Do they thermally de-couple?
Does the gas-to-dust ratio stay constant as a function of the radius?

Several young stars with transition disks are emission-line stars that exhibit signs of accretion 
\citep[e.g.,][]{Najita2007,Muzerolle2010,Cieza2012b}. 
As a consequence, 
the cavities imaged in the sub-mm should have gas
as the material from the optically thick outer disk flows 
through the cavity to be accreted by the central object
\citep[e.g.,][]{LubowAngelo2006}.
In fact, 
several transition disks display emission of CO at 4.7$~\mu$m \citep[e.g.,][]{Salyk2009} a common tracer of
warm gas in the inner disk \cite[e.g.,][]{NajitaPPV}.
Furthermore, in a few transition disks,
CO ro-vibrational emission has been spatially resolved 
up to distances of tens of AU \citep[][]{Pontoppidan2008}.
Moreover, recent studies of dust-scattered light in the near-IR have detected 
emission from small dust grains inside the cavities that were previously imaged with sub-mm interferometry, 
with the additional and remarkable property 
that no sharp edge is seen at the location of the sub-mm cavity's inner radius \citep[e.g.][]{Muto2012,Garufi2013}.
This suggests that what is observed is likely a change in the dust size distribution inside the cavity \citep[][]{Dong2012}.
Near-IR interferometry provides evidence of dust emission (and inhomogeneities) inside the dust cavity of transition disks \cite[e.g.,][]{Kraus2013}. 
Furthermore, recent ALMA observations have spatially resolved gas inside the cavity of transition disks \citep[][]{Casassus2013,Bruderer2013}.

In summary, 
the observational evidence indicating that the dust cavities of transitions disks are indeed {\it not empty} 
has been steadily growing during past years.
This opens interesting questions such as 
how much gas and dust are present in the cavity;  
what is their distribution and chemical composition;
if dust is present, what is its size distribution;
if there is a gap in the gas or in the dust, how large is it?

{ The goal of this paper is to address the problem of the gas and dust structures in transition disks
by performing a detailed study of the transition disk HD~135344B (SAO 206462).
Our aim is to constrain the disk structure by modeling simultaneously and in a coherent way
multiwavelength \& multi-instrument gas and dust observations of the inner and the outer disk.
Most specifically,
we investigate gas and dust disk content and structure inside the cavity by 
simultaneously modeling the SED, 
new VLTI-PIONIER near-IR interferometry data, 
the CO~4.7~$\mu$m emission, and constraints from sub-mm continuum interferometry.
Furthermore, 
we seek to constrain the gas mass and the gas-to-dust ratio in the outer disk
by employing the SED, the [\ion{O}{I}] $63~\mu$m and $145~\mu$m lines, and (sub-)mm observations}.

HD~135344B  is an accreting \cite[$2\times10^{-8}$ M$_{\odot}$/yr,][]{Sitko2012} F4V young star \citep[8$^{+8}_{-4}$ Myr,][]{vanBoekel2005}
that has a ``pre-transition disk" (i.e., a transition disk with near-IR excess).
It has the remarkable characteristic of exhibiting CO~4.7~$\mu$m emission extending tens of AU inside the 45 AU sub-mm dust cavity 
\citep[][]{Pontoppidan2008} and scattered light emission from small dust grains down to 28 AU \citep[][]{Muto2012,Garufi2013}.
HD~135344B  display spiral structures \citep[][]{Muto2012,Garufi2013} 
and asymmetries in its outer disk dust emission \citep[][]{Brown2009,Andrews2011,Perez2014}.
It is a transition disk that exhibits variability in its near-IR SED 
and in optical and near-IR line emission on time scales of months \citep[][]{Sitko2012}.
It is a source on which [\ion{O}{I}] $63~\mu$m emission has been detected in Herschel/PACS observations  \citep[][]{Meeus2012}.
Finally, it is a transition disk without close-in low-mass stellar companions \citep[][]{Vicente2011}.

We start the paper by a brief summary of current observational constraints on the disk of HD~135344B in Section 2.
Then, in Section 3,  we describe the modeling tools and the general modeling procedure.
In Section 4, we present and discuss the different models that were tested.
In Section 5, we examine the disk constraints derived from the final disk model. 
In Section 6, we discuss our results in the context of the study of transitional disks and 
planet formation. 
Finally a summary and conclusions are presented in Section 7.
\vspace{-0.3cm}
\section{Observational constraints}
HD~135344B has been observed with a diversity of instruments and techniques from UV to mm
wavelengths. 
{ In Table~\ref{obs_cons_table}, we present a concise summary of relevant previous observational constraints on HD~135344B.}
The most important observational constraints for this study are: 
\begin{itemize}
\item[$\bullet$] the SED has a dip in the 10$-$50$~\mu$m region \citep[][]{Brown2007}; 
\item[$\bullet$] the 10~$\mu$m silicate feature is absent \citep[][]{Geers2006}; 
\item[$\bullet$] the inner radius of the sub-mm cavity is 46$\pm$5 AU \citep[][]{Brown2009,Andrews2011};
\item[$\bullet$] the 3$\sigma$ continuum flux upper limit at  870 $\mu$m is 10.5 mJy for a  0.24"$\times$0.5" beam centered on the star  \citep[][]{Andrews2011}; 
\item[$\bullet$] CO ro-vibrational emission 
extends at least to 15 AU inside the sub-mm dust cavity (25 AU for $d$=140 pc)
\citep{Pontoppidan2008}\footnote{\citet{Pontoppidan2008} suggests a minimum extension
of 15 AU assuming a distance of 84 pc. The distance currently used to interpret HD~135344B data \citep{Brown2009,Andrews2011} is 140 pc \citep{vanBoekel2005}.
If $R_{\rm out}$ constraint is scaled to the 140 pc that leads to a minimum outer radius of 25 AU.
%Note that strictly speaking this 
}; 
\item[$\bullet$] near-IR dust scattering images reveal small dust inside the sub-mm cavity down to 28 AU
with a smooth surface brightness and no discontinuity at 45 AU \citep{Muto2012, Garufi2013}.
\end{itemize}

\begin{center}
\begin{table*}
\caption{Summary of most relevant previous observations and disk constraints.}
\label{obs_cons_table}
\begin{tabular}{p{3cm}lp{5cm}p{5.5cm}ll}
\hline
{\bf Technique} & {\bf Instrument} & {\bf Observation } & {\bf Interpretation or constraint} & {\bf Ref.} \\[1mm]
\hline
UV - mm photometry & several  & SED: dip at 15~$\mu$m &  40 AU gap & 1,2\\[2mm]
UV spectroscopy & HST-COS &  shape of the continuum flux  & spectral type F4V; E(B-V) $<$0.129  & 3,4,5\\[2mm]
optical spectroscopy & VLT-UVES &  photospheric absorption lines & T$_{\rm eff}$=6750$\pm$250 K; log$\,g>4.0$ & this work \\[2mm]
 & VLT-UVES &   broad [\ion{O}{I}] 6300~\AA~emission &  hot gas down to 0.1 AU & 6,7\\[2mm]
near-IR high-resolution spectroscopy & VLT-CRIRES & extended CO~4.7 $\mu$m emission; line profile and spectro-astrometry signal & 
gas in Keplerian rotation inside the cavity at least until 15 AU assuming $d$=84 pc (25~AU~@~140~pc); 
i$=14^{\circ}$ for $M$=1.65\,M$_{\odot}$; 
PA$\sim56^{\circ}$& 8\\[8mm]
				  &VLT-CRIRES  & non-detection of H$_2$ lines at 2 $\mu$m & fluxes $<10^{-17}$ W m$^{-2}$ &9 \\[2mm]
near-IR imaging & Subaru-HICIAO  &  dust scattered light emission & material down to 28 AU; no discontinuity at 45 AU; spiral patterns; flat outer disk & 10\\
 &VLT-NACO & & strong drop of polarized dust emission at 28~AU & 11\\[4.5mm]
near-IR interferometry &  VLTI/PIONIER & visibilities & continuum emission inside the silicate sublimation radius ($R<0.2$ AU) &  this work \\
mid-IR spectroscopy  &  Spitzer/IRS  &  no 10~$\mu$m silicate feature & limit to the mass of small silicate grains  & 12 \\
  						&&  weak 11.2 $\mu$m PAH emission & limit to the PAH abundance \\
                                            & &  non detection H$_2$ pure-rotational lines, and the [\ion{Ne}{II}] 12.8~$\mu$m line  & fluxes $<10^{-17}$ W m$^{-2}$ & 13 \\[4.5mm]
far-IR spectroscopy & Herschel/PACS & [\ion{O}{i}] 63$~\mu$m detected,  upper limits to the
&  $F_{[\ion{O}{i}]~63~\mu{\rm m}}$=3.6$-$4.8$\times10^{-17}$ W m$^{-2}$ &14, 20 \\
& & [\ion{O}{i}] 145$~\mu$m and [\ion{C}{ii}] $157~\mu$m lines & \\[2mm]
sub-mm interferometry &  SMA &  
weak  870 $\mu$m continuum at R$<40$ AU;&
cavity radius 46$\pm$5 AU; i$\sim12^{\circ}$; $3\sigma$ flux in a $0.24"\times0.5"$ ($33\times70$ AU) beam centerred on the star 10.5 mJy & 15, 16\\[4.5mm]
 & ALMA & resolved 450 $\mu$m emission & asymmetries in the outer disk observed & 17\\ 
  & &  $^{12}$CO 6-5 detected at R$<$40 AU & gas inside the dust cavity. \\[2mm] 
sub-mm line single dish & JCMT & $^{12}$CO 3-2 detected &  double peak; {\it FWHM}=2.2 km/s; $i\sim11^{\circ}$ & 18\\[2mm]
mm continuum and  & SMA  & extended 1.3 mm emission & R$_{\rm out}=$ 220 AU;  i$\sim11^{\circ}$  & 19 \\
line interferometry &  & extended $^{12}$CO 2-1 emission &  \\[2mm]

\hline
\end{tabular}
\newline
\newline
{\bf References} [1] see Table~\ref{table_photometry} in the Appendix, 
[2] \citet[][]{Brown2007},
[3] \citet{Grady2009}, 
[4] \citet{France2012}, 
[5] \citet{Schindhelm2012}, 
[6] \citet[][]{vanderPlas2008},
[7] \citet[][]{Fedele2008},
[8] \citet[][]{Pontoppidan2008},
[9] \citet[][]{Carmona2011}
[10]  \citet[][]{Muto2012}, 
[11] \citet[][]{Garufi2013},
[12] \citet[][]{Geers2006},
[13] \citet[][]{Lahuis2007},
[14] \citet[][]{Meeus2012},
[15]  \citet[][]{Brown2009},
[16]  \citet[][]{Andrews2011},
[17]   \citet[][]{Perez2014},
[18] \citet[][]{Dent2005},
[19]  \citet[][]{Lyo2011},
[20] \citet{Fedele2013}.
\end{table*}
\end{center}

\section{Modeling}
% ----------------------------------------------------

\subsection{Stellar parameters}

{ We provide a detailed discussion of the derivation of the stellar parameters in the Appendix~\ref{star_parameters}.
For our models we used 
a star with $T_{\rm eff}$=6620~K (F4V), a stellar radius of 2.1 R$_\odot$, a mass of 1.65 M$_\odot$, and $A_{\rm v} = 0.4$.
Owing to uncertainties in the photometry, the spectral type and, especially, the distance ($\pm$20 pc) 
the stellar radius can vary $\pm$0.3~R$_\odot$, and the stellar mass does so $\pm$0.1~M$_\odot$. 
The stellar UV spectrum\footnote{FUSE and HST-COS spectra were obtained from DIANA protoplanetary disk observations and modeling database (http://www.diana-project.com/). Details on the data reduction procedures will be given in Dionatos et al. (in prep.).}
was parametrized with a fractional UV excess $f_{\rm UV}$ = $L_{\rm UV}$/$L_{\star}$ equal to 0.001 
($L_{\rm UV}$ is defined as the luminosity between 91.2 and 250 nm) and $F_{\nu} \propto \nu^{~\gamma}$ with $\gamma=$-2.15. % (see Fig. \ref{FUV_parametrization}).
This parametrized stellar UV spectrum is included in the ProDiMo gas heating calculation in addition to the interstellar UV field.}

\subsection{Disk inclination and position angle}

Several data sets give different inclinations for HD~135344B: 
CO~4.7~$\mu$m emission suggests $i=$14$\degr\pm4\degr$ \citep[][for $M=$1.65 M$_{\odot}$]{Pontoppidan2008};
the CO $J=3-2$ line at 870 $\mu$m indicates $i=$11$\degr\pm2\degr$ \citep{Dent2005}; 
sub-mm continuum imaging points toward $i=$12$\degr$ \citep{Andrews2011} and 21$\degr$ \citep{Brown2009};
mm interferometry suggests $i=$11$\degr$ \citep{Lyo2011};
near-IR imaging sets an upper limit of $i=$20$\degr$ \citet{Grady2009};
mid-IR imaging  suggests $i=$46$\degr \pm5\degr$ \citep{Doucet2006} and 45$\degr \pm 5\degr$ \citep{Marinas2011}.
For our analysis of the SED and line profiles, 
we used $i=14\degr$.
{ We used this value because it is derived  from a simultaneous fit to the spectro-astrometry 
signature of the CO~4.7 $\mu$m line in three-slit position angles and the spectrally resolved CO~4.7 $\mu$m line profile,
and because it is consistent with the inclination derived from the observations of CO gas in the 
outer disk\footnote{
Recent near-IR imaging \citep[][]{Muto2012,Garufi2013} and sub-mm interferometry \citep[][]{Brown2009,Andrews2011,Perez2014} 
have revealed that the disk of HD 135344B has spiral arms and asymmetries in the dust emission.
One possible reason for the discrepancy 
between the inclination derived from mid-IR imaging data (i.e.,  dust emission)
and the inclination derived from  near-IR and sub-mm CO gas observations 
is that disk asymmetries could be present in the mid-IR dust emission,
hence a different estimate of the inclination and PA.}}.

Several estimations exist for the disk's position angle (PA).
\citet{Pontoppidan2008} suggest a PA=56$\degr\pm2\degr$ based on spectro-astrometry of the CO~4.7~$\mu$m emission;
\citet{Grady2009} suggest a  PA=55$\degr\pm5\degr$ based on near-IR scattered light imaging;
\citet{Andrews2011}, \citet{Brown2009}, and  \citet{Lyo2011} based on (sub-)mm interferometry continuum observations  
suggest a PA=64$\degr$, 55$\degr$, and 64$\degr$ respectively.
As we aim to compare our model with the CO~4.7 $\mu$m lines we employed a PA of 56$\degr$.

\begin{figure}[t]
\begin{center}
\includegraphics[width=0.5\textwidth]{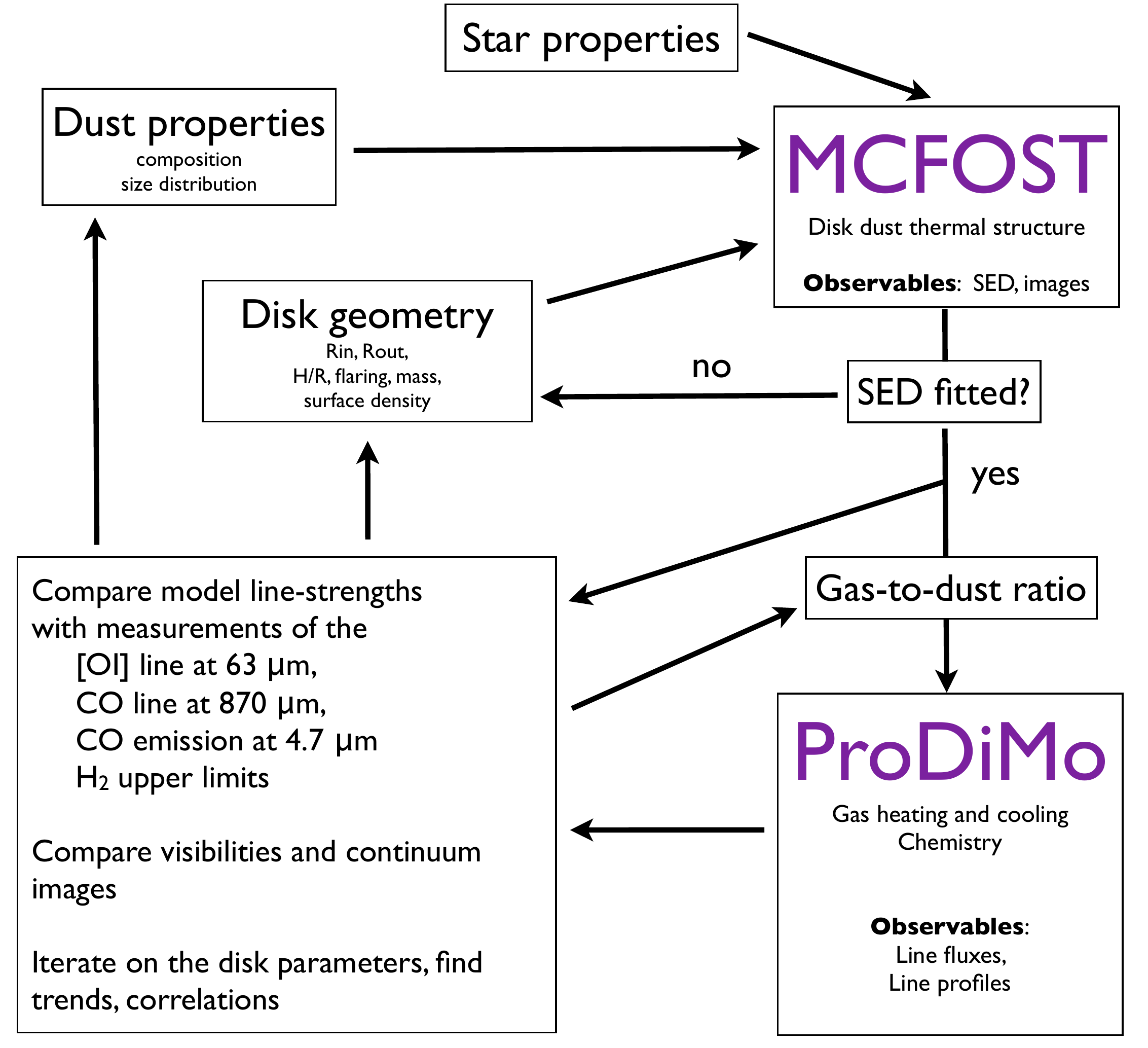}
\caption{Schematic view of the general modeling procedure.}
\label{schema}
\end{center}
\end{figure}

\subsection{Description of the general modeling procedure}
\label{general_description}
We aimed to 
\begin{enumerate}
\item
{\it simultaneously fit} the SED, the line profile of the CO $\nu=1-0$ P(10) line at 4.7545 $\mu$m,
and the near-IR PIONIER visibilities and closure phases;
\item {\it reproduce within a factor of a few}
the detected line fluxes of CO P(10), [\ion{O}{i}] at 63 $\mu$m, 
$^{12}$CO $J=3-2$ at  870 $\mu$m, and
$^{12}$CO $J=2-1$ at 1.27 mm;
\item {\it obtain line fluxes below the upper limits} for [\ion{O}{i}]  at 145 $\mu$m, 
[\ion{C}{ii}] at 157 $\mu$m, 
H$_2$ $1-0$\,S(1) at 2.12 $\mu$m,
H$_2$ $0-0$\,S(1) at 17 $\mu$m, and CO and H$_2$O infrared lines covered by Herschel 
(see Tables \ref{Table_CO_line_fluxes} and \ref{Table_H2O_line_fluxes});
\item {\it obtain} a  870 $\mu$m flux inside a beam of 4\arcsec$\times$0.5\arcsec (33$\times$70 AU) centered on the star
lower than 10.5 mJy.
\end{enumerate}
To be consistent with the sub-mm continuum constraints,
the inner radius of the outer disk's  large and small grains 
was set to 45~AU.
Later in the modeling process, 
the inner radius of the outer disk's small grains was allowed to extend down to 30 AU
to be consistent with the near-IR scattered-light constraints.

In Fig.~\ref{schema}, 
we provide an schematic overview of the modeling procedure.
The modeling starts by assuming 
a dust composition, a dust size distribution, a dust mass, 
and a gas-to-dust ratio, for the inner and the outer disk using a parametric disk (see details in next section). 
Then the Monte-Carlo dust radiative transfer code MCFOST \citep{Pinte2006,Pinte2009} 
is used to calculate the dust's thermal and density structure, $T_{\rm dust}(r,z)$, $n_{\rm dust}(r,z)$, 
and  the mean radiation field $J_\nu(r,z)$. 
MCFOST is employed to compute a synthetic SED, 
and continuum images in the H band and at  870 $\mu$m for comparison with VLTI/PIONIER and SMA continuum observations.

The MCFOST disk structure and radiation field is then used as
input for the thermo-chemical radiative transfer code {\sc ProDiMo} \citep{Woitke2009}. 
{\sc ProDiMo} calculates the disk chemistry, 
the gas and dust heating and cooling, 
and the energy level populations of molecules and atoms.  
{\sc ProDiMo} computes
the synthetic line fluxes and 
channel maps for gas emission lines
from the optical to the mm using escape probability.

Details of the coupling between the codes MCFOST and {\sc ProDiMo}
and brief summaries of technical details of both codes  
are provided in \citet{Woitke2010}, \cite{Pinte2010}, and \cite{Kamp2011}.
Details of the implementation of CO ro-vibrational emission within {\sc ProDiMo} 
are given in \citet[][]{Thi2013}.

The modeling was performed without computing the hydrostatic equilibrium.
The disk gas and dust density  distribution (thus the gas-to-dust ratio) 
was set as input.
We chose this approach to explore a large fraction of the parameter space.
Computing the hydrostatic equilibrium would have resulted in
a running time an order of magnitude longer and would have prevented us from
performing a large exploration of the parameter space.
We have verified that our last model, Model 5, is consistent with hydrostatic
equilibrium; i.e., the scale height we used as input is consistent with
$h_{\rm hydro} = \sqrt{kT_{\rm gas\,(z=0)}\,r^3 / GM\mu}$.

We started our exploration with simple models,
and complexity
in the model was only added when we were not able to fit some of the observations.
The model was refined sequentially following this protocol:
\begin{quote}
\begin{enumerate}
\item fit the SED,
\item fit the CO P(10) ro-vibrational line-profile,
\item check consistency with SMA 870 $\mu$m continuum upper limits inside the cavity,
\item describe near-IR visibilities,
\item produce an [\ion{O}{i}] 63 $\mu$m line flux consistent with the Herschel observations,
\item refine the model to account for the latest near-IR polarization images \citep[][]{Garufi2013}.
\end{enumerate}
\end{quote}

For each family of models, a large portion of the parameter space was explored by
a combination of hand exploration, grid modeling, and simplex optimization algorithms.
In total we tested $\sim$60\,000 MCFOST models and $\sim$4\,000 {\sc ProDiMo} models.

\subsection{Disk structure \& MCFOST parameters}
The philosophy of MCFOST is to model the disk as composed 
of individual zones that may overlap. 
Each zone is a disk extending from an inner radius $R_{\rm in}$ to an outer radius $R_{\rm out}$,
with a scale height $h_0$ at a reference radius $r_0$, flaring exponent ($\beta$), 
and surface density exponent ($q$).
Each zone has a Gaussian vertical density profile
$\rho(r,z)= \rho_0(r)~exp{(-z^2/2 h(r)^2)}$,
a power-law surface density  
$\Sigma(r) = \Sigma_0 (r/r_0)^q$,
and a scale height 
$h(r) = h_0 (r/r_0)^\beta$,
where $\beta$ is the flaring exponent, $r$ the radial coordinate in the equatorial plane, and
$h_0$ the scale height at the reference radius $r_0$.

For each zone,
the dust composition, the dust size distribution, the dust mass, and the gas-to-dust ratio are defined independently.
One zone can include several dust components, such as silicate grains, carbonaceous grains, or PAHs. 
{ Zones can have Gaussian inner edges of size five times the parameter ``$edge$".}
{ In our models, for each zone, the gas-to-dust ratio is constant in the vertical direction}.

Dust grains are defined as homogeneous and spherical 
particles with sizes distributed 
according to the power law $dn(a) \propto a^p~da$, with $a_{\mathrm{min}}$
and  $a_{\mathrm{max}}$ 
the minimum and maximum sizes of grains. 
We used the standard value $p=-3.5$.
Extinction and scattering opacities, scattering
phase functions, and Mueller matrices are calculated using the Mie theory.
We employed the astronomical silicates opacities of \citet{DraineLee1984}, 
the neutral PAH opacities computed by B. T. Draine 
\citep[compiled from][]{LaorDraine1993,DraineLee1984, LiDraine2001},
and the amorphous carbonaceous dust optical constants derived by 
\citet[][]{LiGreenberg1997}. 

In summary, the free parameters of a model are the number of zones used to describe a disk,
and for each zone
$R_{\rm in}$,  $R_{\rm out}$, 
$p$, $\beta$, $q$, $h/r$, $edge$, $M_{\rm dust}$, gas-to-dust ratio, and dust composition, and for each dust species 
$a_{\rm min}$,  and $a_{\rm max}$.

Sub-mm continuum images were convolved with a $0.24\arcsec \times 0.5\arcsec$ beam for comparison with
SMA observations.

\begin{table}[t]
\caption{HD~135344B stellar and geometrical parameters used for the modeling.}
\begin{center}
\begin{tabular}{rcl}
\hline
$T_{\rm eff}$ & = & 6620 K \\
log$~g$         & = & 4.5  \\
$R$		      & = &  2.1 R$_\odot$ \\
$M$	              & = &  1.65 M$_\odot$ \\
A$_{\rm V} $  &  = & 0.4 \\
$d$                & = &  140 pc \\
$i$	              & = & 14$\degr$ \\
$PA$             & = & 56$\degr$\\
$R_{\rm cav}$ & = & 30$-$45 AU\\
$R_{\rm out}$ & = & 200 AU\\
\hline
\end{tabular}
\end{center}
\end{table}

\subsection{{\sc ProDiMo} parameters}
We used the cosmic ray ionization rate  of H$_2$ of 5$\times10^{-17}$~s$^{-1}$.
The incident vertical UV was set to the interstellar medium value.
Non-thermal broadening was set to 0.15 km s$^{-1}$.
We used the UMIST chemical network \citep[9 elements,
71 species connected through 950 reactions neutral-neutral,
ion-molecule, photo-reactions, cosmic ray reactions, and absorption
\& desorption of CO, CO$_2$, H$_2$O, NH$_3$, CH$_4$, see][]{Woitke2009}.
UV fluorescence is included for [\ion{O}{i}] and [\ion{C}{ii}] but not for CO.
UV CO fluorescence has been included as a test later in the paper.
A list of the species we used is provided in \citet[][]{Woitke2009}.

\vspace{-0.2cm}
\begin{table}[t]
  \caption{{\sc ProDiMo} general parameters.}\label{tab_prodimo_parameters}
\begin{center}
\begin{tabular}{rcl}
  \hline
  \noalign{\smallskip}   
  ISM UV field      ($\chi$, Draine)  & = & 1.0 \\
  Non-thermal broadening  & = & 0.15  km s$^{-1}$\\
  fractional UV excess            & = & 0.001 \\
  UV power-law index   & =  & -2.15 \\
  Cosmic ray ionization rate of H$_2$  $\zeta$        & =  & 5$\times10^{-17}$ s$^{-1}$ \\
  \hline
\end{tabular}
\end{center}
\end{table}

\subsection{CO~4.7 $\mu$m data and slit effects\label{spectro-astrometry}}
For our analysis,
we downloaded the 1D CO~4.7~$\mu$m reduced spectrum used in \citet{Pontoppidan2008}\footnote{Available on the webpage of 
the CRIRES large program 
``{\it The planet-forming zones of disks around solar-mass stars: a CRIRES evolutionary survey"} 
(ESO-program 179.C-0151, {PI: van Dishoeck \& Pontoppidan}) http://www.stsci.edu/$\sim$pontoppi/Pontoppidan\_web\_home/\\CRIRES\_Disks.html}.
We flux-calibrated the spectrum using the Spitzer flux at 4.7 $\mu$m (see Appendix for the photometry)
and measured the CO line fluxes by fitting a Gaussian to their line profile.
{The CO P(1) to P(11) line profiles are available within the CRIRES spectrum.
After subtracting the continuum and normalization by the peak flux, 
the lines have the same line profile within the errors (see Fig.~\ref{CO_lines} in the Appendix).
We selected the CO P(10) line for detailed modeling because the line profile is complete, is weakly affected by nearby strong telluric absorption lines,
and has good S/N. 
The line CO P(10) line flux is 1.5$\times10^{-17}$ W~m$^{-2}$ with an error on the order of  20\% owing to slit losses and systematics.}

To compare the {\sc ProDiMo} predictions with the observed CRIRES  CO~4.7 $\mu$m spectra,
the effects of the slit width and orientation, the observing conditions, 
and spectral resolution needed to be taken into account.

{\sc ProDiMo} generates channel maps data cubes; 
i.e., for each velocity bin, a 2D image of CO emission is generated (see details in Bertelsen et al. in prep.).   
The pixel size of the CRIRES detector in the spatial direction is 0.086",
at a distance of 140~pc that corresponds to a pixel size of $\sim$12 AU.
We generated the  {\sc ProDiMo} data cubes with a pixel size of 2 AU.

The {\sc ProDiMo} data cubes were first convolved in the spatial direction with a Gaussian PSF
of {\it FWHM}=180 mas. 
Then they were convolved with a Gaussian of {\it FWHM}=3.3 km/s
in the wavelength direction to simulate R=90\,000 of the observations.
A mask of 0.2" (28 AU) with the correct PA was applied in each channel 
to mimic the slit.
The fluxes were added in the direction perpendicular to the slit to generate a 2D spectrum.

In the 2D spectrum the centroid of the photo center at line velocity ($\upsilon$) was calculated using \citep{Pontoppidan2008}
\begin{equation}
X(\upsilon)=K\frac{\sum_i (x_i(\upsilon)-x_0)F_i(\upsilon)}{\sum_i F_i(\upsilon)} {\rm~~ (pixels)}
\end{equation}
where $x_i-x_0$ is the center of pixel $i$ relative the continuum centroid position,
$F_i(\upsilon)$ is the flux on the pixel $i$, and
$K$ is a correction factor to take into account that not all of the source flux is enclosed in the aperture \citep{Pontoppidan2008}.
We used the value of $K=1.3$ used by  \citet{Pontoppidan2008} in their analysis of the spectro-astrometry 
signal of HD~135344B to be able to compare our models to their measurements.
A 1D spectrum is further obtained by summing the pixels in the spatial direction.
To compare the observed and synthetic line profiles,
the spectra were normalized by dividing them by the median of the continuum,
then after continuum subtraction, the profile was renormalized by dividing it by its maximum flux. 
In this way the continuum is always at 0 and the line peak is always at 1.
The integrated line fluxes of the model (taking the slit effects into account)  
and the line profiles were compared to the observations separately.

\section{Modeling results}
Before presenting the family of models 5, the model that best describes the observations,
we briefly describe the sequence of families of models we tested
to illustrate the reasoning that lead us to the final model.
A family of models is defined primarily by its dust properties (i.e., composition and location).
{ We have extensively explored the parameter space in each family of models.
However, we limit the discussion to the parameter space on the family of models 5.}
In Table~\ref{Table_models} we present the details of a representative example of each family of models.
The different families of models show us the close relation that exists between the 
properties assumed for the dust (size distribution, composition, and mass) at the beginning of the modeling 
procedure and the gas lines obtained.

\subsection{Family of models 1: A disk only composed of astronomical silicate grains}
{ As starting point we assumed the simplest model, 
which has an inner and outer disk composed of 100\% astronomical
silicates and a gas-to-dust ratio of 100 in the whole disk}. The solutions all converged to the same disk structure,
namely a narrow ring of dust at 0.16 to 0.21 AU with a dust mass of a few $10^{-10}$ ~M$_\odot$ 
{ with grains larger than 10 $\mu$m (to reproduce the lack of 10 $\mu$m silicate feature)},
followed by a large gap of 45~AU and an outer disk from 45 to 200 AU with dust mass of $10^{-4}$~M$_\odot$.
By changing the disk's geometrical parameters, 
we found a disk able to reproduce most of the observed gas line fluxes (see Model~1a in Tables
\ref{Table_models} and \ref{Table_line_fluxes}).
However, because in this model the warm { gas} is located in the narrow innermost region, 
the CO ro-vibrational emission produced has a broad double-peaked profile 
{ with FHWM$\sim$50 km/s} that is inconsistent with the observed { FWHM$\sim$ 15 km/s} (see Model 1a in Fig.~\ref{all_models}).

With the aim of producing CO ro-vibrational emission extending tens of AU
and a narrower line profile, in Model 1b, we introduced gas (and dust) between 0.21 and 45 AU. 
We found that although up to $10^{-8}$ M$_\odot$ of dust and $10^{-6}$ M$_\odot$ of gas
can be introduced inside the cavity and still fit the SED, 
the CO ro-vibrational line always displayed a broad double peaked profile with a FHWM of $\sim$50 km/s. 
We tested several flat and flared disks filling the cavity (see Model 1b in Fig.~\ref{all_models}),
but the same result was found in all cases.
{\it The optically thick innermost disk of silicates required to fit the near-IR SED shields the material inside the cavity 
and does not permit the gas at several AU to be warm enough to contribute significantly to the CO ro-vibrational lines}.
A change in the dust's optical properties of the inner disk is required to enable ro-vibrational CO emission at tens of AU.

\begin{figure} [t]
\begin{centering}
\begin{tabular}{cc}
\includegraphics[width=0.31\textwidth]{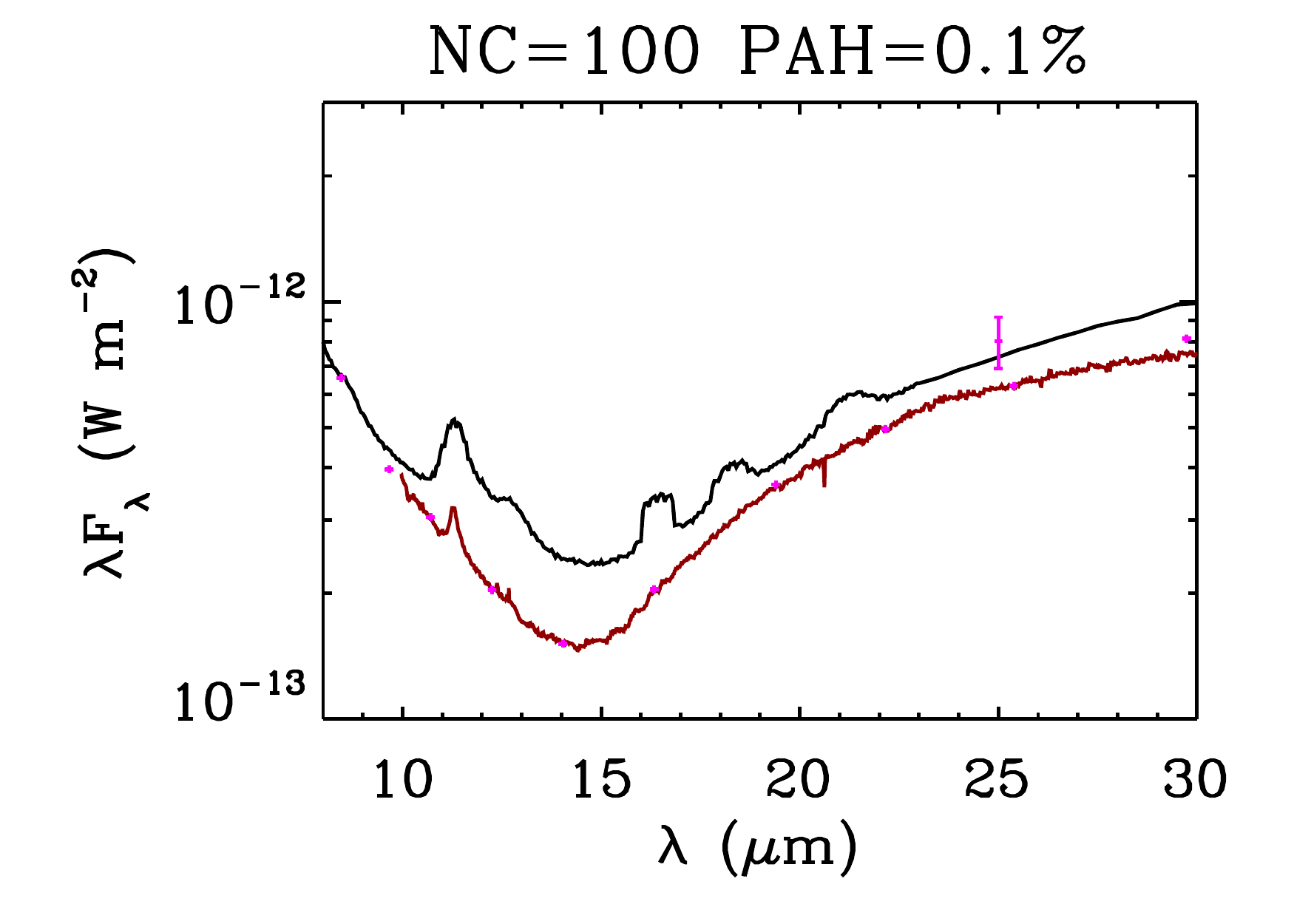}  \\
\includegraphics[width=0.31\textwidth]{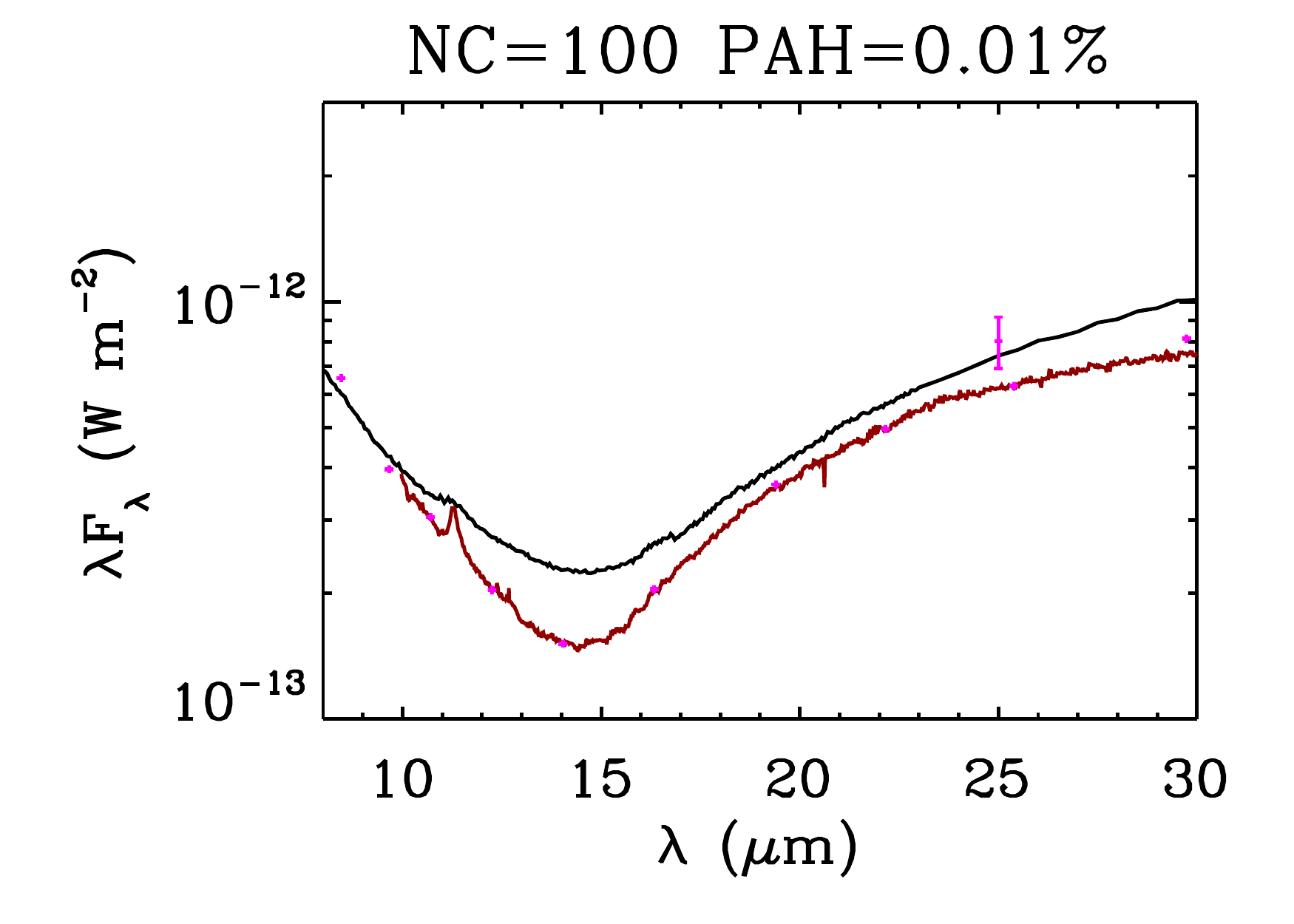} \\
\end{tabular}
\caption{Expected PAH spectrum (black) and {\it Spitzer} IRS spectrum (in red)
for different dust mass fractions of neutral PAHs in the outer disk.
In the rest of the paper, 
we used a 0.01\% PAH fraction. 
NC = number of carbon atoms. 
Magenta points are photometry points (see Appendix).}
\label{PAHspectrum}
\end{centering}
\end{figure}

\subsubsection{PAH content}
PAHs are an important ingredient for calculating the gas heating. 
To constrain the PAHs, 
we employed the {\it Spitzer-IRS} spectrum\footnote{For our analysis, 
we downloaded the Spitzer/IRS observations (AOR 3580672, PI: Houck)  from the Spitzer archive
and reduced the data again. 
The short-low data were reduced using the FEPS pipeline \citep[S18.18.0, see ][]{Bouwman2008}. 
The short-high and the long-high data were reduced with the c2d pipeline \citep[S18.18.0, ][]{Lahuis2006}.
For the high-resolution modules,
we used the PSF extraction method, which includes correction for pointing uncertainties.
The mid-IR spectrum of HD 135334B is characterized by the absence of the silicate feature at 10 $\mu$m,
weak PAH emission at 11.2 $\mu$m, and a lack of other PAH emission features 
\citep[][]{Geers2006, Maaskant2013}.}.
PAHs are implemented as a second dust component in MCFOST. 
Their properties (abundance and size) are passed to {\sc ProDiMo} in order to compute the gas heating 
due to the photoelectric effect.

Initially, we found that the observed 11.2 $\mu$m PAH feature 
could be reproduced with neutral PAHs with 21 carbon atoms (NC=21)
and 0.01\% of the dust in the outer disk in the form of PAHs.  
However, such PAHs generated too much gas heating that translated into 
[\ion{O}{i}]  63 $\mu$m line fluxes that are three to ten  times stronger 
than the observations for almost all the models reproducing the SED.
To have an [\ion{O}{i}] 63 $\mu$m emission compatible within a factor 3 of the observations, 
we increased the PAH size to NC=100 and used a 0.01\% fraction in mass of the dust in the form of PAHs 
($f_{\rm PAH}=6.0\times10^{-3}$ for a gas-to-dust ratio of 100, see Fig.\ref{PAHspectrum}).

\subsection{Family of models 2: An inner disk composed of a uniform mixture of amorphous carbon and astronomical silicates}
{To modify the continuum optical depth of the inner disk 
and to make it possible for the gas at larger radii (R$>>$ 1AU) to contribute to the CO~4.7 $\mu$m emission,
we introduced amorphous carbonaceous grains in the dust mixture in the family of models 2.}
Amorphous carbonaceous grains are commonly employed to fit SEDs of T Tauri and Herbig Ae/Be stars.
We assumed first a carbonaceous/silicate grains ratio constant with radius.
The fraction of carbon grains in the dust mixture sets the extension of the inner disk that fits the near-IR SED.
The greater the fraction of carbon grains, the larger the inner disk that reproduces the near-IR continuum. 
With a 25\% fraction of carbonaceous grains, 
an inner disk extending from 0.18 to 20 AU fits the near-IR SED
(see Model 2a in Table~\ref{Table_models}).
With a 5\% carbon fraction,
an inner disk extending from 0.18 to 3 AU is required (Model 2b).

\begin{figure*}[!ht]
\begin{center}
\begin{tabular}{p{7cm}rr}
\includegraphics[width=0.33\textwidth]{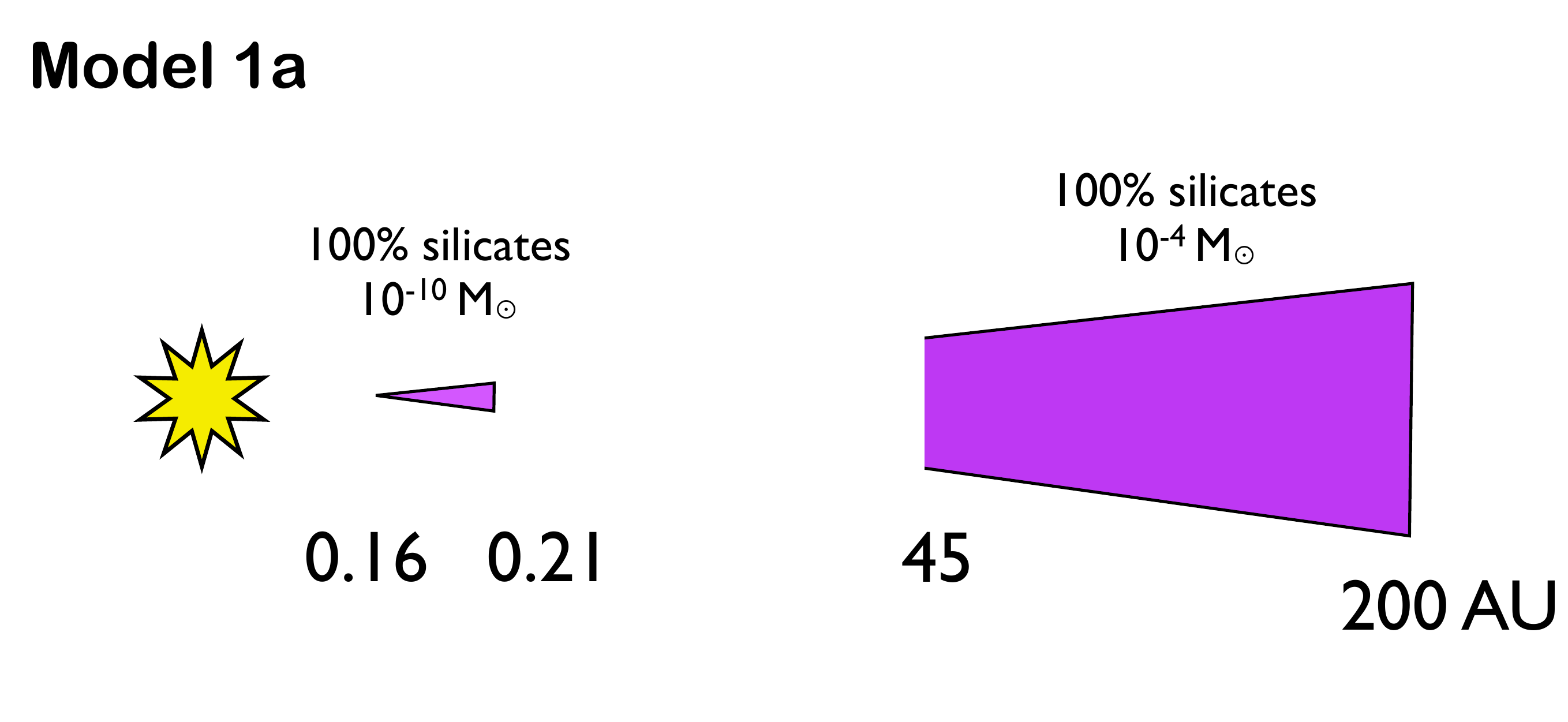} &
\includegraphics[width=0.18\textwidth]{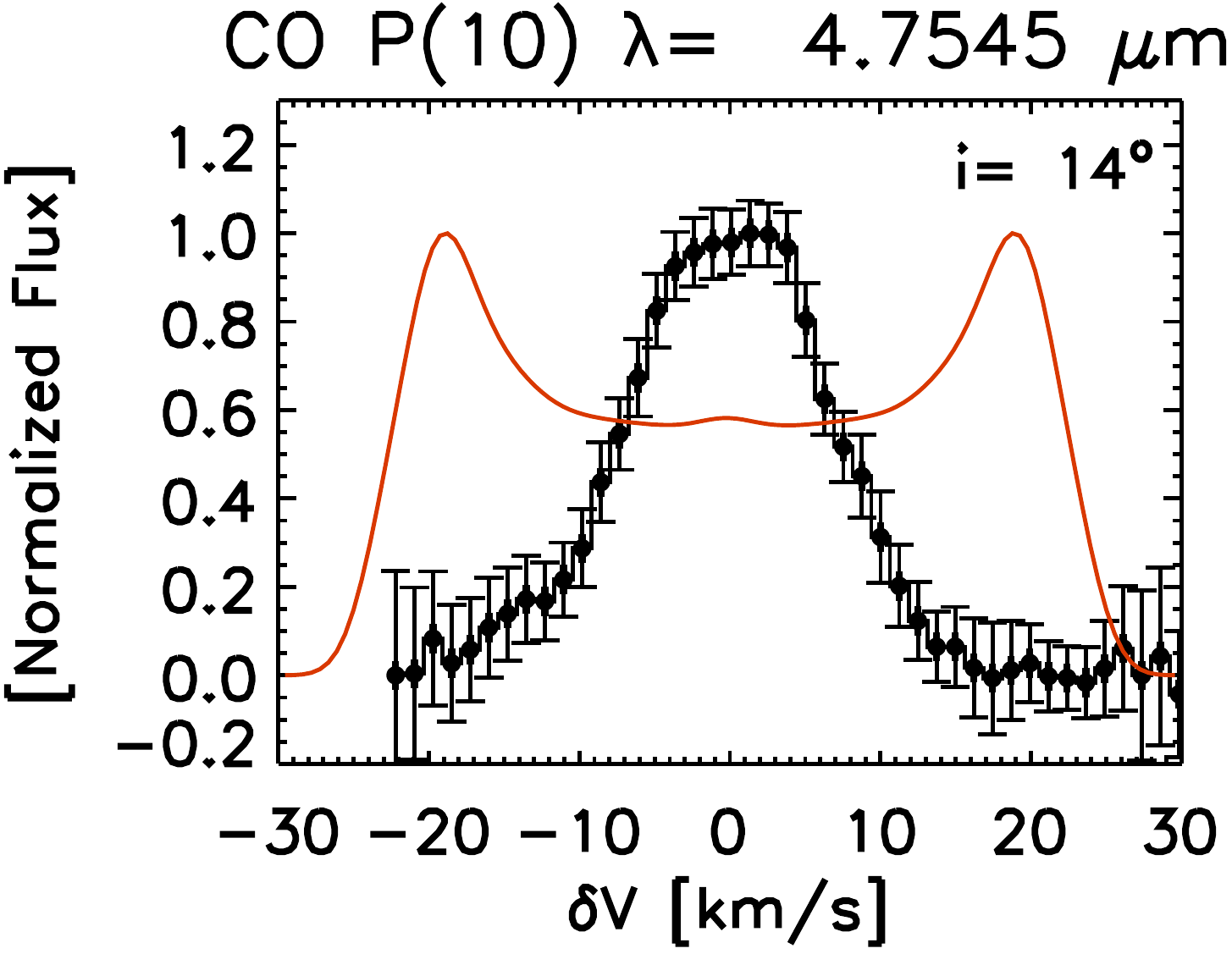} & 
\includegraphics[width=0.2\textwidth]{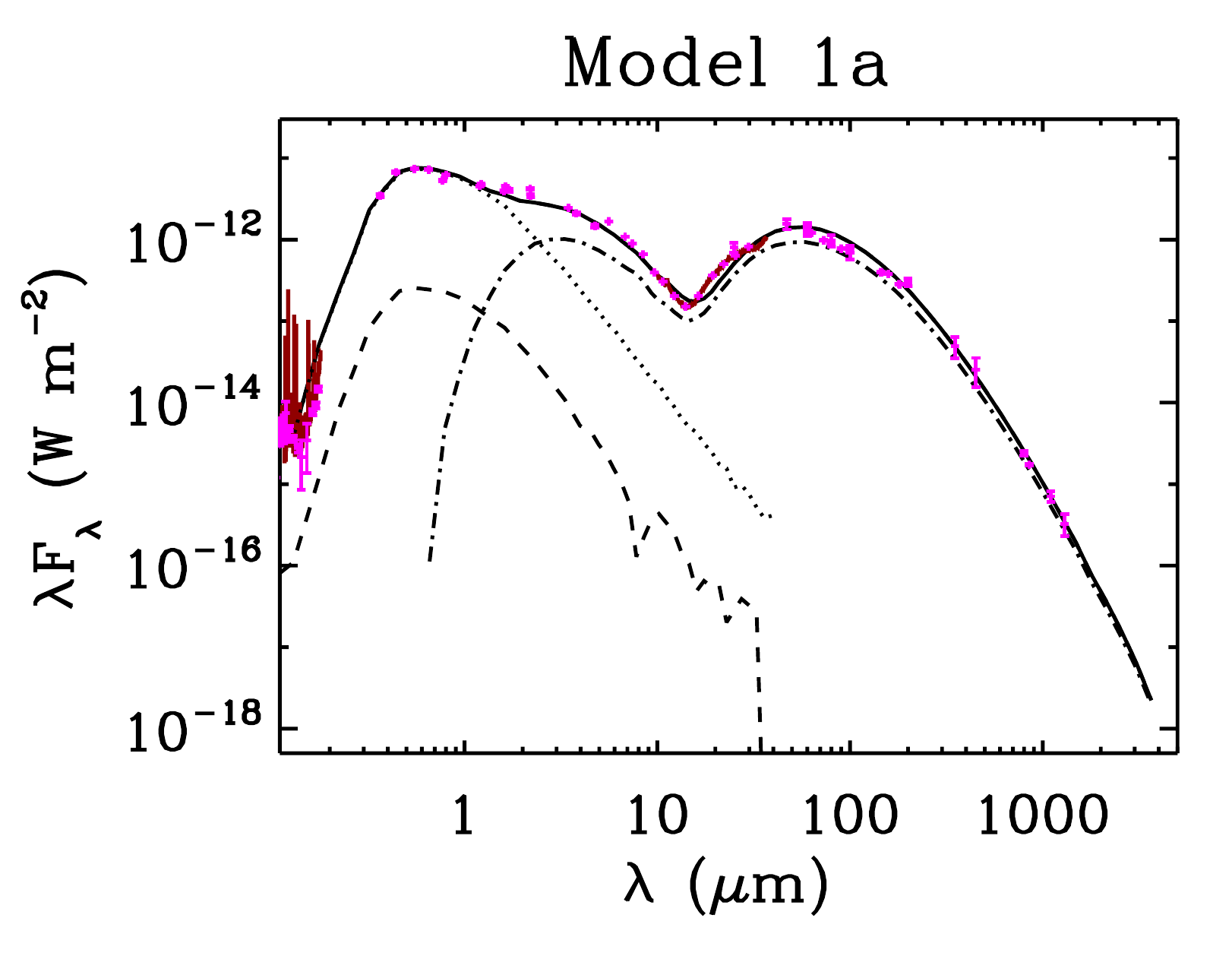} \\[2mm]
\includegraphics[width=0.33\textwidth]{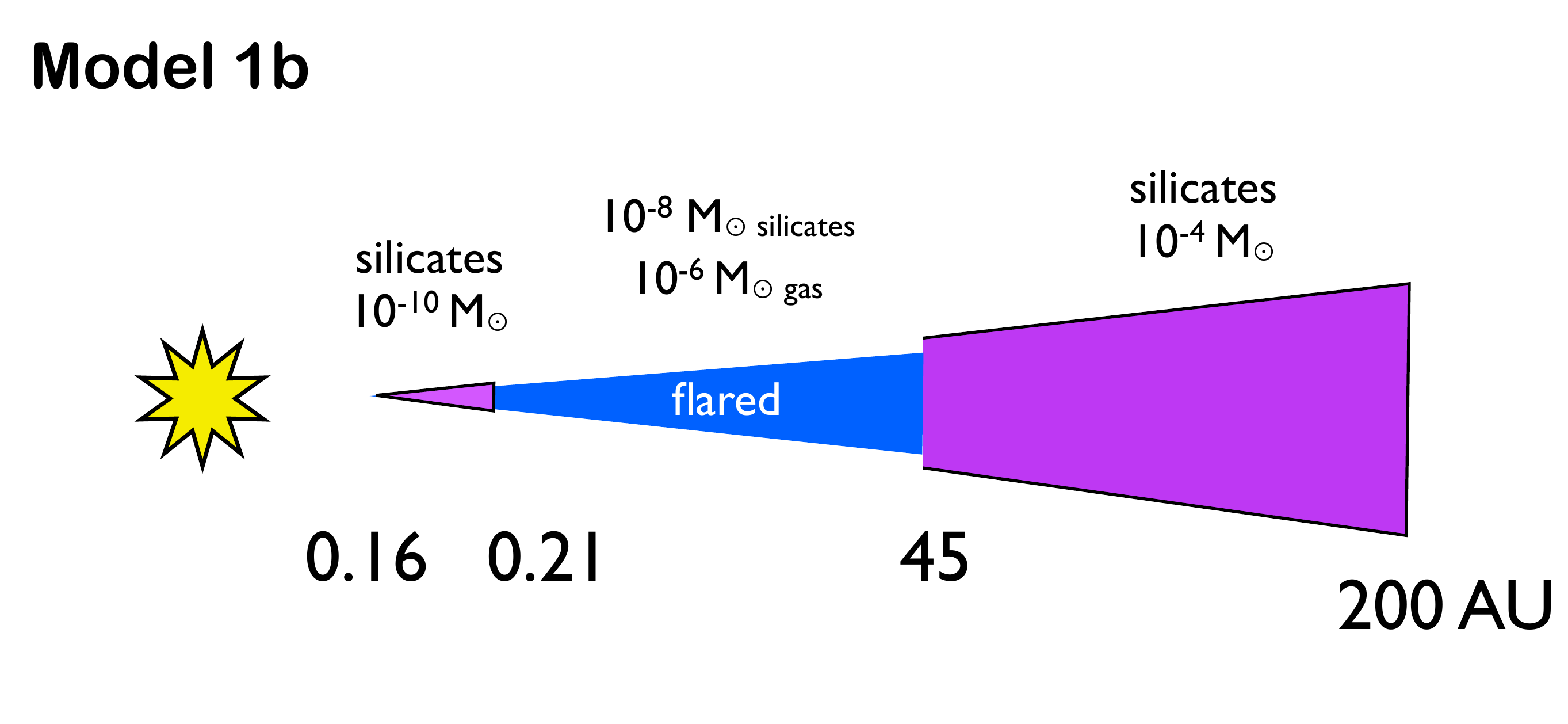} &
\includegraphics[width=0.18\textwidth]{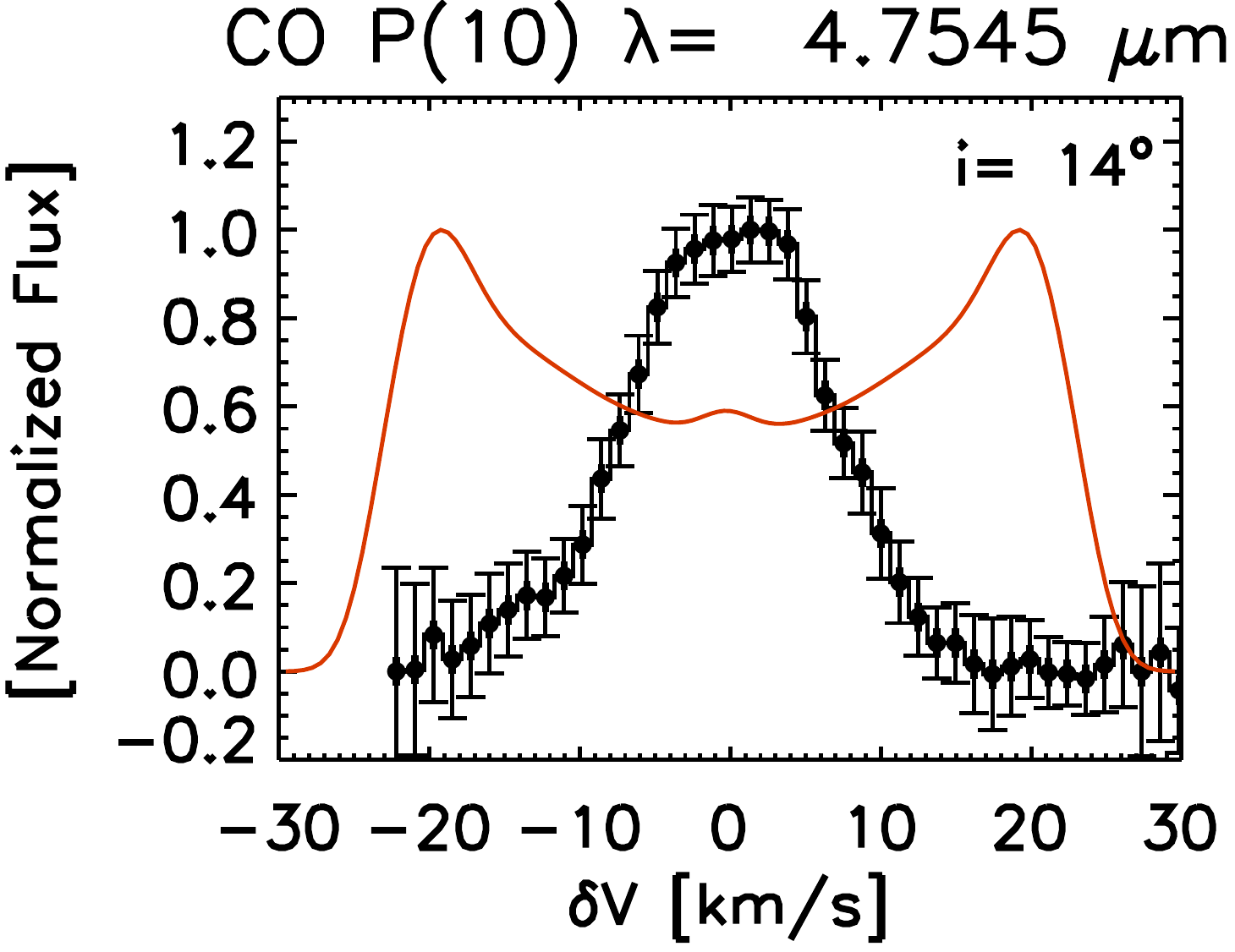} &
\includegraphics[width=0.2\textwidth]{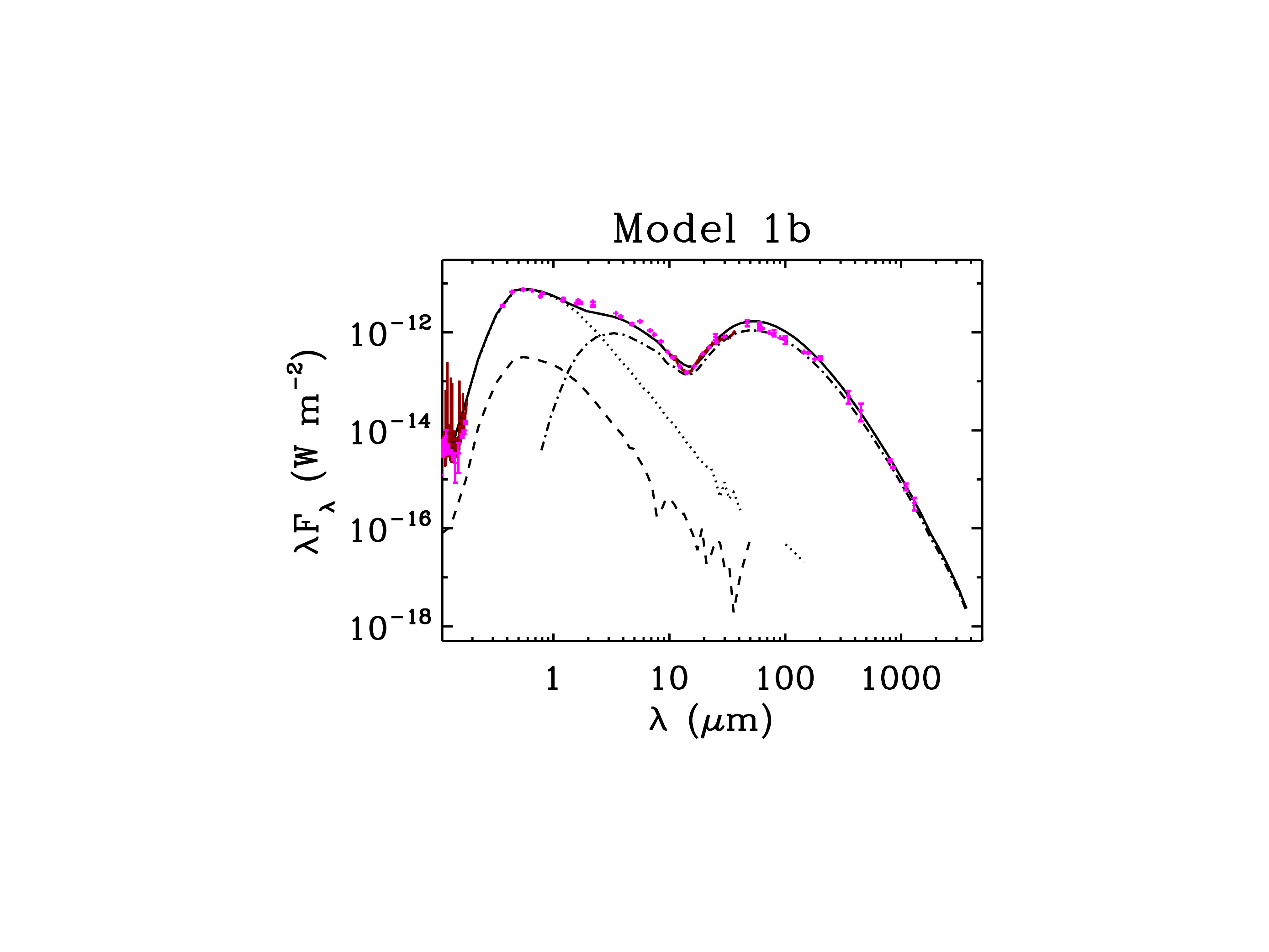} \\[2mm]
\includegraphics[width=0.33\textwidth]{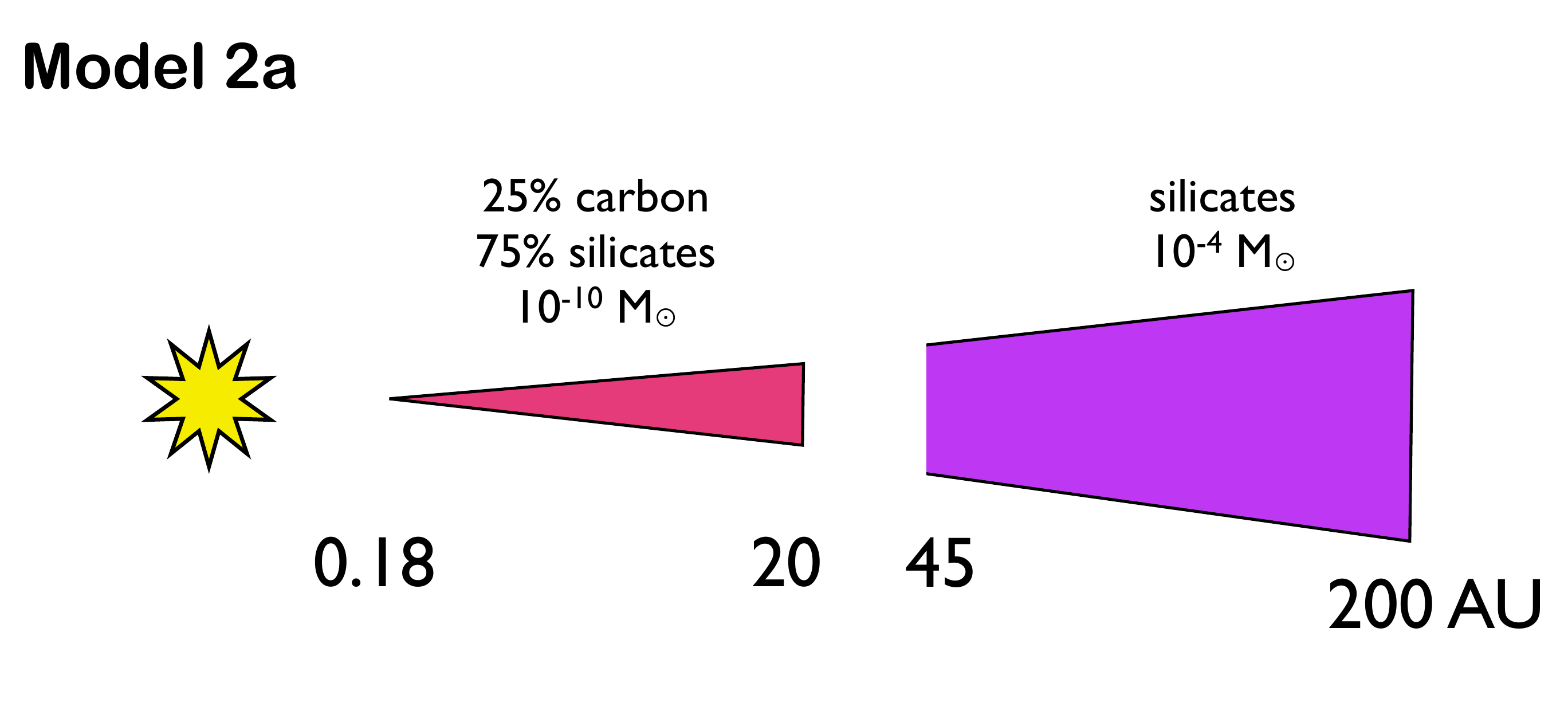} &
\includegraphics[width=0.2\textwidth]{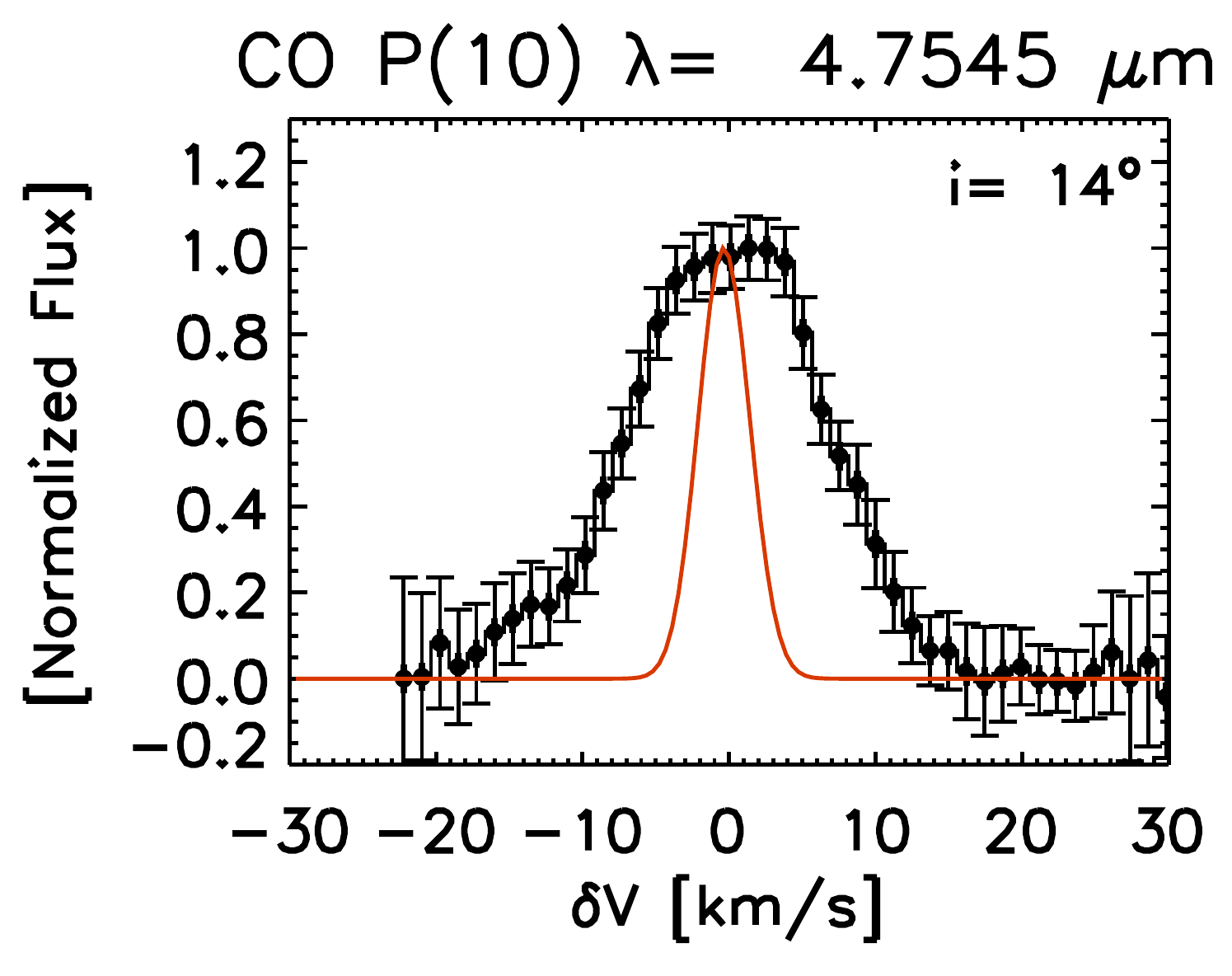} &
\includegraphics[width=0.18\textwidth]{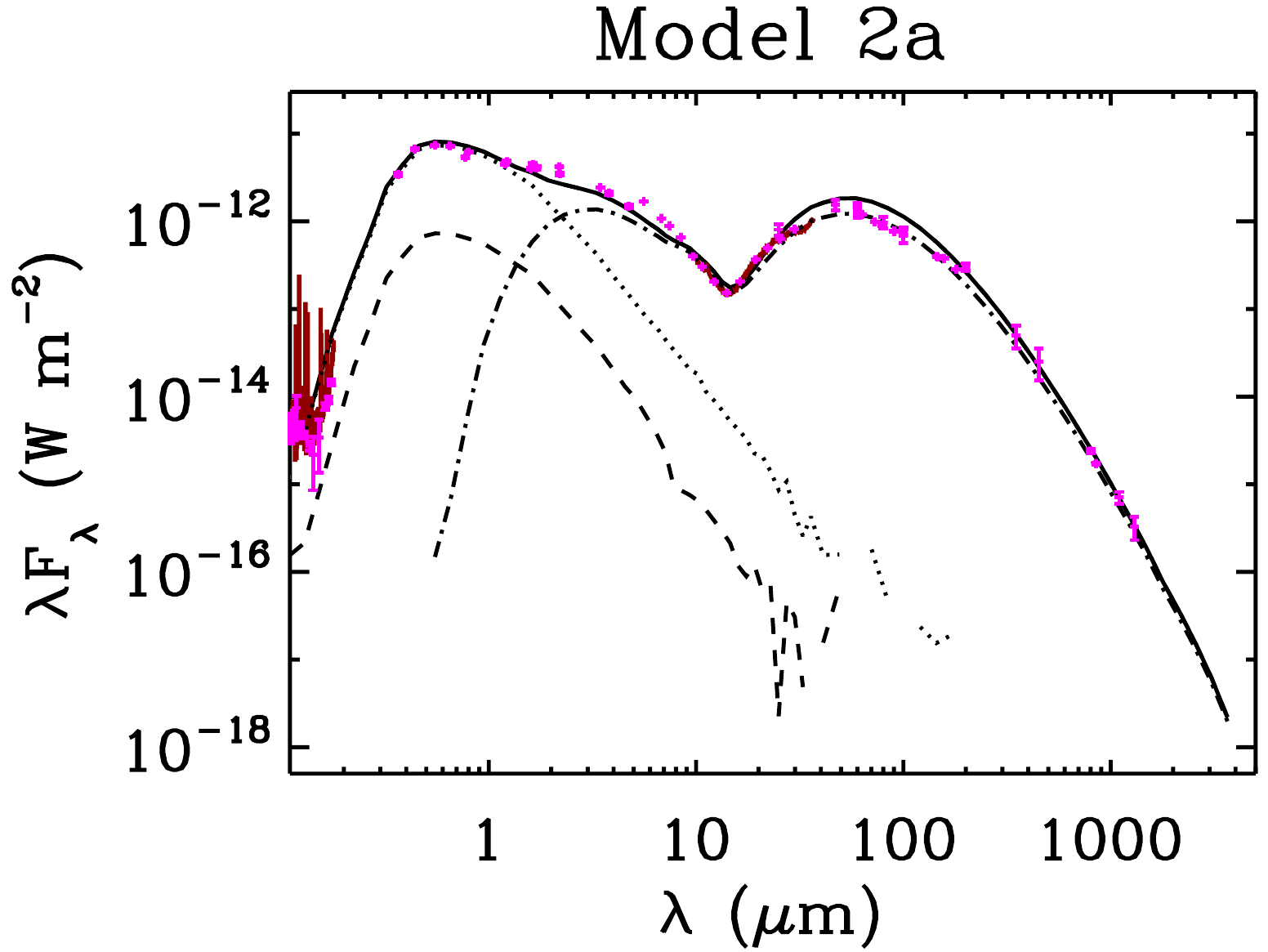}\\[2mm]
\includegraphics[width=0.33\textwidth]{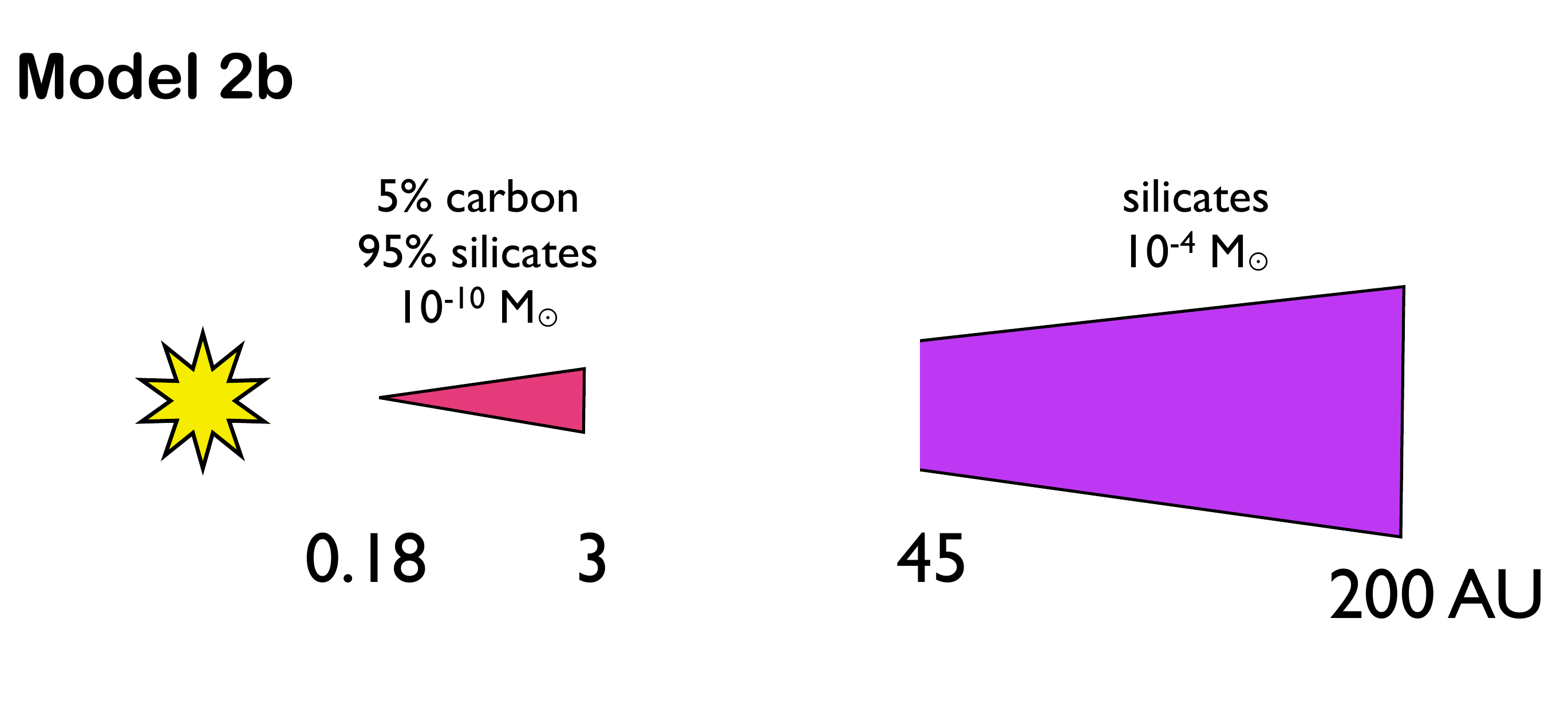} &
\includegraphics[width=0.2\textwidth]{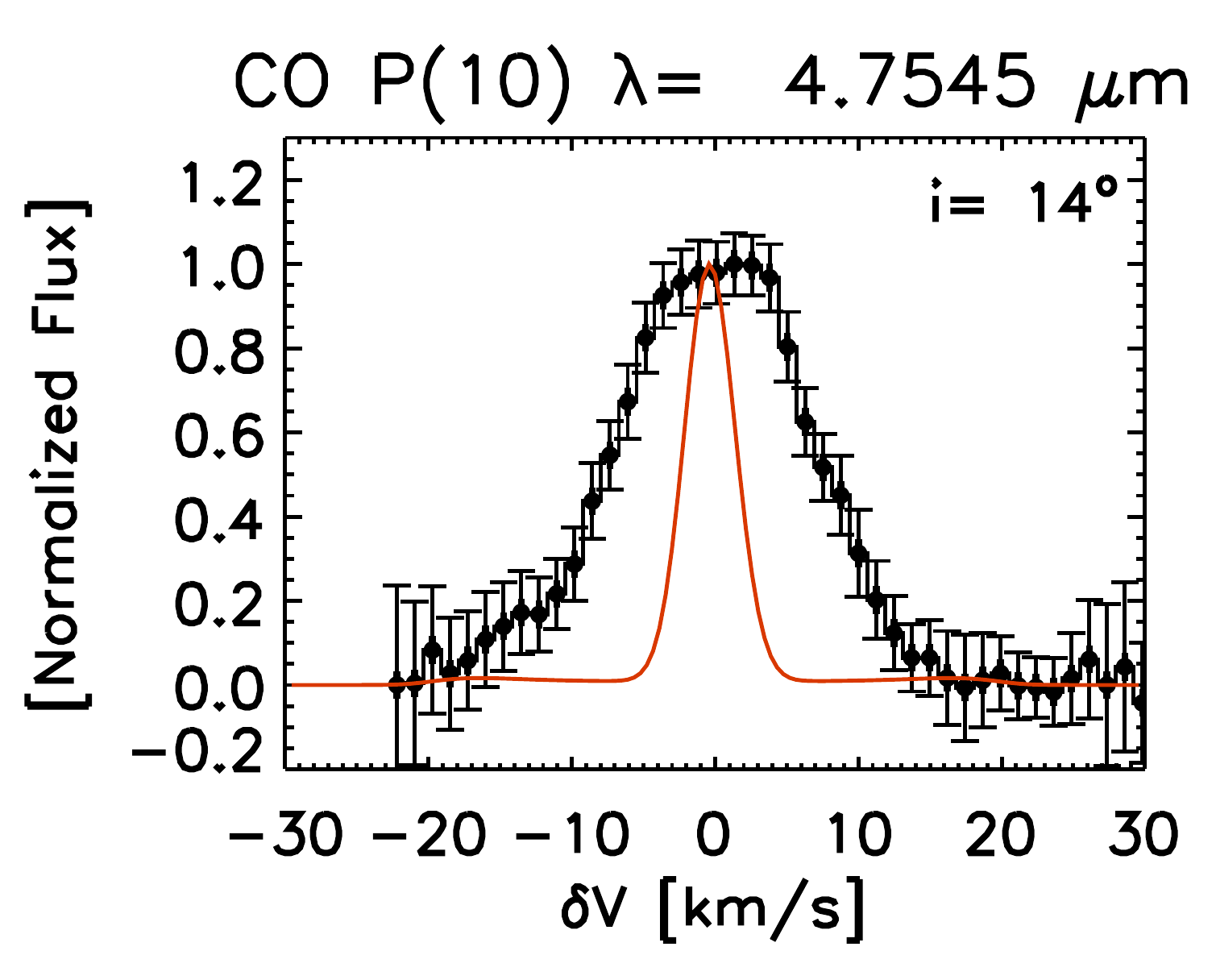}&
\includegraphics[width=0.18\textwidth]{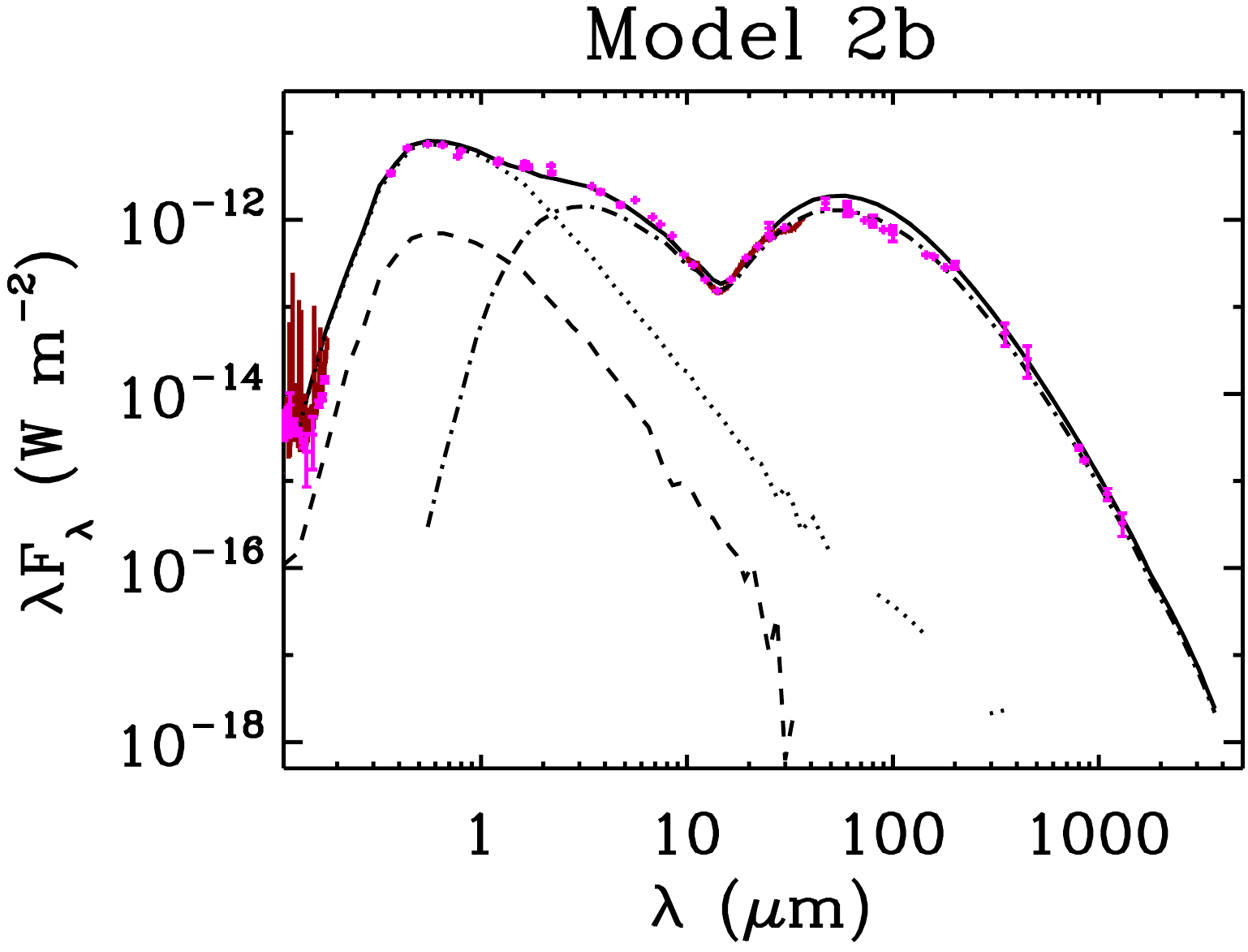}\\[2mm]
\includegraphics[width=0.33\textwidth]{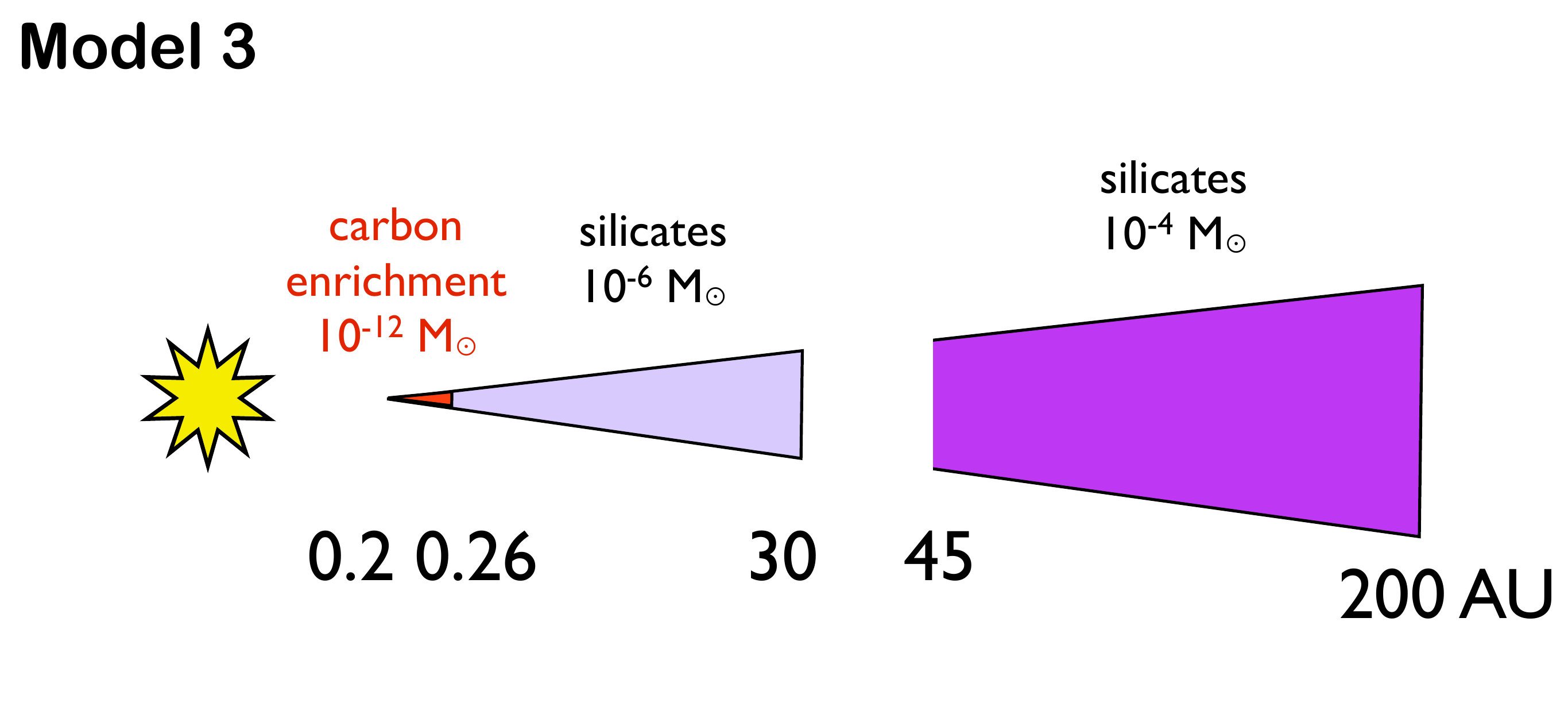} &
\includegraphics[width=0.18\textwidth]{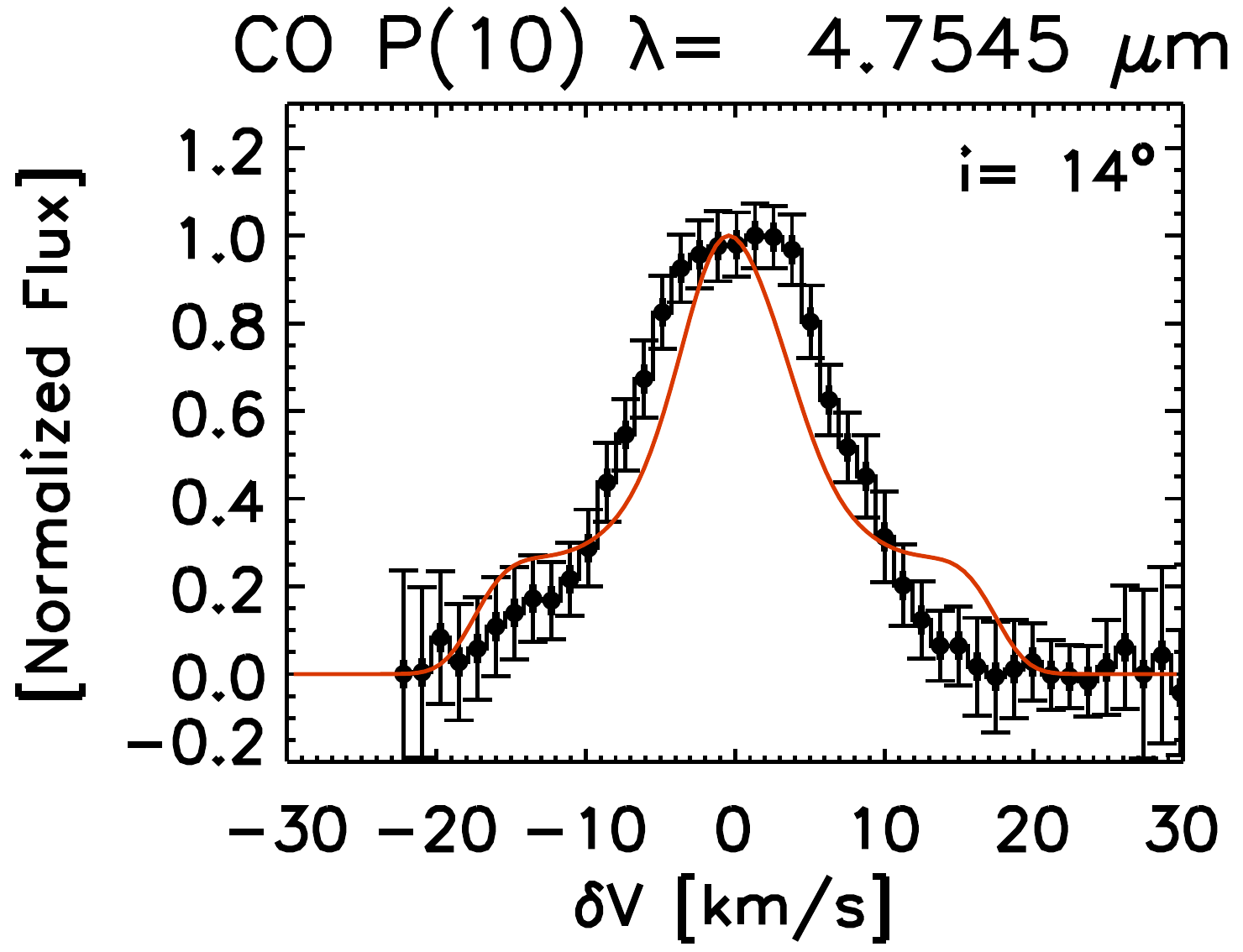}&
\includegraphics[width=0.18\textwidth]{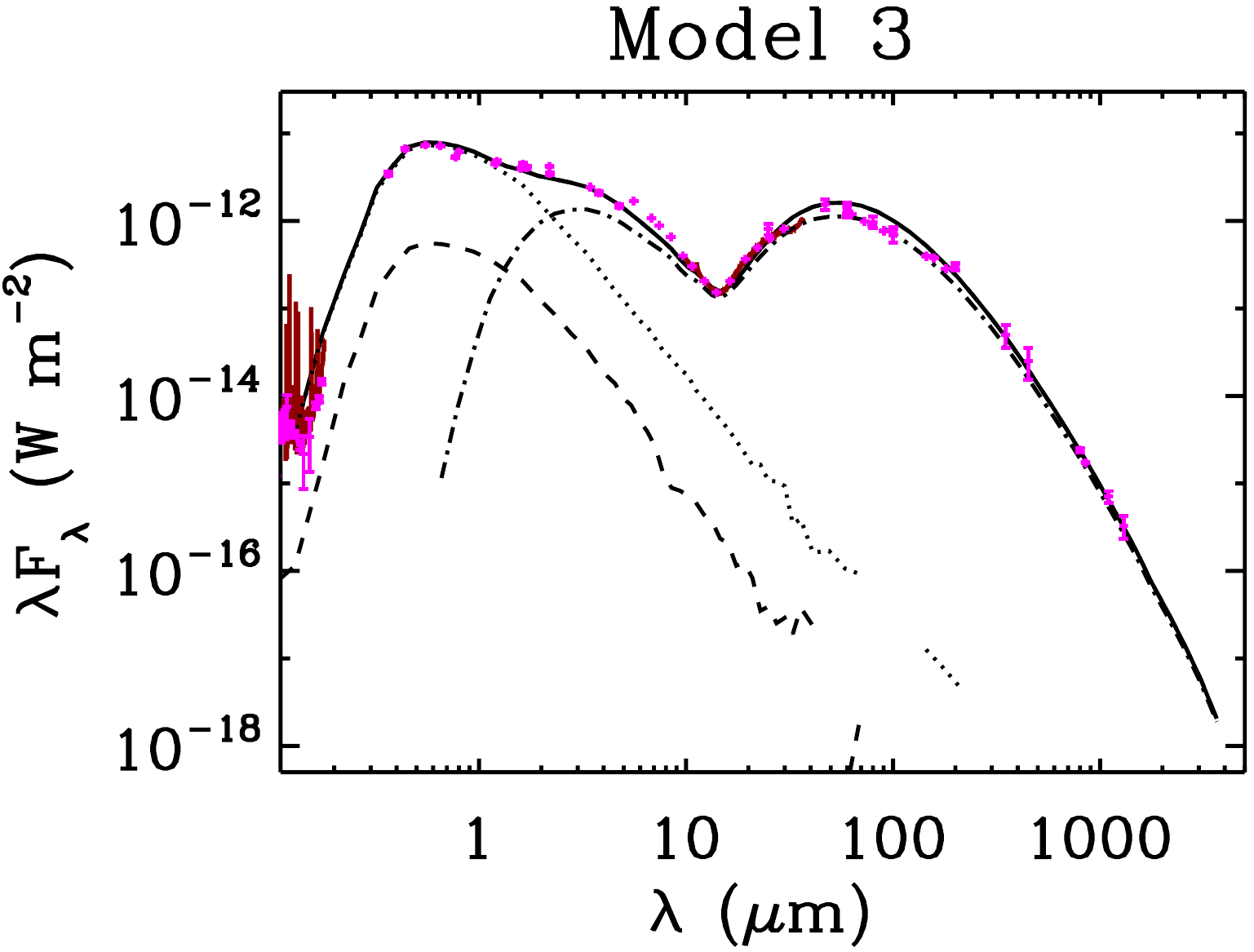}\\[2mm]
\includegraphics[width=0.33\textwidth]{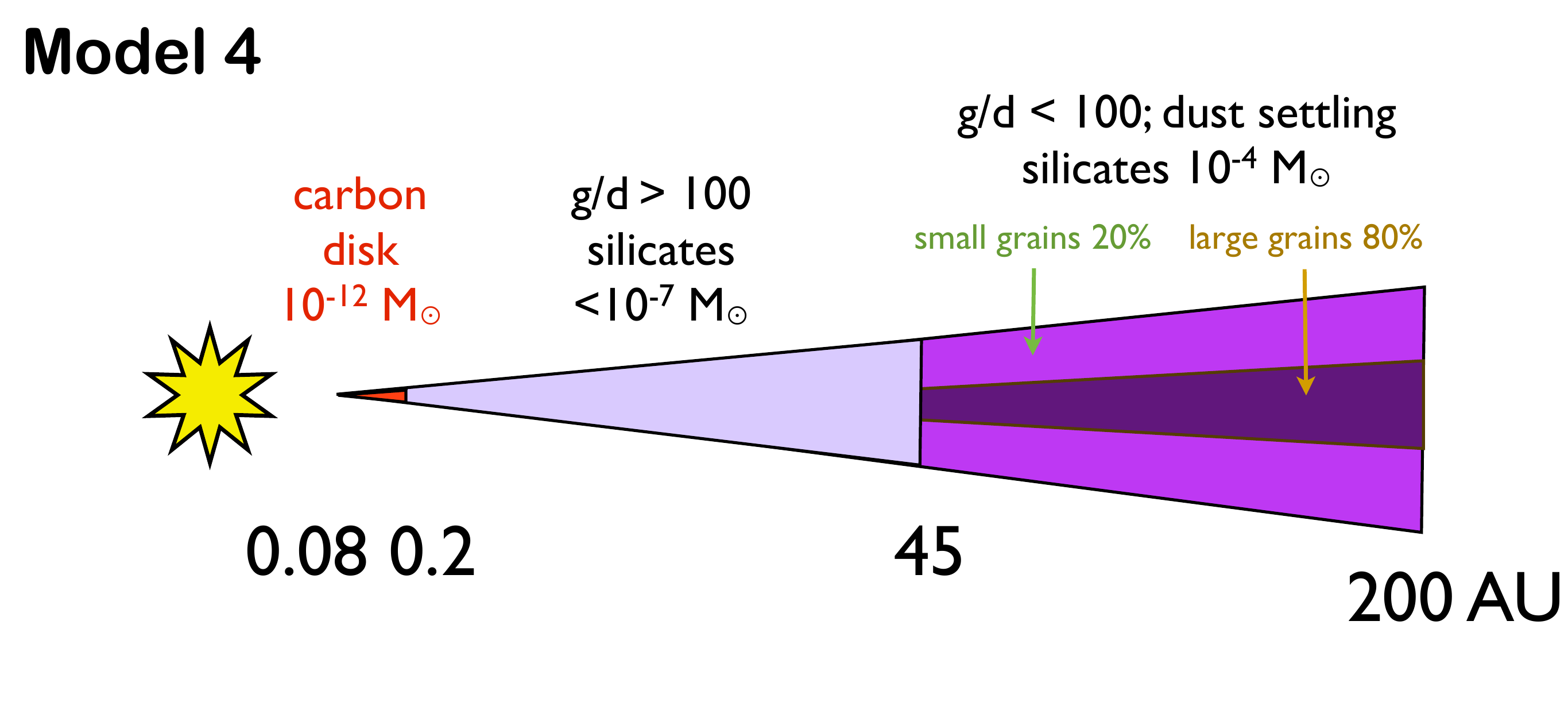} &
\includegraphics[width=0.18\textwidth]{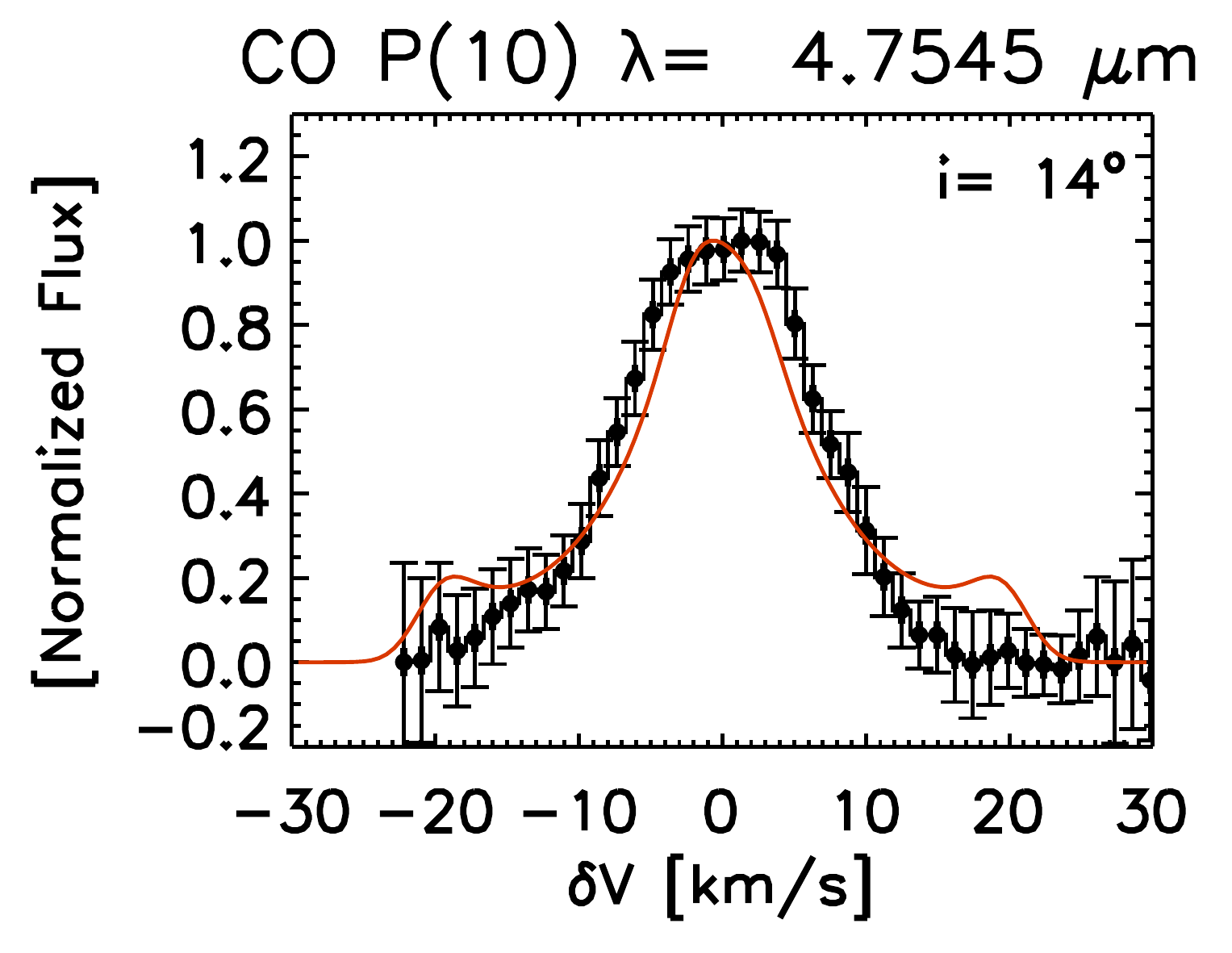}&
\includegraphics[width=0.18\textwidth]{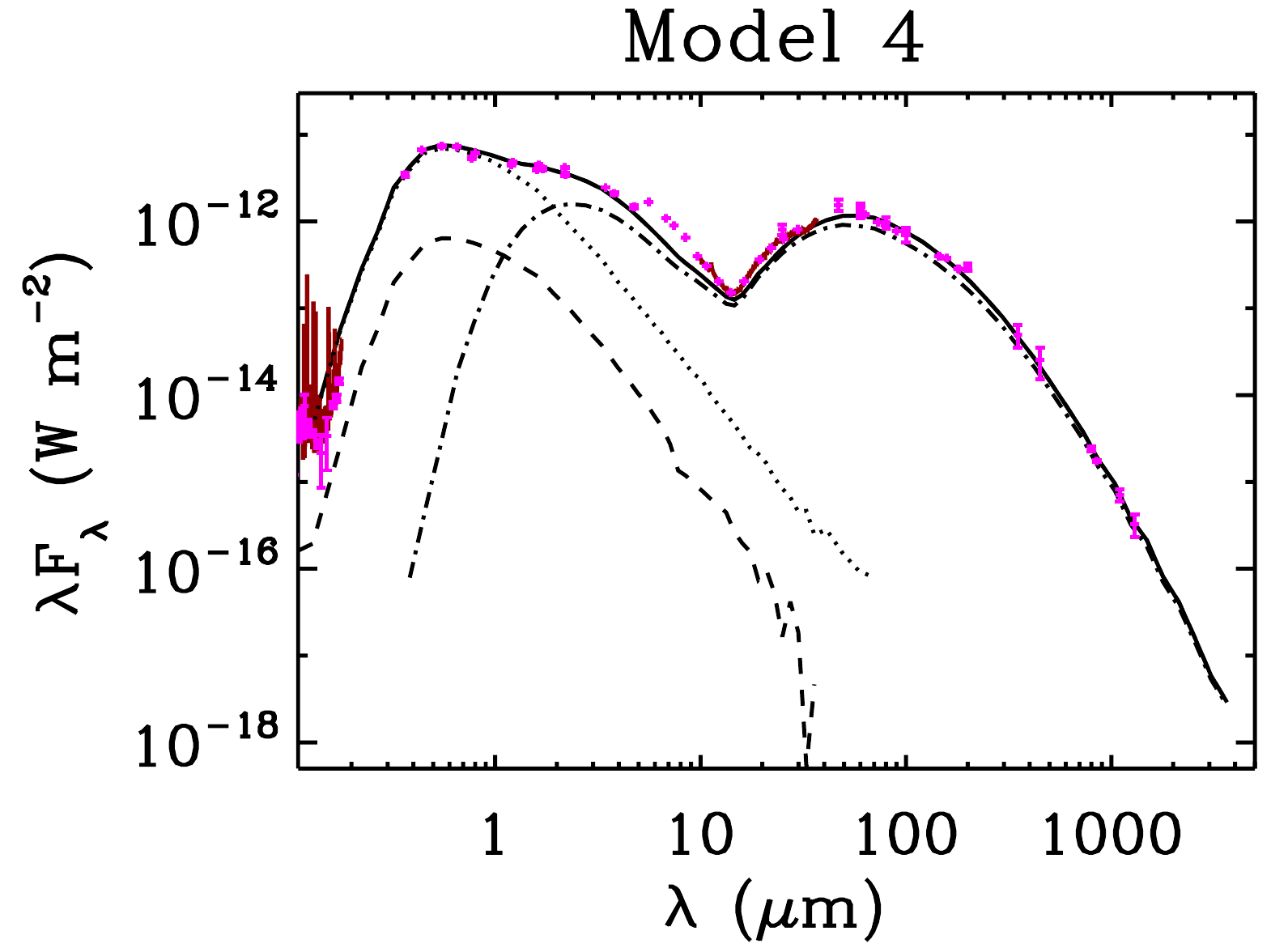}\\[2mm]
\includegraphics[width=0.33\textwidth]{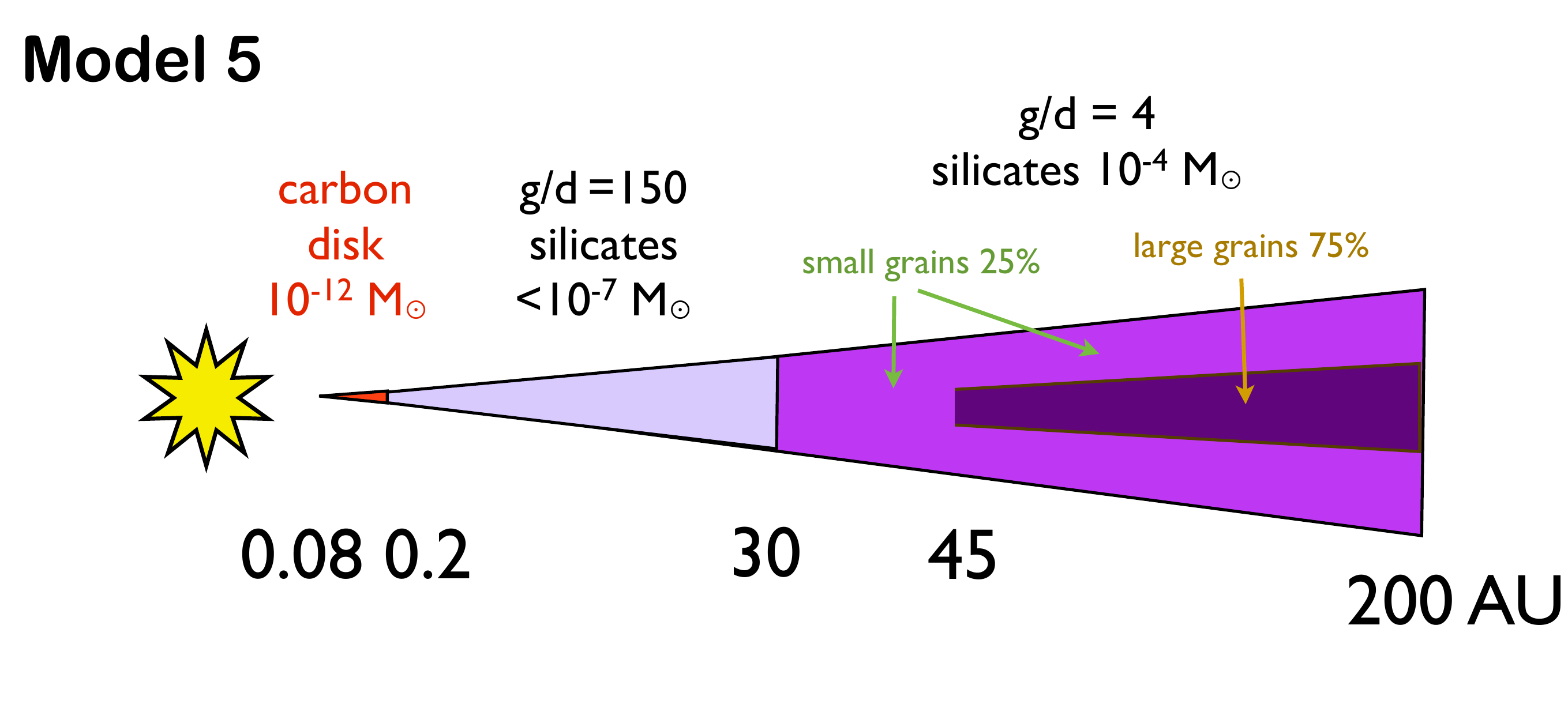} &
\includegraphics[width=0.18\textwidth]{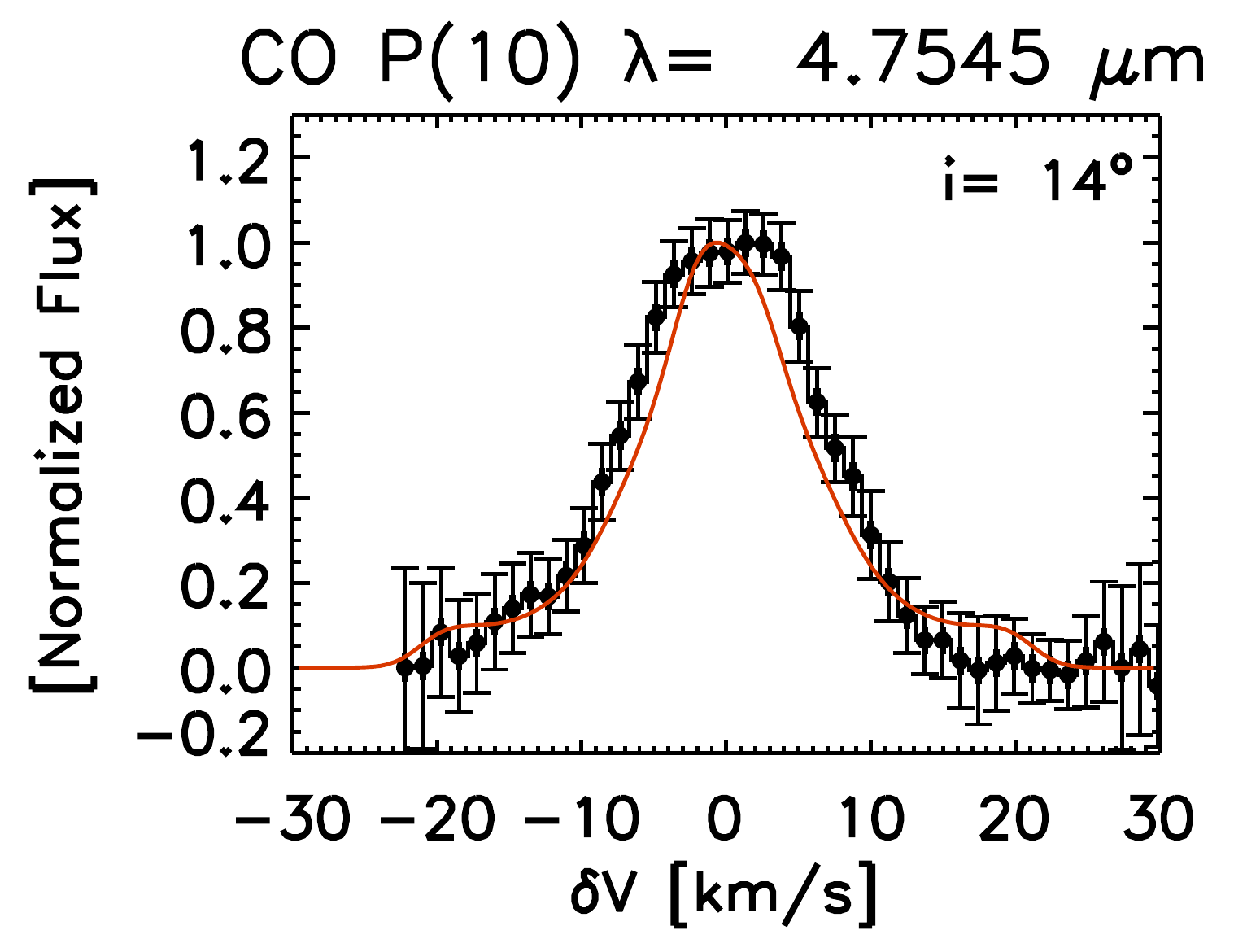}&
\includegraphics[width=0.18\textwidth]{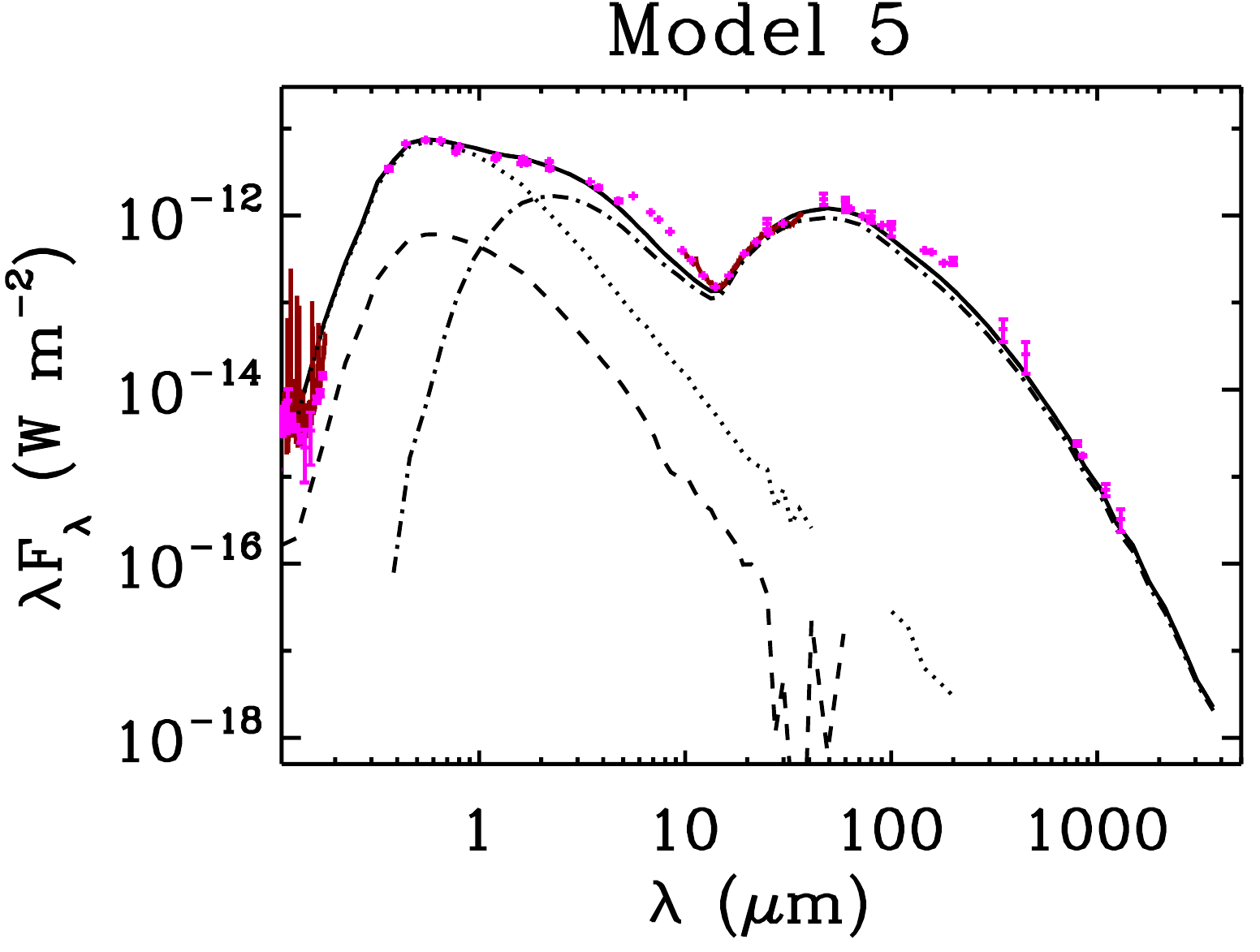}\\[2mm]
\end{tabular}
\caption{Cartoon displaying the disk structure of the family of models tested, together with the 
CO P(10) line profile and SED predicted.
The CO P(10) line profile includes the effects of the slit, and it is displayed for PA of the slit of 180\degr.
A representative example of each family of models is shown. 
In the SED plots, 
the dotted line is the star emission, 
the dash-dotted line is the dust thermal emission, and the dashed line is the 
dust's scattered light emission.
Details of the models are presented in Table~\ref{Table_models},
and the predicted lines fluxes are given in Table~\ref{Table_line_fluxes}.
}
\label{all_models}
\end{center}
\end{figure*}

The introduction of amorphous carbon grains 
reduced  the total dust mass required to fit the near-IR SED by a factor of a few,
allowed for a dust size distribution with smaller grains in the inner disk (and fit the lack of 10~micron silicate feature),  
and drastically changed the inner disk's optical continuum depth.
These changes resulted in an optically thin inner disk at 4.7~$\mu$m,  
thus a CO ro-vibrational emission entirely dominated by emission from
the inner rim of the outer disk,
and a very narrow single peaked CO ro-vibrational line profile with a width of a few km/s,
inconsistent with the observations (see Fig.~\ref{all_models}).

\subsection{Family of models 3: An inner disk with a radial-dependent mixture of carbon and silicates \label{model_selection}}
The results of the families of models 1 and 2 suggested that the solution was an intermediate inner disk structure between
a narrow disk of 100\% silicates and an extended inner disk with a large amount of carbonaceous grains.
{ In the family of models 3, we allowed for a radial-dependent carbonaceous/silicate grains ratio}.

We found that an inner disk of tens of AU composed of silicates, but enriched with a small amount (10$^{-12}$ M$_{\odot}$)
of amorphous carbon grains at small radii ($0.2<R<0.26$ AU) is able to reproduce the near-IR 
SED,  while at the same time it make it possible for the CO located at several AU to contribute significantly to the CO ro-vibrational line flux,
thus reproducing the observed CO P(10) line profile (see Fig.~\ref{all_models}).   

Assuming a gas-to-dust ratio of 100 for the whole disk and a carbon-enriched inner disk,
we tested a large number of models ($\sim$50\,000 MCFOST, $\sim$1\,000 {\sc ProDiMo}) 
by varying the geometry (i.e., $h/r$, flaring, surface density exponent) and dust mass for the inner and outer disk.
We found that the CO ro-vibrational line profile is reproduced by an inner disk extending tens of AU 
in which the {\it surface density is flat or increases as a function of the radius} 
(i.e., a power-law surface density with a positive exponent).
Solutions with flared and anti-flared outer disks were found.

\begin{figure}[t]
\begin{center}
\includegraphics[width=0.50\textwidth]{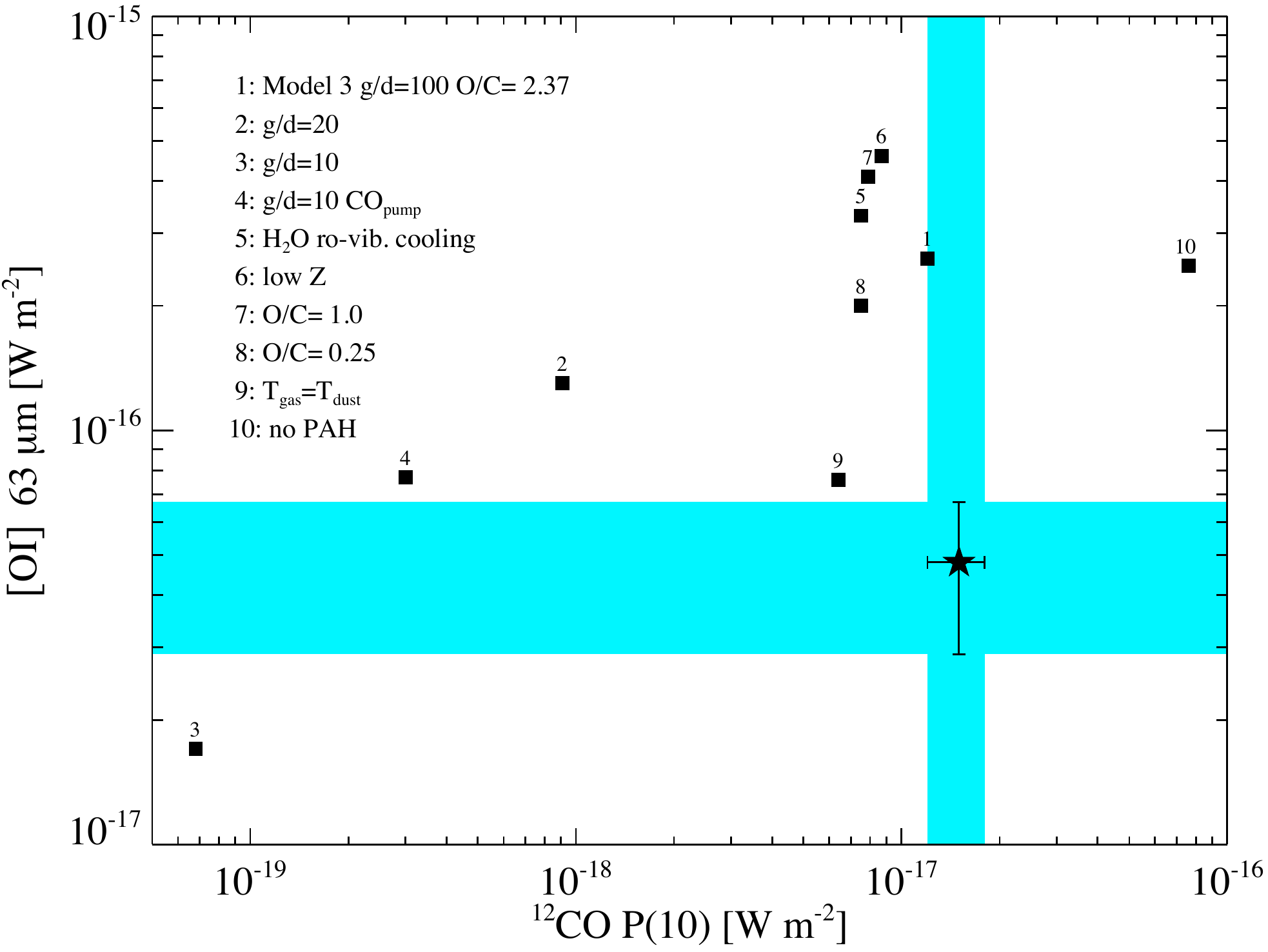}
\caption{Effect of changing different {\sc ProDiMo} assumptions in the  [\ion{O}{I}] 63 $\mu$m  and CO P(10) line fluxes in one representative example of the family of models 3. 
The star indicates the observed line fluxes. 
The cyan intervals represent the 20\% error on the CO P(10) line flux and 40\% error in the [\ion{O}{I}] 63 $\mu$m line flux.
In the legend the symbols mean: 
g/d: gas-to-dust ratio;
CO$_{\rm pump}$: CO ro-vibrational emission calculated including UV fluorescent excitation;
H$_2$O$_{\rm cooling}$: gas temperature calculated including ro-vibrational water cooling (pure rotational H$_2$O cooling is always taken into account);
low Z: low disk metallicity;
O/C : oxygen over carbon abundance ratio;
T$_{\rm gas}$ = T$_{\rm dust}$: maximum possible cooling;
no PAH: gas heating calculated without the effect of PAHs.
}
\label{OIvsCOP10_change} 
\end{center}
\end{figure}

Several Models 3 reproduced the CO P(10) flux and line profile,
the H$_2$ infrared lines upper-limits, and the CO sub-mm and mm line fluxes 
(see one example in Tables~\ref{Table_models} \& \ref{Table_line_fluxes}).
However, in all Models 3 calculated (covering a wide range of geometries),
the emission of [OI] 63 and 145 $\mu$m, and the [\ion{C}{ii}] line at 157~$\mu$m were 
overpredicted by factors over five.
We tested several options available within {\sc ProDiMo} (see Fig.~\ref{OIvsCOP10_change}),
such as global gas-to-dust ratios lower than 100, 
H$_2$O ro-vibrational cooling (pure H$_2$O rotational cooling is calculated by default), 
lower metallicity, no PAHs,
extremely low O/C abundances, but in most of the cases the [\ion{O}{i}] 63 and 145 $\mu$m lines were still too strong. 
We found that the only effective ways 
to significantly decrease the [\ion{O}{i}] 63 and 145 $\mu$m line fluxes were
to assume T$_{\rm gas}$=T$_{\rm dust}$ 
or to {\it significantly} decrease the gas-to-dust ratio.
The first is an extreme case of gas cooling that is unrealistic, 
as we know that T$_{\rm gas}>$T$_{\rm dust}$ in the disk's surface layer
\citep[e.g.,][]{kampdullemond2004}.
The second had the limitation that when
the {\it global} gas-to-dust ratio was low enough to describe the  [\ion{O}{i}]~63~$\mu$m line,
the CO ro-vibrational line was two orders of magnitude weaker. 
{ This last result indicated that the gas-to-dust ratio should be different for the inner and outer disk}.

\begin{figure} [!ht]
\begin{center}
\begin{tabular}{c}
{\bf \large ~~~~~~~\sffamily Model 3}\\[3mm]
\includegraphics[width=0.47\textwidth]{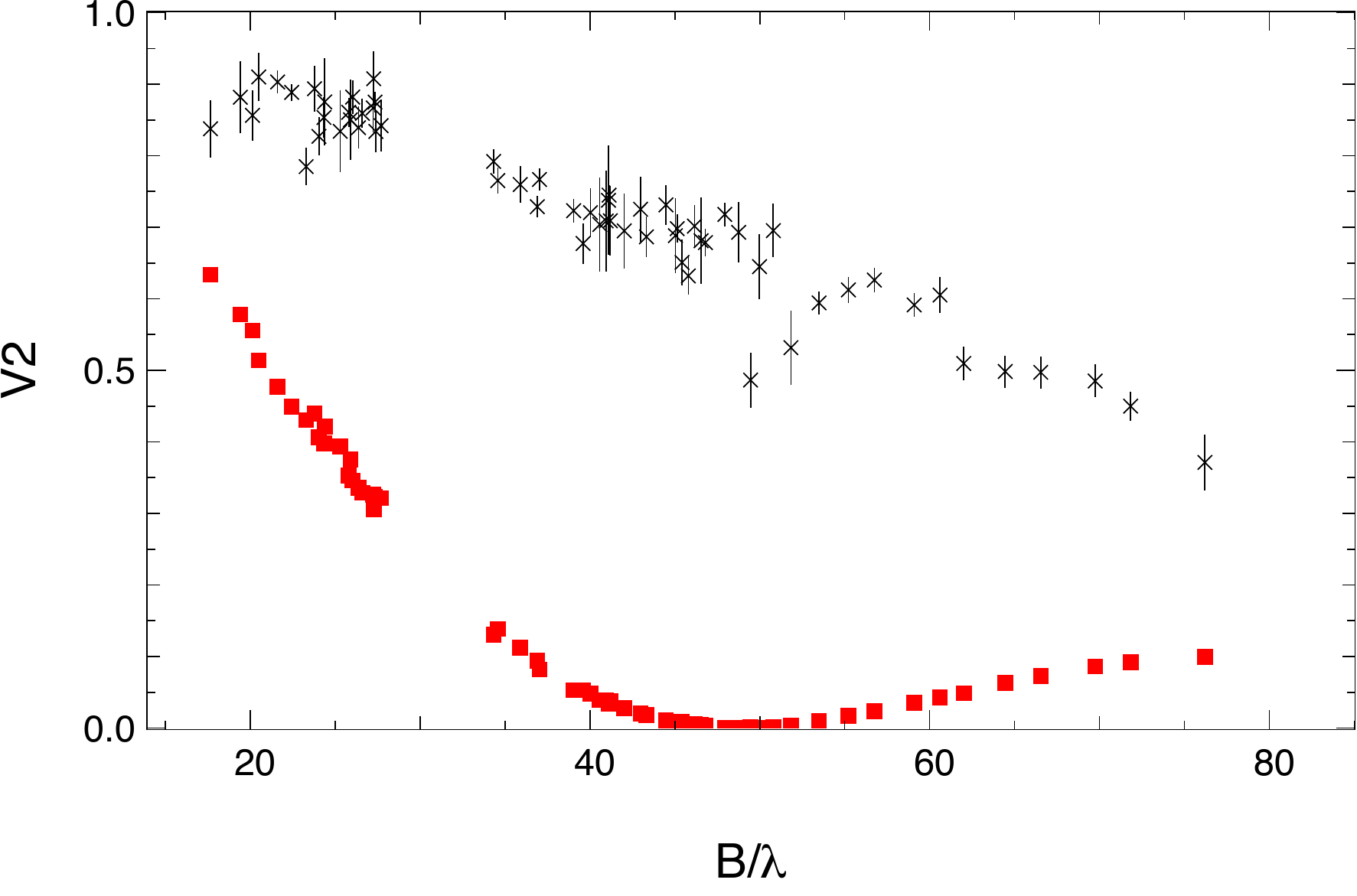}\\
\\[2mm]
{\bf \large ~~~~~~~\sffamily Model 4}\\[3mm]
\includegraphics[width=0.49\textwidth]{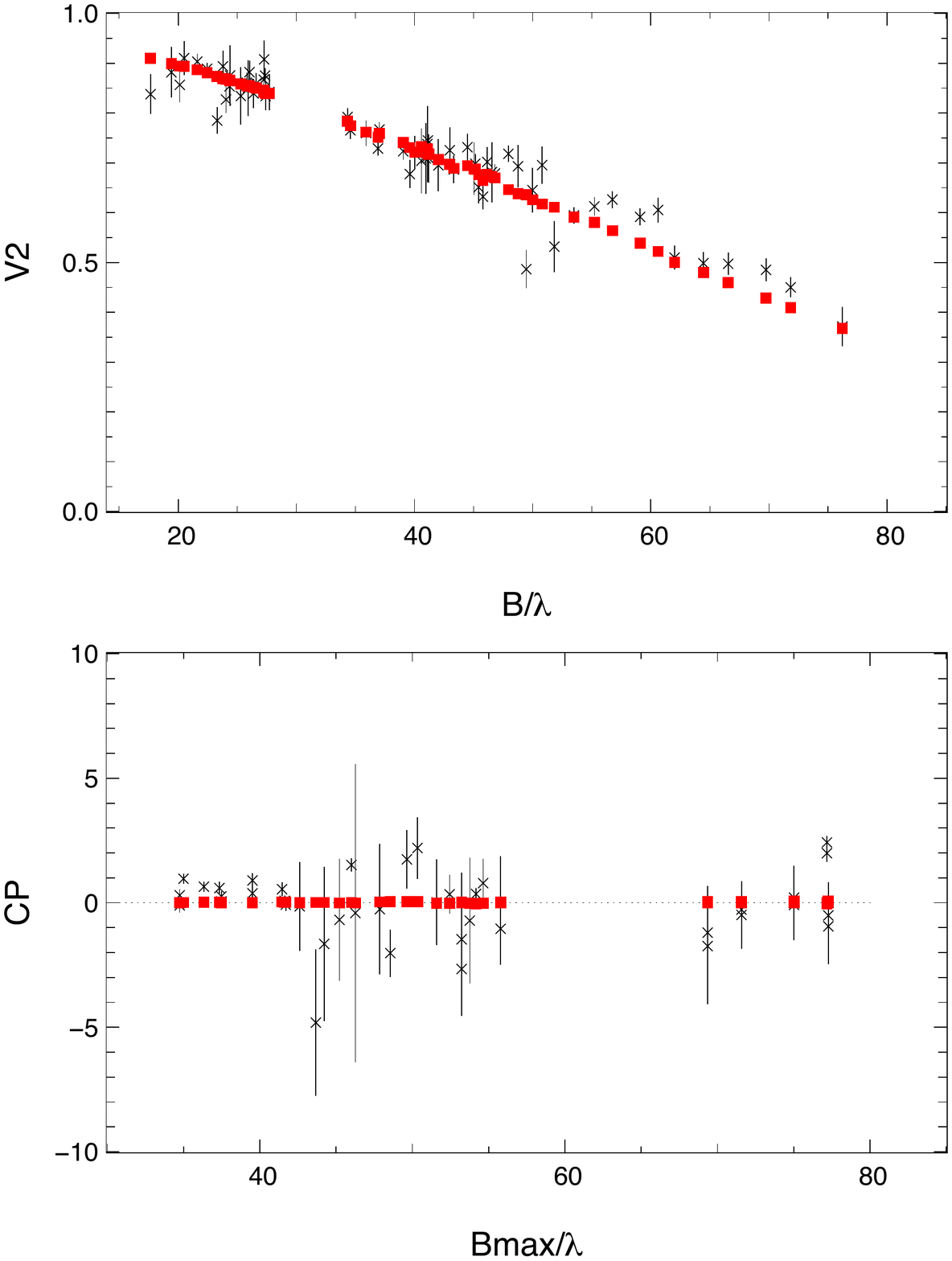}
\end{tabular}
\caption{
{\it Upper panel:}  Squared visibilities {\it (in red)} of Model 3 
overplotted onto the observed  1.6 $\mu$m VLTI/PIONIER  visibilities {\it(in black)}.
An inner disk starting at 0.2 AU is not consistent with observations.
VLTI/PIONIER data clearly shows that there is material at smaller radii (i.e., inside the silicate sublimation radius).
{\it Central panel:} squared visibilities  {\it (in red)} predicted by Model 4 ($R_{\rm in}=0.08$~AU).
 {\it Lower panel:} closure phases (CP) predicted by Model 4 {\it (red)} and  measured by VLTI/PIONIER {\it (black)}.  }
\label{PIONIER} 
\end{center}
\end{figure}

Another limitation of the family of models 3 was the fit to the PIONIER near-IR visibilities. 
In a forthcoming paper (Benisty et al. in prep),
ESO-VLTI PIONIER near-IR observations of HD~135344B
and other transition disks will be presented in detail. 
In Fig.~\ref{PIONIER}, 
we present the observed VLTI/PIONIER squared visibilities at 1.6 $\mu$m
and the predicted visibilities of Model 3.
We see clearly that an inner disk starting at $R_{\rm in}=$ 0.2 AU
is not compatible with the interferometry data. 
The near-IR continuum in Model 3 is produced too far out.

\subsection{Family of models 4: Carbon grains inside the silicate sublimation radius, 
different gas-to-dust ratios for the inner and outer disk, and
dust settling in the outer disk.}
 
\subsubsection{The location of the carbonaceous dust} 

In the family of models 3, the carbonaceous grains were located at R$>$0.2~AU to 
be consistent with the assumption that the temperature in the inner rim 
should not be higher than the silicate grains sublimation temperature 
\citep[T$\sim$1\,500 K, for $n_{\rm H} \gtrsim 10^{16}$ cm$^{-3}$, e.g.,][]{Helling2001}.
However, carbonaceous grains can survive temperatures higher than 
1\,500 K and up to 2\,000K \citep[see for example,][]{Kobayashi2011}.
Therefore, one interesting possibility for physically justifying the inner-disk 
carbon enrichment is that the carbonaceous grains are located inside 0.2 AU, 
the region where silicate grains sublimate. 
The survival of carbonaceous grains in this region is discussed in Section~\ref{carbon}.

Therefore for the family of models 4, 
we located the carbonaceous grain component at $0.08<R<0.2$ AU,
$R_{\rm in}$ equal the corotation radius,
and $R_{\rm out} $ equals the silicates sublimation radius.  
The astronomical silicate-grain component of the inner disk was set to start at 0.2 AU.
With this inner disk configuration we obtained an excellent match to SED 
and the PIONIER visibilities and closure phases (see  Fig.~\ref{PIONIER}).
In our models the temperature at the innermost radius is $\sim$2\,000K.

\subsubsection{Decreasing the [\ion{O}{I}] 63~$\mu$m emission}

One limitation of the family of models 3 was that it produced [\ion{O}{i}] 63~$\mu$m line fluxes that are too strong.
The [\ion{O}{i}] 63 $\mu$m line is produced partly by gas inside the cavity between 10 and 45 AU, 
but principally by gas in the outer disk between 45 and 60 AU. 
Since the surface density and gas mass in the inner disk are set by CO ro-vibrational line,
to lower the [\ion{O}{i}] line fluxes, three modifications to the family of models 3 were 
introduced in the family of models 4:
\begin{enumerate}
\item the gas-to-dust ratio in the outer disk was allowed to be lower than 100,
\item the surface density of the outer disk needed to be shallower, (We obtained good solutions with $q=$-1.0, 
similar to the surface density power law exponent found by \citet[][]{Andrews2011} from SMA sub-mm continuum observations.)
\item the scale height of the outer disk needed to be lower than 10\%.
\end{enumerate}

These changes in the outer disk geometry led to a bad fit of SED at $\lambda>15~\mu$m.
To resolve this, a fourth modification was introduced in the outer disk by splitting it into two superposed disks.
A first disk with lower H/R (5\%), 
with 80\% of the gas and dust mass, and with a dust population of large grains ($0.01<a<1000~\mu$m);
and a second disk with higher H/R (0.8$-$0.13), 
with the 20\% remaining gas and dust mass, and  with a dust population of
smaller grains ($0.01<a<10~\mu$m).
This modification aims to keep most of the gas mass of the outer disk at low H/R (to fit the [\ion{O}{i}]~63~$\mu$m line),
while allowing some small dust particles to be present in an extended outer disk atmosphere at higher H/R to fit the SED at $\lambda>15~\mu$m.
This two-layered outer disk echoes the expected effect by dust coagulation and sedimentation 
(i.e., large grains closer to the midplane).  

Finally, for the inner disk, 
to address the fact that scattered light imaging \citep[][]{Muto2012, Garufi2013} revealed material inside the cavity down to 28 AU, 
we set, as a first approximation, the inner disk to have an outer radius of 45 AU, such that the whole cavity is replenished with gas,
and we allowed the gas-to-dust ratio to be over 100 in the inner disk to be able to have sufficiently high CO~4.7~$\mu$m emission.
 
With this ensemble of modifications, we found a family of disk models able to simultaneously describe  
the  SED, the CO P(10) line profile, the line fluxes of [\ion{O}{I}] 63 $\mu$m (within a factor 2),
CO P(10) (within a factor 4),  $^{12}$CO 3-2, $^{12}$CO 2-1, 
and the upper limits of  [\ion{O}{I}] at 145 $\mu$m, [\ion{C}{ii}] at 157 $\mu$m, and the H$_2$ lines in the near and mid-IR.
The properties of a representative model of the family of models 4 is  described in Table~\ref{Table_models},
and the line fluxes predicted are presented in Table~\ref{Table_line_fluxes}.

The family of models 4 confirmed the result that to describe the CO P(10) line profile, 
the gas in the inner disk should be distributed
with a {\it surface density increasing as a function of the radius}, 
and indicated that to describe the [\ion{O}{I}] 63 $\mu$m line flux observed by Herschel, 
{\it the gas-to-dust ratio in the outer disk should be much lower than 100}.
The best match to the Herschel lines [\ion{O}{I}] 63 $\mu$m was obtained by a gas-to-ratio below 10;
nevertheless, gas-to-dust ratios up to 40 provided  [\ion{O}{I}] 63 $\mu$m fluxes within a factor 2 of the observations.

\subsection{Family of models 5: Introducing recent constraints from polarized and mid-IR imaging.}

In a recent work \citet{Garufi2013} has shown that polarized scattered light in HD~135344B
drops significantly at a radius of 28 AU.
A lack of polarized emission can be the signature of either a lack of material or 
a change in the illumination of material inside the cavity.
Similarly,  \citet{Maaskant2013} propose a cavity of size 30 AU based on modeling the SED and mid-IR imaging.
The detection of scattered light down to 28 AU, inside the sub-mm dust cavity of 45 AU,
indicates a different spatial location for the small and large dust grains in the outer disk.

To account for the recent results from \citet{Garufi2013}, 
we slightly modified the disk structure of the family of models~4.
First, 
we shortened the outer radius of the inner disk to 30 AU and
decreased its dust mass to keep it consistent with the SMA 870~$\mu$m upper limit. 
Second, we extended the small particle component of the outer disk down to 30 AU.
The large grain component of the outer disk was kept at 45 AU to account for the sub-mm 870~$\mu$m imaging constraints.

\begin{figure*}[ht!]
\begin{center}
\begin{tabular}{c}
{\bf \sffamily \Large Model 5}\\[3mm]
 \includegraphics[width=0.8\textwidth]{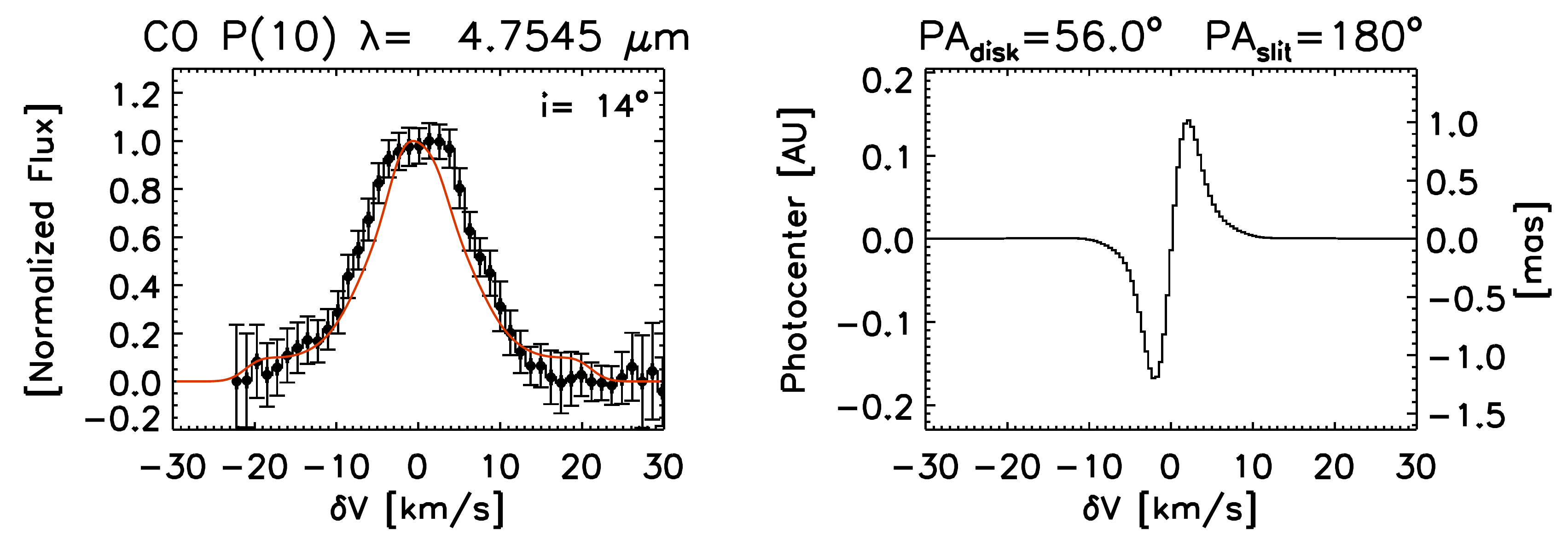}\\[2mm]
\end{tabular}
\begin{tabular}{cc}
\includegraphics[width=0.4\textwidth]{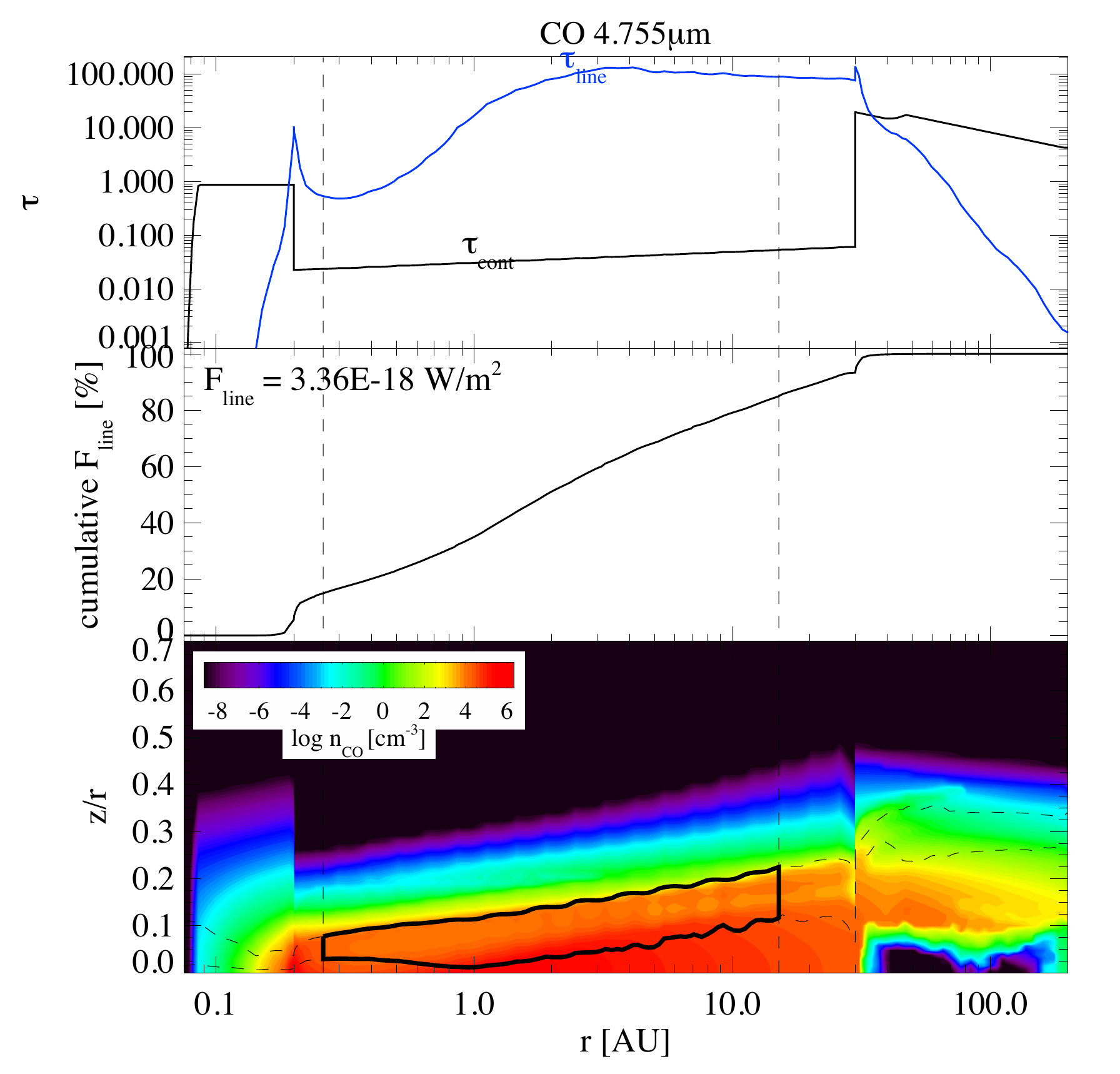}&
\includegraphics[width=0.4\textwidth]{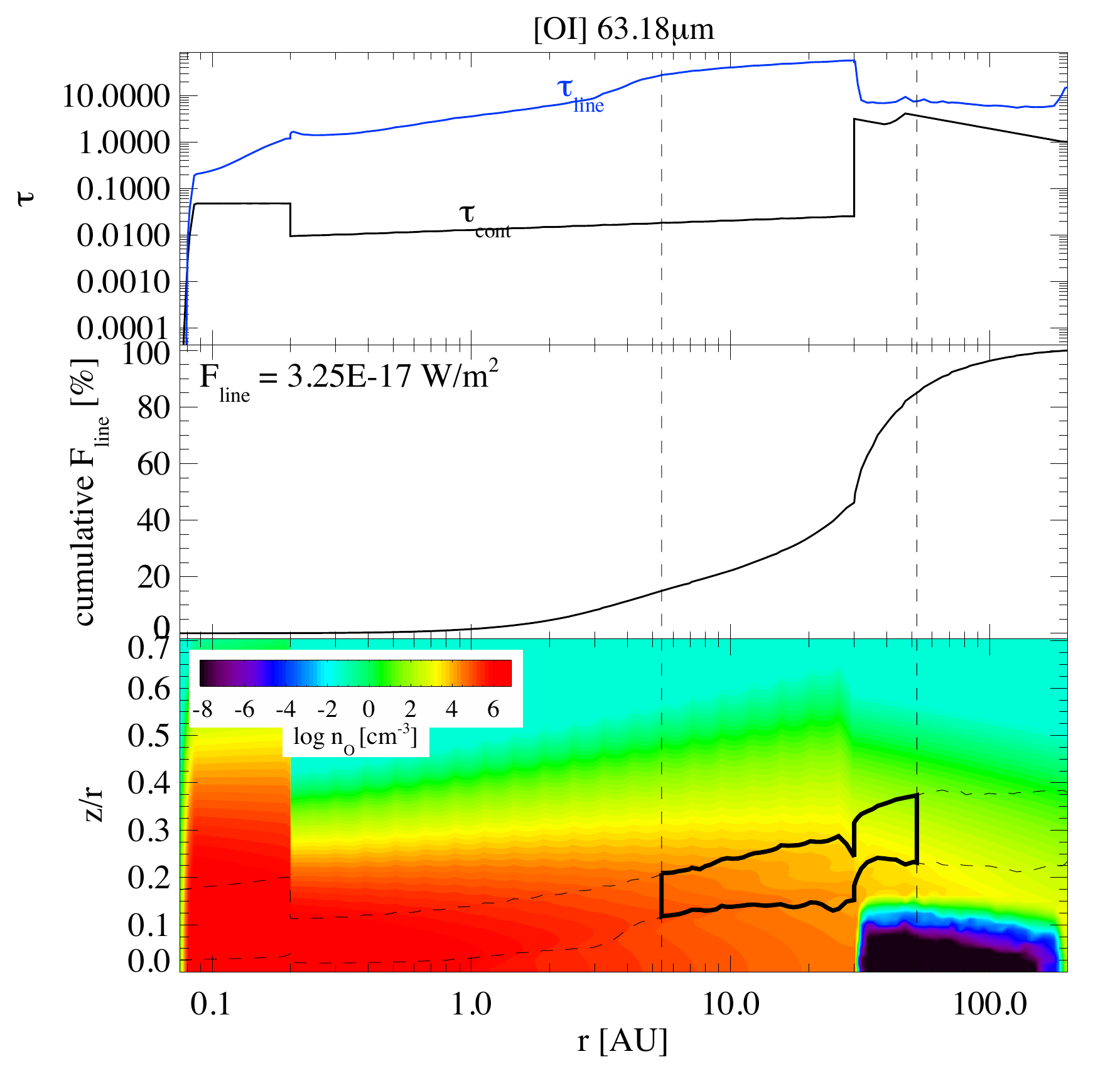}  \\[1mm]
\includegraphics[width=0.4\textwidth]{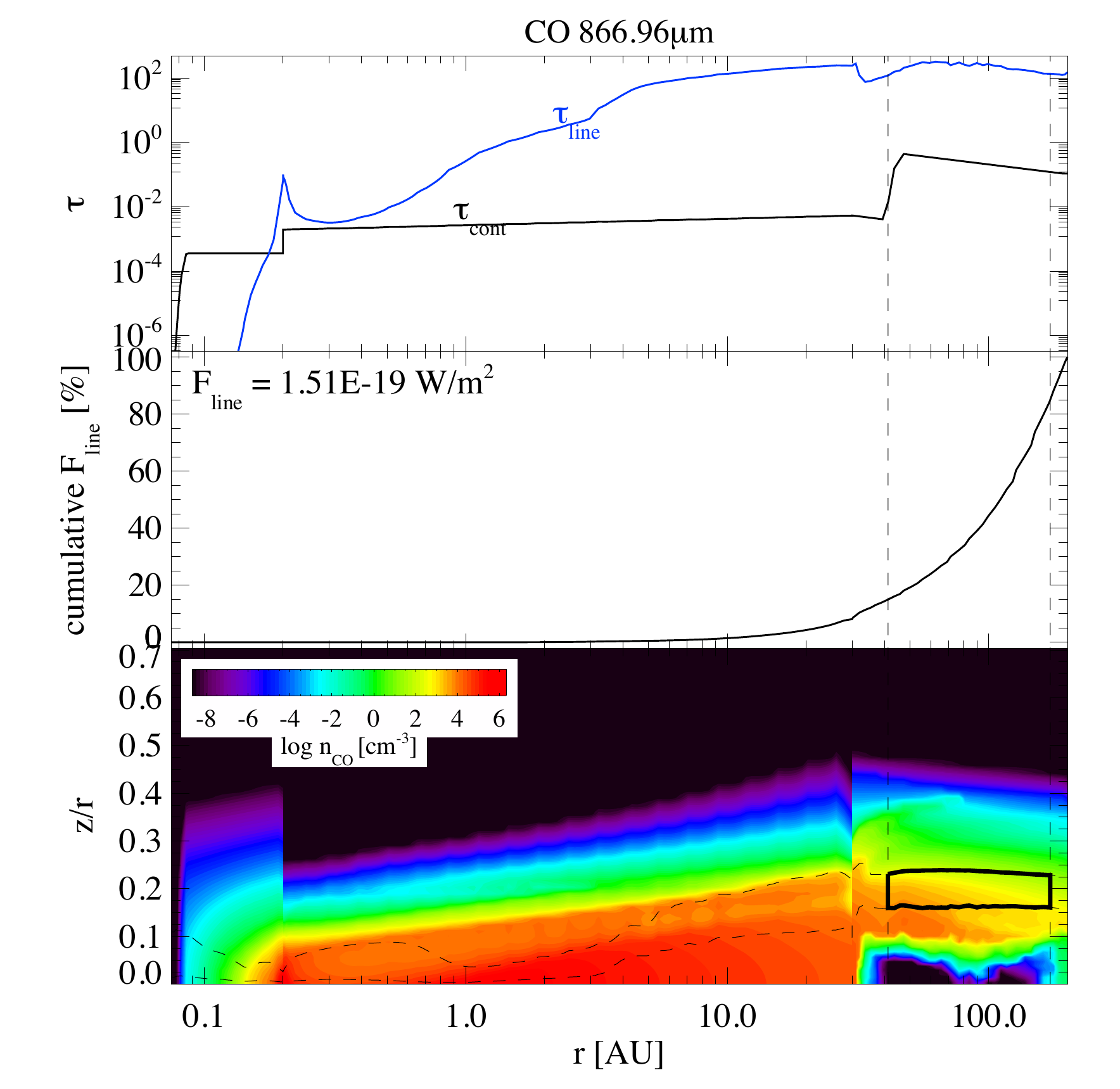}&  
\includegraphics[width=0.4\textwidth]{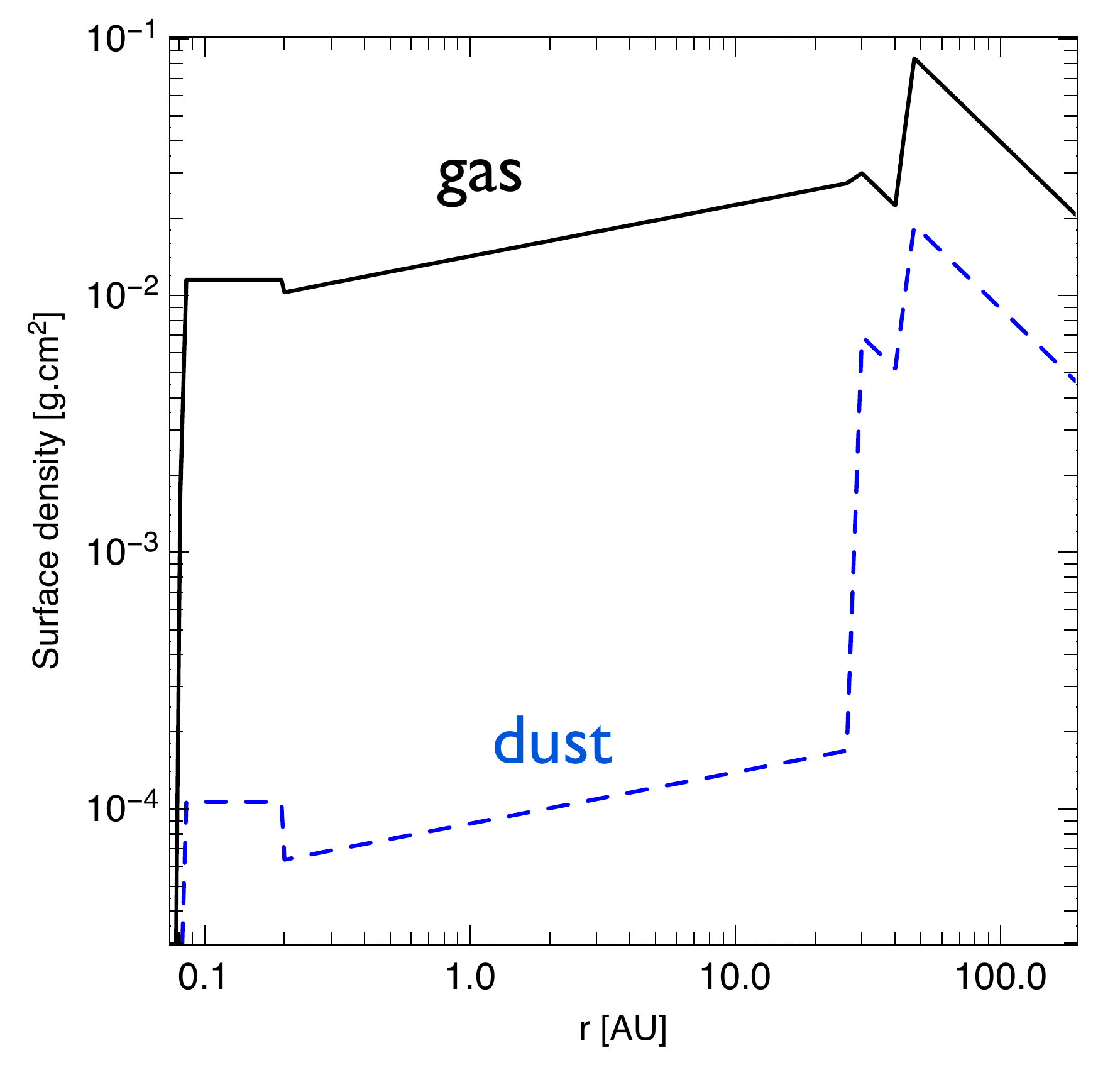}\\[3mm]
\end{tabular}
\end{center}
\caption{{\it Upper panels:} CO P(10) profile (left) and expected spectro-astrometry signature (right) for Model 5. 
{\it Central panels} optical depth of the line and of the continuum, cumulative line flux, number density, and
emitting region diagrams for the CO P(10) (left) and [\ion{O}{I}] line at 63 $\mu$m (right) lines.
The box in thick black lines represents the region in the disk that emits 70\% of the line radially and 70\% of the line vertically,
thus approximately $\sim$ 50\% of the line flux.
{\it Lower panels:} ({\it left)} Similar plots for the $^{12}$CO 3-2 line at  870 $\mu$m,
{\it (right)} surface density of the gas (black continuous line),
and the dust (blue dashed line) as a function of the radius.
The line fluxes quoted in the panels are the integrated line fluxes from the whole disk.
For the CO~P(10) integrated line flux taking the slit losses into account see Table \ref{Table_line_fluxes}.
More information about the model is presented in Fig.~\ref{Model5_appendix} in the Appendix.
}
\label{Model5_ref} 
\end{figure*}

With these modifications, 
we obtained a disk structure compatible with the Garufi disk structure,
while keeping the fit to the SED and the gas lines. 
Our model has gas and (some) dust inside 30 AU to account for the extended CO ro-vibrational line
observed. 
However, the amount of dust at R$<30$~AU ($<$10$^{-7}$ M$_\odot$) 
is much less than the amount of dust in the outer disk ($2\times10^{-4}$ M$_\odot$).
The properties of a representative model of the family of models 5 is 
described in Table~\ref{Table_models},
and the line fluxes predicted are presented in Table~\ref{Table_line_fluxes}.
In Fig.~\ref{Model5_ref}, 
we present the synthetic SED, 
the predicted CO P(10) line profile and spectro-astrometry signal,
and the plots describing the optical depth of the line and of the continuum, the cumulated line flux intensity, 
and the number density of the species as a function of the radius 
for the CO P(10), [\ion{O}{I}] 63 $\mu$m,
and CO 3-2 lines at 870 $\mu$m lines.
Figure~\ref{Model5_ref} also presents  a plot of a surface density of the gas and the dust as a function of the radius.
In Fig.~\ref{Model5_appendix} in the Appendix, we present the plots of the number density of H as a function of the radius, the dust and gas temperature in the disk,
and similar plots to those in Fig.~\ref{Model5_ref} but for the [\ion{O}{I}] 145 $\mu$m, [\ion{C}{ii}] 157 $\mu$m, and H$_2$ 0-0 S(1) 17 $\mu$m lines.

\begin{table*}[t]
\caption{Summary of model parameters}
\label{Table_models}
\scriptsize
\begin{center}
\begin{tabular}{l|cccccccllllllll}
&Zone & $\beta^{\dag}$ &$M_{dust}$  &$R_{in}$ & $R_{out}$ & edge &  $q^{\dag\dag}$ & H/R  &Dust type & $a_{\rm min}$ & $a_{\rm max}$ &   f$_{\rm PAH}$ & g/d     \\
                   &        &[M$_\odot$] & [AU]      & [AU]          & [AU]   &    [AU]          &  & &&$[\mu$m] & $[\mu$m] \\
\hline\\[-1mm]   
{Model 1a} &\multicolumn{12}{l}{\bf Only astronomical  silicates. No gas in the gap} \\
%\hline
        &   1 &  1.10 &  7.0E-10 &  0.16 &   0.21 & 0.002 &  -1.5 &   0.016/    0.10 & Astro-Silicates & 10.00 & 1000.00 & &100  \\
                &   2 &  1.00 &  1.1E-04 & 45 & 200 & 1.0 &  -2.5 &   4.5/   45 & Astro-Silicates  &   0.05 & 1000.00   & 6.5E-3 &100\\[3mm]%Model 085
 		
Model 1b & \multicolumn{5}{l}{\bf Only astro. silicates. Gas in the gap.} \\
          &   1 &  1.10 &  7.0E-10 &  0.16 &   0.21 & 0.002 &  -1.5 &   0.016/    0.10 & Astro-Silicates   &    10.00   & 2000.00      & & 100  \\ %Model 085/23
          &   2 &  1.10 &  1.0E-08 &  0.21 &  40 & 0.0 &  -1.0 &   0.015/    0.10 & Astro-Silicates   &  10.00   & 2000.00    &  & 100 \\
          &   3 &  1.00 &  1.1E-04 & 45 & 200 & 1.0 &  -2.5 &   4.5/   45 & Astro-Silicates  &    0.05   & 1000.00     & 6.5E-3 & 100\\[3mm]
          
Model 2a &  \multicolumn{5}{l}{\bf 25\% carbon grains inner disk}\\  %grid carbon32  
          &   1 &  1.25 &  1.8E-10 &  0.18 &  20 & 0.001 &  -1.5 &   0.012/    0.10 & Amor. Carbon    &    0.01    &     10  && 100\\ 
          & & & & & & & & & Astro-Silicates   & 0.01    &   1000     &   \\
          &   2 &  1.00 &  1.0E-04 & 45 & 200 & 1.0 &  -2.5 &   5.5/  45 & Astro-Silicates  &    0.05   &   1000    & 6.5E-3 & 100\\[3mm]

Model 2b & \multicolumn{5}{l}{\bf 5\% carbon grains inner disk} \\%grid carbon13
&   1 &  1.00 &  3.0E-10 &  0.18 &   3 & 0.001 &  -2.0 &   0.010/    0.10 & Am. Carbon    &    0.01   &      10    & & 100\\ 
& & & & & & & & & Astro-Silicates  &    10     &  1000     &   &100 \\
          &   2 &  1.00 &  1.1E-04 & 45 & 200 & 1.0 &  -2.5 &   6.0/   45 & Astro-Silicates   &    0.05    &   1000    & 6.5E-3 & 100\\[3mm] 
      
Model 3 & \multicolumn{7}{l}{\bf carbon-enriched innermost disk ($0.2<R<0.25$ AU), g/d = 100 in the whole disk}\\%grid 51139 
          &   1 &  1.00 &  6.0E-12 &  0.21 &  30 & 0.002 &  -5.0 &   0.010/    0.10 & Am. Carbon   &       0.01    &     10   & &100\\
          &   2 &  1.00 &  1.0E-06 &  0.21 &  30 & 0.002 &   0.25 &   0.015/    0.10 & Astro-Silicates &    100   &    2\,000 &&100 \\
          &   3 &  0.60 &  1.0E-04 & 45 & 200 & 1.0 &  -2.0 &   6.3/   45 & Astro-Silicates  &   0.05   &   1000    & 6.5E-3 &100 \\[3mm]
    
Model 4 & \multicolumn{14}{l}{\bf inner disk: carbon grains $0.1<R<0.2$, silicates 0.2 to 45 AU, (g/d)$_{\rm  inner}$=150; outer disk: dust settling,  (g/d)$_{\rm  outer}$=2; continuos gas surface density at 45 AU}\\
          &   1 &  1.10 &  2.5E-12 &  0.085 &   0.20 & 0.002 &   0.0 &   0.012/0.10 & Am. Carbon      &   0.01    &     10     & & 150 \\
          &   2 &  1.12 &  3.5E-07 &  0.20 &  44  & 0.0 &   0.30 &   1.2/10 & Astro-Silicates  &   0.10    &   1\,000        & & 150\\
          &   3 &  1.0 &  0.4E-04 & 45 & 200 & 0.0 &  -1.0 &   1.2/10 & Astro-Silicates   &    0.01    &   10      &  0.18      & 2 \\
          &   4 &  1.0 &  1.6E-04 & 45 & 200 & 0.0 &  -1.0 &   0.5/10 & Astro-Silicates   &   0.01      & 1\,000    &  0.18   & 2\\[3mm] 
Model 5 &  \multicolumn{13}{l}{\bf inner disk: carbon grains $0.1<R<0.2$; silicates 0.2 to 30 AU; outer disk: small dust dust down to 30 AU, large dust down to 45 AU} \\
          &   1 &  1.10 &  2.5E-12 &  0.085 &   0.20 & 0.002 &   0.00 &   0.012/ 0.10 & Am. Carbon    & 0.01  &    10   & & 100 \\
          &   2 &  1.12 &  1.0E-07 &  0.20 &  30.00 & 0.0 &   0.20 &   1.1/10 & Astro-Silicates  &  0.10 &   1\,000  &  & 150 \\
          &   3 &  1.00 &  0.5E-04 & 30.00 & 200.00 & 0.0 &  -1.00 &   1.1/10 & Astro-Silicates &   0.01  &    10  &  0.09 &  4 \\
          &   4 &  1.00 &  1.5E-04 & 45.00 & 200.00 & 1.0 &  -1.00 &   0.7/10 & Astro-Silicates   &  0.01  &  1\,000 & 0.09 & 4 \\[3mm]               
 \hline  
\end{tabular}
\\[0.3cm]
\end{center}
{\bf Notes:} $^{\dag}$ $\beta$ is the flaring exponent; $^{\dag\dag}~q$ is the surface density exponent.\\[0.5cm]
\normalsize
\caption{Observed and modeled line fluxes.}
\label{Table_line_fluxes}
\begin{center}
\scriptsize
\begin{tabular}{l|cccccccc}
 \hline
 & [\ion{O}{I}]  & [\ion{O}{I}] & [\ion{C}{ii}] & $^{12}$CO J3-2  & $^{12}$CO J2-1 & $\nu=$1-0 P(10)$^a$ & H$_2$ 1-0S(1) &  H$_2$ 0-0S(1) \\
  & 63 $\mu$m & 145 $\mu$m & 157 $\mu$m  & 866 $\mu$m  & 1.3mm  & 4.7545 $\mu$m& 2.12 $\mu$m& 17.03$\mu$m \\
\hline
{\bf Observed }& 3.6$-$4.8E-17 & $<$4.6E-18  & $<$6.4E-18 & 1.2E-19 & 8.0E-20 & 1.5E-17 & $<$1.6E-17 & $<$ 1.0E-17   \\
\hline
Model 1a & 5.4E-17 & 4.4E-18 & 3.7E-18 & 2.3E-19 & 7.4E-20 & 1.6E-17  74\% & 8.2E-20 & 1.9E-19 \\%Ring 085
Model 1b & 7.2E-17 & 5.5E-18 & 4.7E-18 & 2.5E-19 & 7.8E-20 & 1.7E-17  74\%  & 1.8E-19 & 4.4E-19 \\ %Model 085 23
Model 2a & 2.7E-16 & 1.8E-17 & 8.7E-18 & 2.6E-19 & 8.0E-20 & 1.5E-17  21\% & 5.4E-18 & 9.4E-18 \\ %grid carbon32
Model 2b & 3.4E-16 & 2.8E-17 & 1.2E-17 & 3.3E-19 & 1.0E-19 & 7.4E-17  26\% & 8.7E-18 & 1.2E-17 \\ %grid carbon13
Model 3 & 2.6E-16 & 2.0E-17 & 1.2E-17 & 2.8E-19 & 8.5E-20 & 1.2E-17  66\% & 1.5E-18 & 4.6E-18 \\%grid 51139
Model 4 & 4.0E-17 & 2.4E-18 & 2.2E-18 & 1.6E-19 & 5.0E-20 & 3.7E-18  72\% & 5.4E-19 & 2.6E-19 \\%Model4\_pp\_PAH 
Model 5 & 3.2E-17 & 1.5E-18 & 1.7E-18 & 1.5E-19 & 4.7E-20 & 2.5E-18  72\% & 6.0E-19 & 1.8E-19 \\%Model5\_pp\_PAH 
\hline
\end{tabular}
\end{center}
{\scriptsize {\bf Notes:} $^a$ The percentage after the CO P(10) line-flux corresponds to the fraction of line flux retrieved inside the slit.} \\
\end{table*}

\begin{figure*}[t]
\begin{center}
\includegraphics[width=0.8\textwidth]{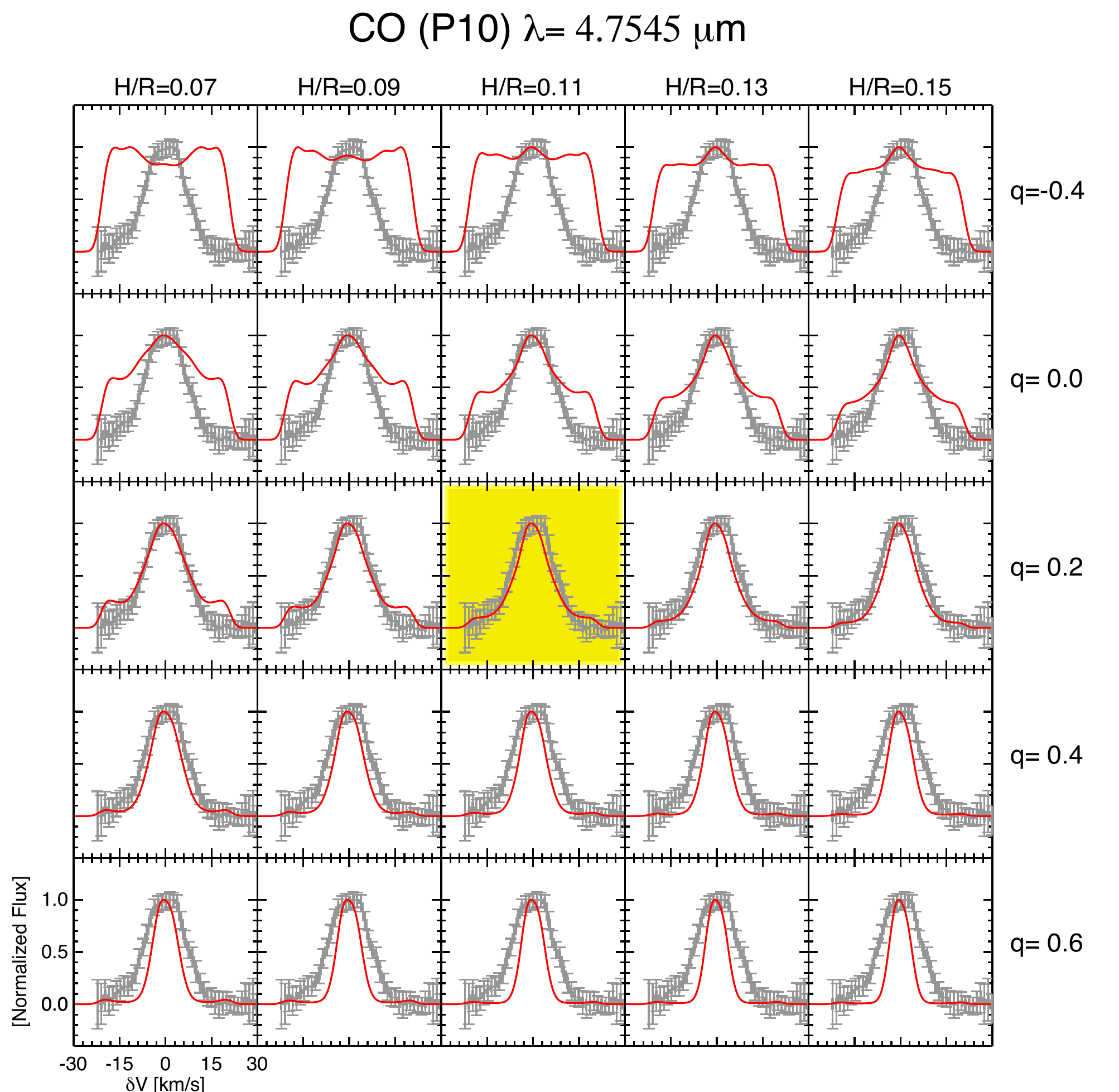} \\[0.1cm]
\caption{CO P(10) line profiles for diverse values of H/R and the surface density exponent of the inner disk $q$ in the family of models 5.
The reference radius of H/R s 10 AU. H/R is the same for the inner and the outer disk.
All other parameters of Model 5 are kept constant. The best fit is highlighted.}
\label{OICO_vsHR_profile} 
\end{center}
\end{figure*}

\begin{figure*}[t]
\begin{center}
\begin{tabular}{c}
\includegraphics[width=0.69\textwidth]{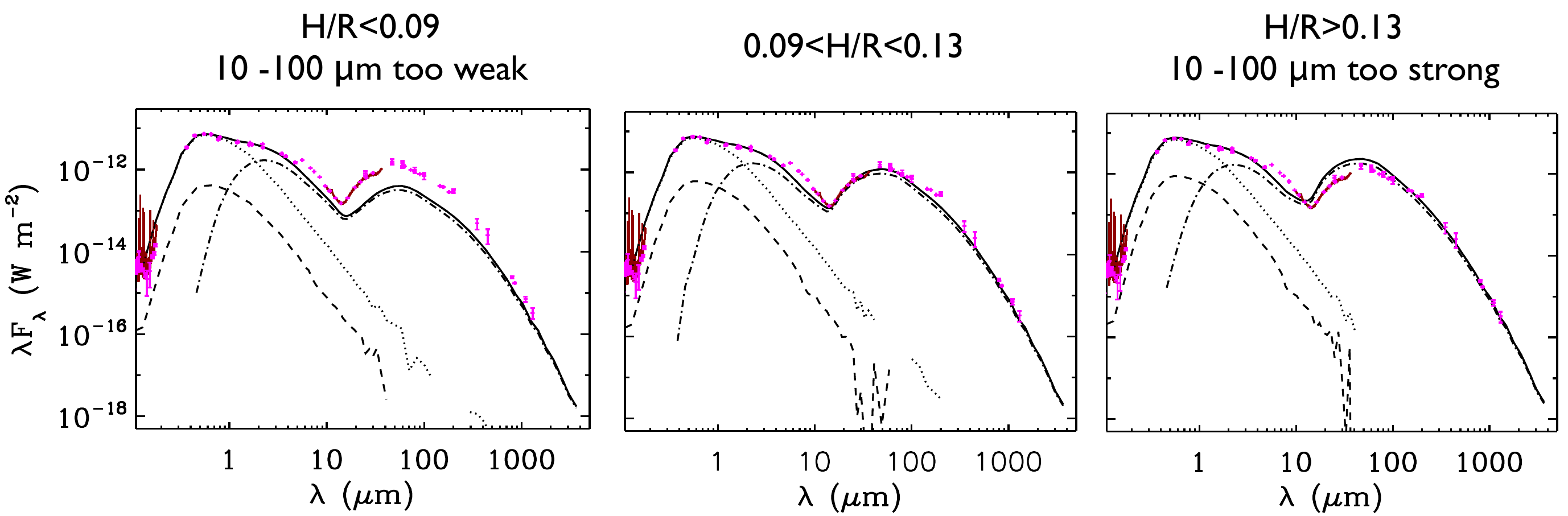} \\[2mm]
{\bf \Large a)} \\[1mm]
\includegraphics[width=0.55\textwidth]{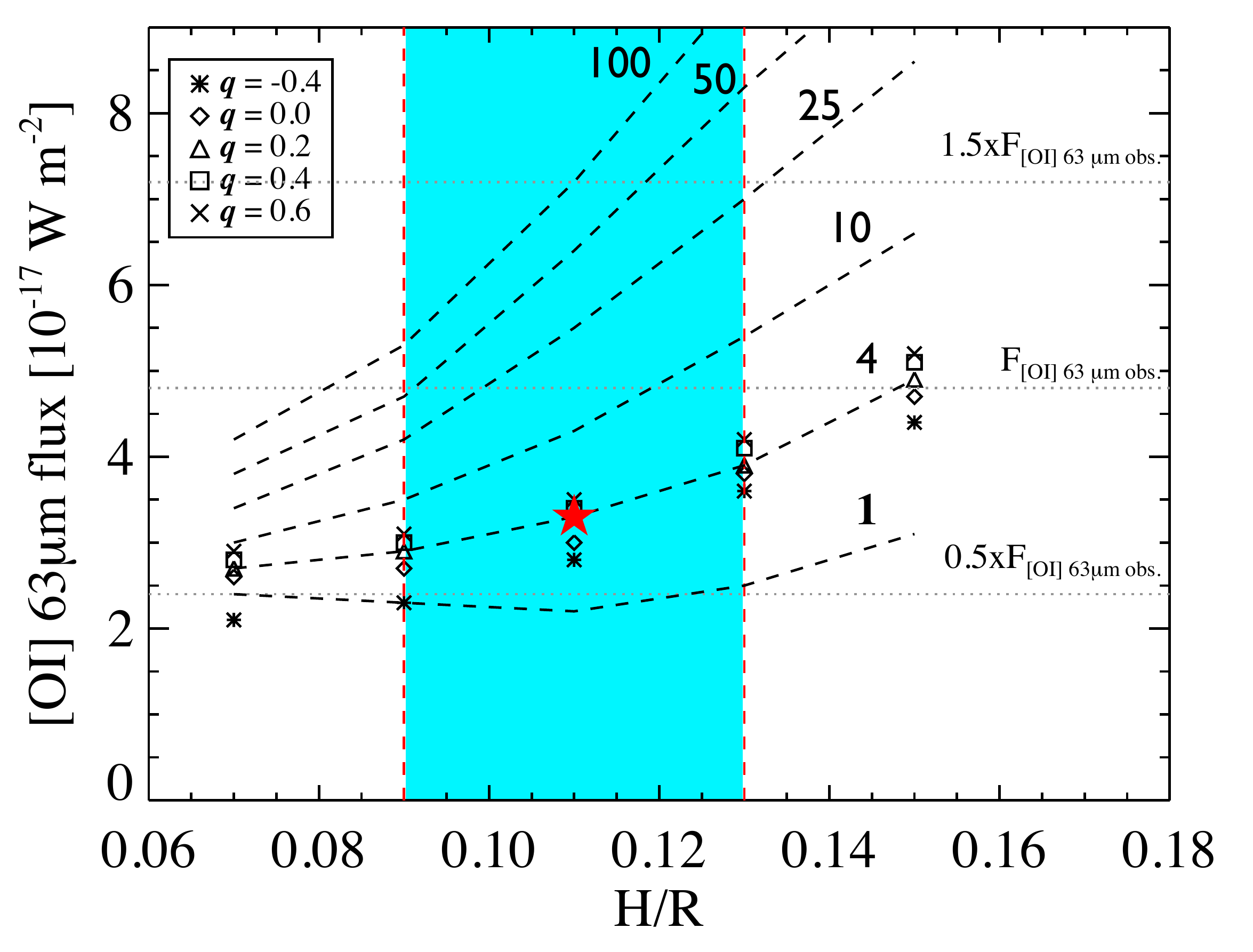} \\[2mm]
{\bf \Large  b)} \\[1mm]
\includegraphics[width=0.77\textwidth]{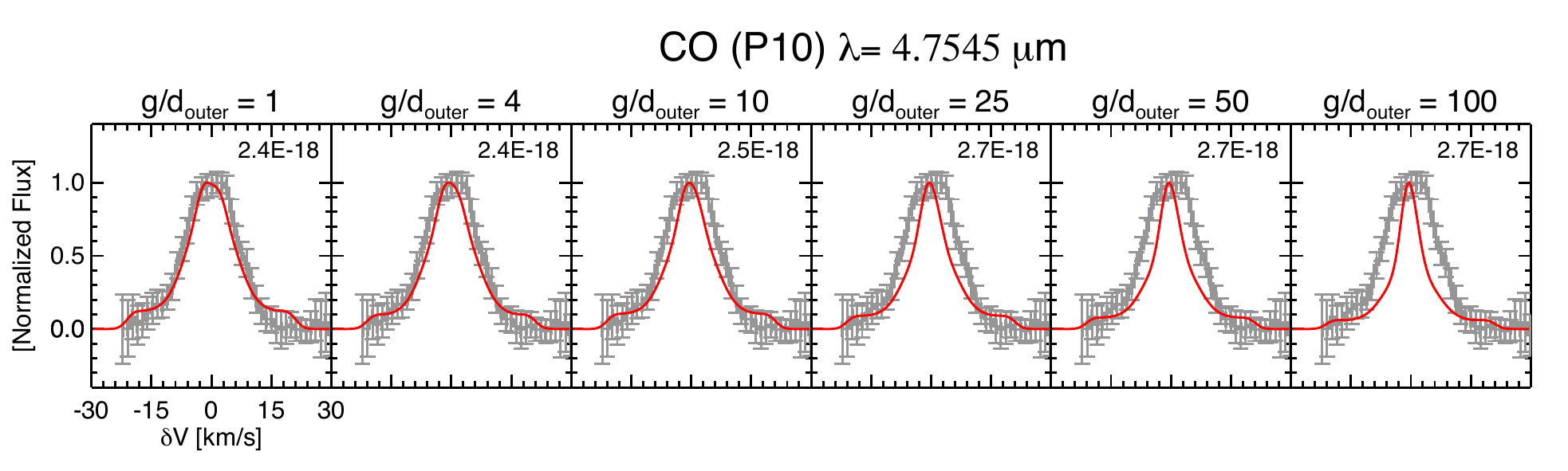}\\%[2mm]
{\bf \Large c)} \\
\end{tabular}
\caption{{\it Top panels:} SED for disks with different values of H/R in the family of models 5. 
The reference radius used is 10 AU, and H/R is set to be the same for the outer and the inner disk.
The color cyan shows the allowed range of H/R consistent with the SED.
{\it Central Panel:}
Flux of the [\ion{O}{I}]  line at $63~\mu$m as a function of H/R for surface density exponents of the inner disk ranging from -0.4 to +0.6 (see legend)  
for a gas-to-dust ratio of the outer disk equal to 4.
The dashed lines show the line flux for different gas-to-dust ratios of the outer disk for a constant inner disk surface density exponent $q=0.2$. 
{\it Bottom Panels:} Expected CO P(10) profiles for different values of the gas-to-dust ratio in the outer disk for the family of models~5 
assuming H/R=0.11 and $q=0.2$.
 The number in the top right of each panel is the CO P(10) line flux in W m$^{-2}$.
The star represents the values of Model 5 in Table~\ref{Table_models}  and Fig.~\ref{Model5_ref}.
The best combined fit of the [\ion{O}{I}] $63~\mu$m line flux and the CO P(10) line profile is given by gas-to-dust ratios below 10 in the outer disk.}
\label{OICO_vsHR_flux} 
\end{center}
\end{figure*}

\begin{figure*}[t]
\begin{center}
\includegraphics[width=0.7\textwidth]{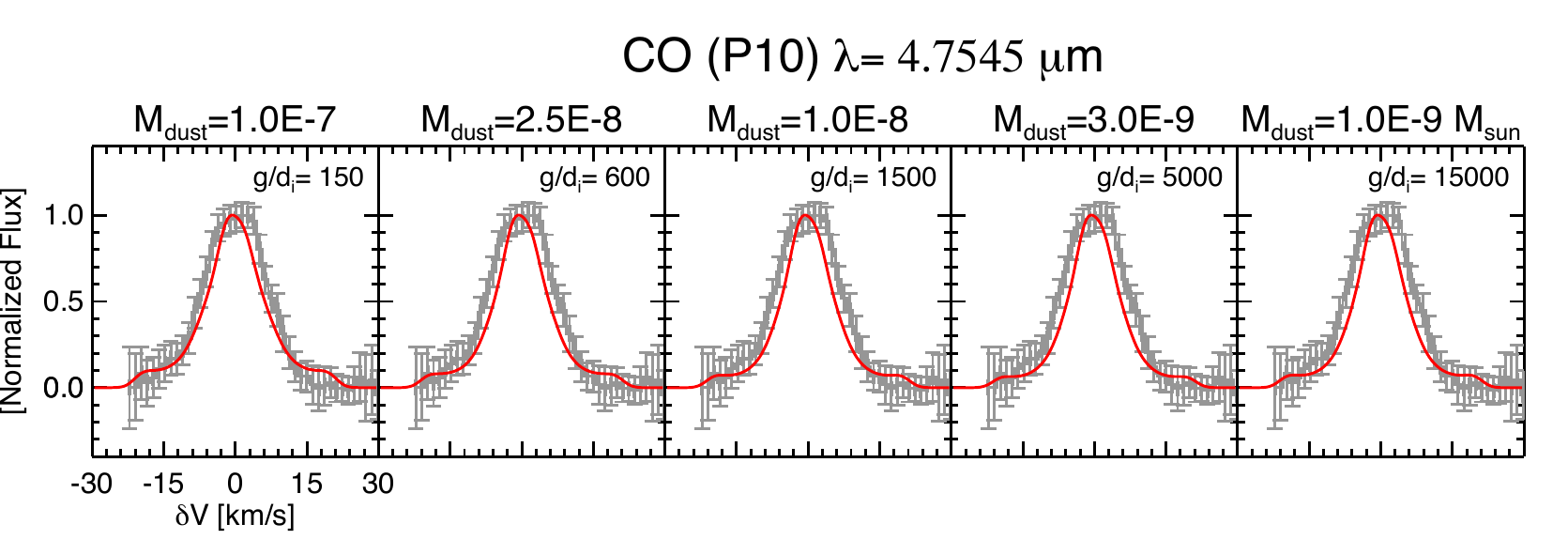}\\
\caption{CO P(10) line profiles as a function of the gas-to-dust ratio in the inner disk in Model 5. 
The gas mass (1.5$\times10^{-5}$ M$_{\odot}$) and gas surface density (see Fig.~\ref{Model5_ref}) 
at R$<30$~AU is kept constant. The mass of carbon grains in the innermost disk is kept constant. 
The outer disk is kept constant. We decreased the mass of silicates at $0.2<R<30$ AU up to a factor 100 from left to right.}
\label{Figure_g2d} 
\end{center}
\end{figure*}

\section{Disk structure constraints derived from Model 5.}

The interest of performing multi-instrument modeling is to use the constraints obtained
from different gas and dust tracers to break the model degeneracies
and narrow down the parameter space of possible solutions.
Ideally, a Bayesian analysis of a large number of models
covering a significant fraction of the parameter space should be performed.
However, 
owing to the prohibitive amount of computing time that this kind of analysis 
would require when the heating and cooling balance and the chemistry calculation are included, 
we limit our discussion to the parameter space surrounding Model 5, the best solution found.
Our solution is a model that reproduces most of the constraints imposed by observations;
however, there is the possibility that the solution is {\it not} unique.

\subsection{Compromises  during the modeling procedure}

Ideally, a model should be able to reproduce all the observations available.
However, to be able to reproduce most of the observations simultaneously,
we needed to relax the perfect fit for a few of them.
ProDiMo includes a larger number of physical processes, 
but, not all the physics are included in the code.
Furthermore, our models are axisymmetric,
HD~135344B is known to display a spiral structure \citep[][]{Muto2012,Garufi2013}.
Not including these spiral structures may have an impact, or not, on the integrated line fluxes that 
we are fitting.
But, we are modeling integrated quantities
and spiral arms produce local changes, 
so, it is not clear whether their impact would be dramatic. 

The first compromise is the fit to near-IR  SED at  8-10 $\mu$m.
Our model is fainter than the observations at those wavelengths.
For an innermost carbon disk that reproduces the near-IR interferometry data and the SED near-IR excess,
one can construct a disk between 0.2 to 30 AU with sufficient dust mass
for a given silicate dust size distribution,
such that the 8 to 20 $\mu$m excess is well reproduced (see for example Model 3).
However, the upper limit of the continuum emission at  870~$\mu$m inside 30 AU
sets a stringent limit on the amount of dust mass that can be put in the inner disk.
Higher masses can be achieved using smaller grain size distributions;
however, when the dust size distribution is dominated by dust with $a_{\rm min}<10~\mu$m, then the silicate feature appears.
A solution for better fitting the near-IR SED might be to introduce an additional zone at higher H/R in the
inner disk with small grains ($a_{\rm min}<10~\mu$m) with 1\% to 10\% of the inner disk's dust mass. 

The second compromise is the absolute flux of the CO ro-vibrational lines. 
A cavity without discontinuity in the gas implies that around half of the [\ion{O}{I}] 63 $\mu$m and 145 $\mu$m lines will be 
emitted {\it inside} the cavity (see Fig.~\ref{Model5_ref}).
Therefore, when the gas mass or temperature is increased to better fit the CO P(10) flux, 
the flux of the [\ion{O}{I}] lines will also increase.
Since the [\ion{O}{I}]  63 and 145 $\mu$m excitation is better understood and tested than CO ro-vibrational excitation 
(the collision rates are known only within a factor ten \citep[][]{Thi2013}),
and because the [\ion{O}{I}]  63~$\mu$m line can be used to constrain the gas mass,
we gave priority to fit the [\ion{O}{I}]  63 $\mu$m and 145 $\mu$m line fluxes and upper limits simultaneously with the CO P(10) line profile, over reproducing the CO P(10) absolute line flux. Our models underpredict the CO P(10) line flux, hence the spectro-astrometry signature.

\subsection{Inner disk surface density exponent}
The CO P(10) line profile strongly depends on the surface density power law exponent of the inner disk gas.
In Fig.~\ref{OICO_vsHR_profile}, we display 
the calculated CO P(10) line profile after taking the slit effects into account
for  disks with inner and outer disk H/R ranging from 0.07 to 0.15,
and $q$ ranging from -0.4 to +0.6.
The CO P(10) line profile is best described by slightly positive power-law surface density exponents,
in other words, by {\it a surface density that increases as a function of the radius}.
Power law exponents $q$ smaller than 0.0 produce line profiles that are too broad,
and $q$ larger than +0.4 produce line profiles that are too narrow.
Keeping in mind the uncertainties and simplifications in the {\sc ProDiMo} code, 
we can exclude gas surface density profiles with negative exponents for the inner disk. 
A steep surface density profile with a negative 
exponent will assign too much gas to the inner few AU of the disk. 
This would result in too much hot CO gas close 
to the star emitting at high-velocities;
hence, a broad line-profile (see upper row in Fig.~\ref{OICO_vsHR_profile})
that is inconsistent with 
the line profile and spectro-astrometric signal observed, 
which show that the CO ro-vibrational emission comes from radii greater than a few AU.

This result is relatively independent of the scale height assumed for the inner disk.
Here we have assumed that the CO~4.7~$\mu$m line is produced  entirely by disk emission.
Although a contribution from a weak disk wind cannot be ruled out 
\citep[the object displays asymmetric oxygen emission at 6300 \AA,][]{vanderPlas2008}, 
the symmetric spectro-astrometry signature detected by \citet[][]{Pontoppidan2008} at three slit positions
favors a dominant contribution from disk emission.

Even though that all the cavity is filled with gas, 
the predicted spectro-astrometry signature of our models is a factor 2 lower 
than the observations by \citet[][]{Pontoppidan2008}, and the line flux is weaker than the line flux observed. 
Since we fit the line profile, the relative contributions to the flux by the different radii should be correct. 
The missing flux might be related to a CO heating mechanism or CO physics not yet included in ProDiMo.

\subsection{Inner and outer disk scale height}
We have set the scale height at the reference radius of 10 AU to be the same
for the inner disk and the small dust grain component of the outer disk.
In Figure~\ref{OICO_vsHR_flux}b, 
we display the effect of changing the scale height, 
the power law exponent of the surface density of the inner disk, 
and the gas-to-dust ratio of the outer disk
in the flux of the [\ion{O}{I}] 63 $\mu$m line.

The  [\ion{O}{I}] 63 $\mu$m line flux changes little with a change in the surface density of the inner disk,
and is slightly sensitive to changes in the scale height of the inner and outer disk. 
In general, higher H/R values produce larger [\ion{O}{I}] 63 $\mu$m fluxes.
However, the mild sensitivity in the H/R of the outer disk is 
because most of the gas mass in the outer disk is in its midplane layer that has lower H/R and large grains.
Changes in the scale height of the small dust extended outer disk's atmosphere change 
the  [\ion{O}{I}] 63 $\mu$m line flux little because its mass is low with respect to that of the midplane.

{\it The scale height of the disk is mainly constrained by the SED fit}.
In Figure~\ref{OICO_vsHR_flux}a, we show
 three insets displaying the SEDs for Models 5 with H/R 0.7, 0.11, and  0.15. 
The SED is compatible with Models 5 with an H/R between 0.9 and 0.13.
Disks with H/R lower than 0.9 underestimate the flux at $\lambda>$ 10 $\mu$m region,
and disks with H/R over than 0.13 overestimate the flux in the 10 - 100 $\mu$m region.

\subsection{Outer disk's gas-to-dust ratio}\label{g2d_outer}
{ The fit to the SED provides a relatively robust estimate of the mass of mm-size grains in the outer disk.
This value changed little in all the models tested and is on the order of 2$\times10^{-4}$~M$_\odot$.
The gas mass, hence the gas-to-dust ratio, in the outer disk is constrained by the simultaneous 
fit to the [\ion{O}{i}] 63 $\mu$m line flux and the CO P(10) line profile.

Figure~\ref{OICO_vsHR_flux}b displays the flux of the [\ion{O}{i}] 63~$\mu$m as a function of H/R 
for gas-to-dust ratios of the outer disk ranging from 1 to 100.
We can see that the flux of the [\ion{O}{i}] 63~$\mu$m line is very sensitive to the value of the 
gas-to-dust ratio in the outer disk.
We find that the [\ion{O}{i}] 63 $\mu$m line flux tends to be described by models
with gas-to-dust ratios in the outer disk that are much lower than 100.
Gas-to-dust ratios between 25 and 4 provide the best fit to 
line flux for the scale heights that are compatible with the SED.

In all the models, 
the [\ion{O}{i}] 145 $\mu$m line flux is below the Herschel flux upper limits.
In all the models,
CO sub-mm lines are so optically thick ($\tau>100$ see Fig.~\ref{Model5_ref}) 
that decreasing the gas mass does not affect the line fluxes, 
so these cannot be used to trace the gas mass.

The amount of gas in the inner rim of the outer disk influences the shape of the CO P(10) profile, 
making it more centrally peaked when more gas is present in the outer disk (see Fig.~\ref{OICO_vsHR_flux}c). 
The best fit to the CO P(10) line is given by 
gas-to-dust ratios in the outer disk below 10.

The exact value of the gas-to-dust ratio in the outer disk is model dependent.
However, since most of the models that  simultaneously describe the  [\ion{O}{i}] 63 $\mu$m line flux, 
the SED, and the CO P(10) line profile
require gas-to-dust ratios for the outer disk that are smaller than 10, 
we believe that a gas-to-dust ratio much lower than 100 in the outer disk is a robust result.

Finally, the simultaneous modeling of the CO P(10) line profile and [\ion{O}{i}] 63 $\mu$m line flux
favors models in which the the {\it gas} surface density contrast between the inner and the outer disk
at the inner rim of the outer disk is less than a factor 20.} 

\begin{table*}[!t]
\caption{Predicted continuum fluxes for Model 5 at R$<30$ AU for diverse values of the astronomical silicates dust mass at 0.2$<R<$ 30 AU for different grain sizes.
The gas mass at $R<30$ AU is kept constant and equal to 1.5$\times10^{-5}$ M$_{\odot}$.}
\label{dust_inner_disk}
\begin{center}
\begin{tabular}{cc|cc|cc|cc}
\hline
\multicolumn{2}{c}{$0.2<R<30$ AU} &\multicolumn{2}{c}{$0.1<a<1000$} & \multicolumn{2}{c}{$0.1<a<100$} & \multicolumn{2}{c}{$0.1<a<10$} \\
Silicate dust mass & gas/dust$\,_{\rm inner}$ & 430 $\mu$m & 870 $\mu$m & 430 $\mu$m & 870 $\mu$m & 430 $\mu$m & 870 $\mu$m \\
M$_{\odot}$ &  & mJy & mJy & mJy & mJy & mJy & mJy \\
\hline
$1.0\times10^{-7}$ & 150 & 13.4 & 1.6 & 17.1 &  0.4 & 3.5 & 0.2 \\
$5.0\times10^{-8}$ & 300 & 8.2 & 0.9 & 10.3 & 0.3 & 3.0 & 0.3  \\
$2.5\times10^{-8}$ & 600 &  5.5 & 0.6 & 6.7 & 0.2  & 2.8 & 0.2 \\
$1.0\times10^{-8}$ & 1\,500 &3.8 & 0.3 & 4.3 & 0.2 & 2.7 & 0.2 \\
$3.0\times10^{-9}$ & 5\,000 & 3.0 & 0.2 & 3.2 & 0.2 &  2.7 & 0.2 \\
$1.5\times10^{-9}$ & 10\,000 & 2.8 & 0.2 & 3.0 & 0.2  & 2.7 & 0.2\\
$1.0\times10^{-9}$ & 15\,000 &2.8 & 0.2 & 2.8 & 0.2 & 2.6  & 0.2\\
\hline
\end{tabular}
\end{center}
\end{table*}

\subsection{Inner disk's gas mass, dust mass, and gas-to-dust ratio}
For a given dust composition and size distribution,
the SMA  870~$\mu$m photometry upper limit of 10.5 mJy 
in a 34$\times$70 AU beam centered on the star \citep[][]{Andrews2011}
sets the maximum amount of dust that can be located inside 30 AU almost 
independently of the surface density power law exponent and the scale height of the inner disk.
With the currently available data, the dust size distribution cannot be constrained.

{ Assuming $a_{\rm min}$=0.1~$\mu$m and $a_{\rm max}=1000~\mu$m for the inner disk
(similar to that of the outer disk, except for $a_{\rm min}$ that is slightly larger to avoid the 
presence of a strong 10 $\mu$m silicate feature),
and an inner disk extending from 0.2 to 30 AU,
a dust mass in the inner disk of 1$\times10^{-7}$ M$_{\odot}$ astronomical silicates 
reproduces the dip in the SED at 15 $\mu$m and 
generates a flux of 1.6 mJy at 870 $\mu$m inside 30 AU.  
The disk model convolved with a 34$\times$70 AU beam produces 6.7 mJy of continuum flux  centered on the star,
which is a flux below the 3$\sigma$ upper limit of the SMA observations.
Using this dust mass, we found that gas-to-dust ratios between 100 and  200 can 
describe the CO P(10) line profile for a surface density power law exponent of 0.2.
Lower gas masses in the inner disk (i.e., lower gas-to-dust ratios) generate CO P(10) profiles that are too narrow, 
and higher gas masses (i.e., higher gas-to-dust ratios) produce line profiles with high-velocity wings
that are too broad to be compatible with the observations.
The best fit was achieved with a gas-to-dust ratio of 150, hence, a gas mass inside the cavity of 1.5$\times10^{-5}$ M$_{\odot}$.

There is a degeneracy between the gas mass of the inner disk and the exponent of the power law of the surface density.
By increasing the power-law exponent (i.e., more positive), 
higher gas masses in the inner disk are allowed while reproducing the CO ro-vibrational line profile.
However, 
there is an upper bound to the gas mass in the inner disk,
because the surface density of the inner disk should be equal to or lower than the surface density of the outer disk at 30 AU 
and the flux of the [\ion{O}{i}] line at 63 $\mu$m should be reproduced.
We found that up to $10^{-4}$ M$_{\odot}$ of gas is possible in the inner disk.}

An alternative way of exploring the gas-to-dust ratio in the inner disk is to leave 
the gas mass and surface density constant 
and to decrease the dust mass in the disk at $0.2<R<30$ AU.
In Fig.~\ref{Figure_g2d}, 
we display the expected CO P(10) line profiles for Model 5,
keeping the inner disk's gas mass constant (1.5$\times10^{-5}$ M$_{\odot}$) 
and decreasing the amount of dust down to a factor 100 (i.e., gas-to-dust ratios from 150 up to 15\,000).
We found that if the gas surface density inside the cavity is constant, a lower (astronomical silicates) 
dust mass does not significantly affect the CO~4.7~$\mu$m profile and line flux, 
or the flux of the [\ion{O}{i}] 63 $\mu$m line.
A lower silicate dust mass inside the cavity leads to weaker emission in the 5-10 $\mu$m band.
However, 
since the SED near-IR emission is dominated by the carbon grains in the innermost disk, the SED fit is still compatible with the observations.
The lower bound to the dust mass inside the cavity is not constrained by current observations.
To constrain the dust mass inside the cavity, observations at high spatial resolution are required, for example with ALMA.
In Table~\ref{dust_inner_disk}  we predict the continuum fluxes at 430 and 870 $\mu$m inside 30 AU for different silicate dust sizes and masses in Model 5.

\subsection{Outer radius of the inner disk}
In the family of models 5, 
the inner disk has  gas up to 30 AU.
We have tested models with outer radii of the inner disk ranging 
from 15 to 30 AU.
Solutions to the CO ro-vibrational line profile can be found with a similar gas mass to Model 5,
by allowing for a more positive exponent of the power law of the surface density distribution.
For example, for $R_{\rm out}$= 15 AU, $q$ increases to +0.8.
Models with an inner disk outer radius smaller than 30 AU 
produce slightly better fits to the top of the CO ro-vibrational line;
however, they produce lower spectro-astrometry signatures
and generate a surface density in the inner disk's outer radius that is slightly higher than the surface density at the outer disk's  inner radius. 
Current data do not allow us to set the exact value of the outer radius of the inner disk.
We favor a model with a cavity replenished with gas up to 30 AU because
it generates a larger spectro-astrometry signature  (although even with $R_{\rm out}$=30 AU, the spectro-astrometry signature is underpredicted)
and because surface densities with power-law exponents larger than 0.5 might result in an unstable disk.
{ In summary, the presence of a gap in the gas of a few AU {\it in the gas}
is consistent with the current data, but a large gap in the gas of tens of AU does not appear likely}.

\section{Discussion}

\subsection{ The disk gas structure of HD 135344B and the origin of the transition disk shape.}

{ The growing observational evidence indicates that the gaps suggested
by the SEDs and spatially resolved sub-mm observations are in fact structures that reflect 
the distribution of large grains, but not necessarily the distribution of small dust, and, especially the gas.

As noted by \citet{Pontoppidan2008} the presence of gas in the cavities disfavors photoevaporation as the possible origin
of the transition disk shape in HD 135334B.
A gas-to-dust ratio over 100 in the inner disk indicates that the mechanism depleting the material in the cavity 
is more efficient at depleting dust than the gas. This could potentially favor the grain-growth scenario.
But, as noted by \citet[][]{Birnstiel2012}, grain-growth alone has difficulties explaining the lack of sub-mm emission in the gap. 

Positive and flat power law surface density profiles ($q$=0 and 1) for the inner disk of transition disks 
have previously been suggested by \citet[][]{Dong2012} in the framework of their general disk model to explain 
the H-band scattered light  present inside the sub-mm dust cavities. 
Our results for the modeling of the CO ro-vibrational lines in HD~135344B provide an independent
argument suggesting that the surface density profile inside the cavity can increase with radius in transition disks.

What implications does a positive surface density profile have? 
In most protoplanetary disks the surface density distribution can be described by a power law with a negative exponent, typically -1.0
(i.e., a surface density that {\it decreases} with radius).
A surface density profile with a positive power law exponent (or at least flat),
indicates that the disk's gas structure has {\it significantly} changed in the inner disk.
It is not entirely clear what mechanism will generate a surface density increasing with radius.

One interesting possibility is that we are indeed observing the effects of a Jovian planet { inside the inner cavity}.
Although the exact results of simulations depend on the inner boundary conditions set, 
for example the accretion rate onto the star, and the evolution time of the simulation,
models studying the interaction of a Jovian planet and the disk show
that a disk with an initial negative power-law surface density profile exponent
could evolve into a disk that has a flat or a positive power-law surface density profile exponent at $R<R_{\rm planet}$
(see for example Fig.~2 in  \citet{LubowAngelo2006}, Fig.~1 in \citet{Varniere2006}, or Fig.~5 in \citet{Tatulli2011}).
Furthermore, 
a single Jovian planet is expected to open a gap typically of a few AU width \citep[see review by ][]{KleyNelson2012}.
Such a small gap is compatible with current data and models of HD~135344B.
Another predicted effect of the presence of a planet in the disk is a higher surface density for the gas in the outer disk 
with respect to the surface density for the gas in inner disk. 
We retrieve such a gas surface density structure in our models, although, 
we find a surface density difference that is smaller than found by the migration calculations \citep[e.g.,][]{LubowAngelo2006}.

Taking all this together, the characteristics found in HD~135344B evoke the effects of a migrating 
Jovian planet present in the disk.
However, at the present time there is no firm evidence for such a companion.
Our models disfavor the presence of gaps of tens of AU in the gas inside the cavity of HD~135344B,
therefore,
disfavor the hypothesis that a multi-Jovian planet system   
\citep[e.g.,][]{Zhu2011,Dodson-RobinsonSalyk2011} is what is responsible for the transitional disk shape in HD~135344B.
This conclusion could be extended to other accreting pre-transition disks with CO ro-vibrational emission extending tens of AU
or with CO emission sub-mm emission inside the cavity.

Finally, the result that the total the amount of gas in the disk of HD 135344B  (a few 10$^{-3}$ M$_\odot$) is much lower than expected for a gas-to-dust ratio 100
indicates that HD 135344B is an evolved protoplanetary disk that has already lost a large portion of its gas mass. 
A low gas-to-dust has been reported for another transition disk, IRS 48 \citep[][]{Bruderer2013}. 
Further detailed studies with a larger sample of objects are required to test whether a lower gas mass is a common characteristic of 
the transition disk population.
}

\subsection{Sublimation of dust grains in the inner disk, carbon grains, and optically thick ``inner walls" in pre-transition disks}
\label{carbon}
One crucial ingredient for the simultaneous fitting of the SED and the CO~4.7~$\mu$m emission
that extends tens of AU is the hypothesis of an inner most disk (R$<$0.25 AU) enriched with carbonaceous grains.
This feature is essential for making it possible for the CO warm gas at tens of AU to contribute to the CO P(10) line profile while
having an inner disk that reproduces the near-IR SED.

The VLTI/PIONIER data clearly shows that the H-band excess is located inside 0.16 to 0.20 AU,
the silicate sublimation radius of HD~135344B (the exact value depends on the mass of dust and $n_{\rm H}$).
This provides a plausible physical explanation for the carbon enrichment in the innermost disk:
{\it carbon grains are present at $0.08<R<0.2$ AU because silicate grains have sublimated}.
The mass of carbonaceous grains required to fit the SED and the near-IR visibilities is 2.5$\times$10$^{-12}$ M$_\odot$,
which is a small amount of dust when compared to the total dust mass of the disk that is around 10$^{-4}$ M$_\odot$
and the dust mass at $R<30$ AU that is 10$^{-9}$ to 10$^{-7}$ M$_\odot$.
In Model 5, the carbon abundance relative to $n_{\rm H}$ required in carbon 
grains is around two times the solar abundance.

We assumed that the size distribution of the carbonaceous grains in the innermost disk is 0.1 to 10 $\mu$m.
Future detailed work on the process of dust evaporation in the inner disk of HD~135344B
(for example, as performed in the inner rim of Herbig Ae stars by \citet[][]{Kama2009}, 
or in the warm debris disk of Formalhaut  by \citet[][]{Lebreton2013}
including the effects of gas)
would be required to constrain with a physical basis the dust size distribution of carbon and silicates
as a function of the radius in the inner 1 AU of HD~135344B.

The presence of carbonaceous grains in inner disks to fit near-IR visibilities 
has been suggested in a few circumstellar disks,
such as \citet[][]{Absil2006} and \citet[][]{Lebreton2013} in the cases of the warm debris disks of Vega and Formalhaut.
\citet[][]{Kraus2013} analyzed the (pre-) transitional disk V1247~Ori, a source that is relatively similar to HD~135344B.
To fit V1247~Ori near-IR and mid-IR interferometry data,
Kraus et al. suggest a disk structure where the innermost disk is composed of a mixture of carbon (50\%) and silicate (50\%)
grains at $0.19<R<0.34$ AU, and a mass of 10$^{-7}$ M$_\odot$ carbon dust grains inside $0.3<R<46$ AU.  
The gap composed of carbon was suggested to avoid the presence
of the 10 $\mu$m silicate feature and also to avoid a strong excess at $\lambda>8~\mu$m.
Furthermore, 
Kraus et al. suggest the presence of some inhomogeneities in the ``gap" material. 
At the time of writing, there have been no observations of CO ro-vibrational emission for V1247 Ori.
In our model of HD~135344B, we ruled out any large percentage of carbon grains ($>$25\%) at 0.2$<R<$30 AU
on the basis of the CO ro-vibrational profile.
A gap with a large amount of carbon grains produces a narrow single-peaked CO ro-vibrational line profile 
because the CO emission is dominated by the inner rim of the outer disk (see Model 2).

{The carbonaceous grains innermost disk inside the silicate sublimation radius proposed here for HD~135344B 
provides an alternative disk structure to the ``optically thick-wall" suggested for pre-transition disks
\citep[e.g.,][]{Espaillat2007,Espaillat2010,Brown2007}.
We have shown here that the ``optically thick wall disk structure" is {\it not} 
compatible with detecting CO ro-vibrational emission extending tens of AU.  
{Therefore,
the carbon-rich {(or refractory grains rich)} inner disk structure suggested for HD~135344B 
might be applicable to other transition disks displaying near-IR excess and CO~ 4.7~$\mu$m emission extending several 
AU~\citep[e.g., SR 21,][]{Pontoppidan2008}}.
Further near-IR interferometry observations of (pre-) transition disks are required to establish    
what fraction of them have dust material inside the silicates sublimation radius.
Moreover, since 
CO ro-vibrational emission has been detected in several transition and pre-transition disks \citep[e.g.,][]{Pontoppidan2008,Salyk2009},
it would be of great interest to extend the analysis presented here to them.

The models described here have used amorphous carbon grains. 
We tested graphite grains using a mixture of
50\% parallel and 50\% perpendicular opacities. 
The fit to the CO P(10) line profile did not change significantly. 
The SED displayed a slightly weaker emission at 3 - 10 $\mu$m.

{A potential challenge for the presence of carbonaceous grains in the innermost disk (0.08$<R<$0.2 AU) 
is an oxygen-rich atmosphere.
Carbon grains would react with the oxygen and be destroyed \citep[see for example the models of][]{Gail2001, Lee2010}.
The presence of carbon grains in the innermost disk depends on the balance between 
the destruction and the replenishment timescales.

To estimate the destruction timescale by oxygen atoms (i.e., chemical sputtering), 
we take one carbon grain of density $\rho$, initial size $a_0$, and radius $a$.
Atomic oxygen reacts with individual carbon atoms at the surface of the carbon grain,
producing a change in the carbon grain's size with time \citep[][]{Evans1994}:
\begin{equation}
\frac{\Delta a }{\Delta t}=\frac{a_0}{t}\simeq\dot{a}=-\frac{n_{\mathrm{O}}m_{\mathrm{C}}Y}{\rho}\left(\frac{kT_{\mathrm{gas}}}{2 \pi m_{\mathrm{O}}}\right)^{1/2},
\end{equation}
where 
$t$ is the time for the grain to be destroyed,
$n_{\mathrm{O}}$ the number density of oxygen ($n_{\mathrm{O}}=n_{\mathrm{H}}\chi_{\mathrm{O}}$ and $\chi_{\mathrm{O}}=3\times10^{-4}$),
$Y$ the yield of this Eley-Rideal type reaction, 
$k$ the Boltzmann constant,
and $m_{\mathrm O}$ the mass of the oxygen atom ($m_{\mathrm{O}}=16\times m_{\mathrm{H}}$).
Here we assume that the destruction rate is constant.
Solving for $t$ we obtain
\begin{equation}
t \sim a_0 \frac{\rho}{n_{\mathrm{O}}m_{\mathrm{C}}Y}\left(\frac{2 \pi m_{\mathrm{O}}}{kT_{\mathrm{gas}}}\right)^{1/2}.
\end{equation}
If  the grain's mass density $\rho$ is 2.2 g cm$^{-3}$, then the  life time of a carbon grain of radius $a_0$ is
\begin{equation}
t \sim 0.28 \times \left(\frac{0.1}{Y}\right) \left(\frac{a_0}{0.01\ \mu\mathrm{m}}\right) \left(\frac{2000\ \mathrm{K}}{T_{\mathrm{gas}}}\right)^{1/2}\left(\frac{10^{10}\ \mathrm{cm}^{-3}}{n_{\mathrm{H}}}\right)\ \mathrm{yr}.
\end{equation}
The value of the yield is not well constrained.
Assuming $Y=0.1$ \citep[][but it can be lower]{Vierbaum2002}, $n_{\rm H}= 2\times10^{10}$ cm~$^{-3}$ (mid-plane value
at 0.1 AU in Model 5, see Fig.~\ref{Model5_appendix} in the Appendix),
this simple calculation shows that carbon grains with sizes 0.01, 0.1, and 1 $\mu$m survive
at 2000~K time scales of around a month, a year, and ten years, respectively.  

To calculate the carbon replenishment time scale at R$<0.2$~AU,
three input parameters are needed:
the gas mass accretion rate, the gas-to-dust ratio, and the fraction in mass
of dust grains in the form of carbonaceous grains at R$>0.2$~AU.
\citet{Sitko2012} suggest a gas accretion rate of 2$\times10^{-8}$~M$_{\odot}$/yr.
Our model indicates that the gas-to-dust ratio at $R<30$ AU is 150 (it can be higher).
The mass fraction of carbon grains with respect to silicates at 0.2$<R<$30 AU can be constrained observationally
by the maximum mass of carbon that can be added and still keep the fit to the SED and the CO ro-vibrational line.
We have found that up to 3\% of the dust mass (10$^{-9}$ M$_{\odot}$) at 0.2$<R<$30 AU can be in the form of
carbon grains and still fit the SED and CO lines.
Taking the gas accretion rate, the gas-to-dust ratio, and the fraction in mass of carbon together, 
we obtain a supply rate of carbon grains to the innermost disk of 4$\times10^{-12}$~M$_{\odot}$/yr.
The life time of carbon grains larger than 0.1 $\mu$m is years,
thus, the disk can have around $10^{-12}$~M$_{\odot}$ transient carbon grains at R$<0.2$~AU.
A carbon mass similar to that is required to fit the near-IR SED in our models.

In summary, taking  the supply rate via dust mass accretion
and the destruction timescales in the innermost disk together, 
we find that a carbon innermost (R$<0.2$~AU) disk could indeed be plausible in HD 135344B.

In a recent work, \citet[][]{Lee2010} have studied the survival of carbon grains in a T Tauri disk.
They find that carbon grains can be destroyed by oxygen atoms in the warm (T$\sim$500 K)
upper layers of the disk on time scales ranging from tens to thousands of years depending
on the carbon grain size, porosity, and distance of the star.
The potential destruction of carbon grains at R$>0.2$ AU could present a challenge to the interpretation
presented here. Lee et al. models are not dynamical models in the sense that the dust components are
not followed with the gas in a simulation. Thus, it is unknown what fraction of the total amount of carbon grains 
will effectively reach the upper layers of the disk and be destroyed, what fraction of the carbon grains will survive 
in the mid-plane, and with what rate  carbon grains at a few AU disk are replenished by the outer disk in an accreting disk.
Future, more detailed models that include the dynamics of the gas, the interaction between the gas and the dust,
the photochemistry of the disk, and carbon destruction would be of great help in further testing of
the scenario proposed here. 
 
From the point of view of interpreting of the observations,
other refractory grains in the innermost disk, such as titanium, calcium, or 
aluminum-bearing minerals (e.g., titanium or aluminum oxide), 
could be responsible for the near-IR excess observed.  
Since the current data do not constrain the inner disk composition to this 
degree of refinement (i.e., multiple refractory dust species),  we did not
attempt to fit our data with mixtures of carbon and/or other refractory components. 
 We retained the amorphous carbon solution since it is the simplest refractory grain that can be assumed.

  The conclusion that the innermost disk of HD 135344B could be
  carbon-rich potentially presents a problem in a wider cosmochemical
  context, because the Earth and other terrestrial planets \citep[as well
  as some exoplanetary systems, e.g,][]{XuJura2013} display a
  carbon to silicon deficit in comparison with meteorites, the Sun, and
  the Interstellar Medium.  Perhaps HD 135344B is an unique object,
  or the processes responsible for the carbon depletion on Earth are
  particular to the Solar System, or perhaps there is much wider
  variety of the possible chemical compositions that terrestrial
  planets can have in different planetary systems.  We stress that the
  carbon-rich innermost disk of HD 135344B is a plausible solution
  that allows us to simultaneously reproduce all available gas and dust
  observables of the inner disk, most notably the near IR visibilities indicating
  continuum emission inside the silicate sublimation radius and the CO
  ro-vibrational emission extending tens of AU. 
  But other refractory grains could provide similar solutions.
  Amorphous carbon grains provide the simplest solution that fits the
  data, but they are not the only possible solution.
  }

\subsection{Double and single-peaked CO-rovibrational line profiles in transition disks}

In a recent study of CO-rovibrational emission in young stars,
\citet[][]{Bast2011} have observed three types of line profiles:
(1) double-peaked, (2) narrow-single peaked, 
and (3) single-peaked with a broad base.
Bast et al. argue that the profiles of type 3 stem from a combination of 
emission from the inner part  of the disk ($R<$a few AU),
and emission from a slow moving disk wind.
In their analysis, Bast et al. suggest line profile parameter P$_{\rm 10}$ 
(the full width line profile at 10\% of its height divided by the full width of the line profile at 90\%)
to distinguish between the different types of line profiles.
In the case of HD~135344B, 
the value of P$_{\rm 10}$, obtained from an average of the lower J-transition lines up to P(14), is 6.9.
Thereby  locating HD~135344B just above the value of 6,  the maximum
line profile parameter for a Keplerian disk model with a power-law temperature profile.

In our models of HD~135344B, we have found that the CO ro-vibrational line profile is the result of the combination
of two contributions: (1) the emission from the gas inside the dust cavity and 
(2) the emission from the gas located in the inner rim of the outer disk.
The relative contribution of both components, hence, the CO ro-vibrational line profile, 
depends on the properties of the dust in the inner disk and the gas mass in the inner and outer disk.

In Models 1 and 2, 
we have shown the effect of the inner disk dust composition on the CO ro-vibrational line profile. 
If the optical of the continuum at 4.7~$\mu$m inside 30 AU is small ($\tau<0.1$, e.g., Model 2), 
then the CO ro-vibrational emission will be dominated by the contribution of the inner rim of the outer disk,
and a narrow single-peaked line profile would be produced.

A single-peaked line profile can also be obtained by increasing the gas mass in the outer disk, 
diminishing the gas mass in the inner disk, or
increasing the scale height of the outer disk with respect to the inner disk.
In the case of Model 5, 
if the gas mass in the outer disk is increased such that the gas-to-dust ratio is more than 10,
then the line profile becomes more and more single-peaked 
(see bottom panel of Fig.~\ref{OICO_vsHR_flux}).

In summary, our radiative transfer calculations of the CO~4.7~$\mu$m emission show that in the case of transition disks,  
the presence of single-peaked CO ro-vibrational line profiles can be explained by Keplerian disk emission 
without recourse to a disk wind component.
This only applies to sources in which the emission line center is not blueshifted with respect to the stellar velocity.

\section{Summary and conclusion}

We conducted a modeling project aimed at constraining 
the gas mass and the gas and dust disk structure of the transition disk HD~135344B 
from multi-instrument and multiwavelength observations of gas and dust.
We found that the previously suggested inner disk structure { \citep{Brown2007}}, 
namely a narrow dust inner disk of from 0.18 to 0.45 AU 
followed by a large 45 AU dust gap replenished with gas,
fails to reproduce the CO ro-vibrational emission observed as the line profile produced 
from this disk model is a broad double peak.

 We have found a disk model that is able to reproduce current observational constraints.
This disk is composed of three zones:
\vspace{-1mm}
 
\begin{enumerate}
\item A first zone between 0.08 and 0.2 AU composed of small carbonaceous grains (and gas) with a total dust mass of a few 10$^{-12}$ M$_{\odot}$ 
{ (a few solar abundances of carbon).}
The presence of this inner carbonaceous grains provides
\begin{itemize}
\item[{\it a)}] {a fit to the near-IR H-band visibilities and closure phases,}
\item[{\it b)}] {a fit of the near-IR SED, while allowing the warm CO at several AU to emit and contribute to the $4.7~\mu$m line profile,}
\item[{\it c)}] an agreement with the higher temperatures ($T>$1\,500 K) expected in this zone.
\end{itemize}

\item { A second zone extending from 0.2 to 30 AU (i.e., the dust cavity) replenished with gas ($10^{-5}$-$10^{-4}$ M$_{\odot}$) 
with a surface density increasing as a function of the radius and dust mass of astronomical silicates of maximum $10^{-7}$ M$_{\odot}$}.
An increasing surface density profile with radius is required to fit the shape of the CO ro-vibrational emission lines. 
The fit to the SED constrained the scale height between 0.09 and 0.13 at 10 AU with a flaring exponent 1.12.
The gas-to-dust ratio in this zone is larger than 100, 
however, the exact value is not well constrained.
We found models up to gas-to-dust ratios of 15\,000 that are 
consistent with the observations either by decreasing the silicate dust mass by a factor 100,
or by increasing both the gas mass by a factor of a few and the power-law exponent of the surface density distribution.
The upper bound to the gas mass at $R<30$ AU is given by the 
flux of the [\ion{O}{i}] 63~$\mu$m, combined with the requirement that the surface density
of the inner disk should be equal or lower than the surface density of the outer disk at 30 AU.
The dust surface density at R$<$30 AU is lower than the one expected from extrapolating the dust surface density from the outer disk.
This zone can contain up to a 3\% in mass ($\sim10^{-9}$ M$_{\odot}$) in carbon grains and keep the  fit to the SED.
 
\item A third zone from 30 AU to 200 AU (the outer disk) with astronomical silicate grains, a dust mass of  $2\times10^{-4}$ M$_{\odot}$,
a gas mass $10^{-4}-10^{-3}$ M$_{\odot}$ { (gas-to-dust ratio $<10$)}, surface density exponent of -1.0, 
and flaring of 1.0. 
In this zone large ($0.1<a<1000~\mu$m) and small ($0.1<a<10~\mu$m) dust grains have different radial and vertical spatial distributions. 
Large grains are located at $45<R<200$ AU in a disk with scale height 0.05 and 75\% of the gas and dust mass.
Small grains are located at $30<R<200$ AU in a disk with higher scale height (0.09 to 0.13,
i.e., the same H/R as zone 2) and 25\% of the gas and dust mass.
The vertical structure of the outer disk echoes the expected effects of dust growth and settling.
A gas-to-dust ratio much lower than 100 in the outer disk is required  to fit the [\ion{O}{i}] 63~$\mu$m 
line flux and to reproduce the CO ro-vibrational line profile simultaneously.\\
\end{enumerate}

{ The models suggest that the best fit to the gas observations in HD~135344B  
is provided by a disk in which the gas surface density and the scale height have no large 
discontinuities at 30 and 45 AU. 
In other words, there is no large gap in the gas distribution of HD~135344B. 
The cavity observed in the near-IR and sub-mm is replenished by gas and (some) dust. 
The presence of a small gap of a few AU in the gas is consistent with current data,
a large gap in the gas of tens of AU does not appear likely.

The global gas-to-dust ratio, i.e., integrated over the full disk,
is much lower than 100.
This provides further evidence that HD~135344B is an evolved protoplanetary disk
that has already lost a large amount of its gas mass. 
The disk structure proposed for HD~135344B could be applied to other pre-transition disks with CO ro-vibrational emission extending to several AU.

The increasing surface density profile of the gas in the inner disk, 
the difference in the radial distribution of large and small  grains in the outer disk,
and a small gap in the gas between the inner and outer disk 
are  compatible with  the dynamical interaction
of a single Jovian planet and the disk.
However, we should be cautious because other mechanisms
could be responsible for the gas and dust distribution observed in HD~135344B.

The current ensemble of observations show that the transitional disk features (i.e., gap) 
observed in the continuum reflect only changes on the distribution of large and small dust particles in the disk.
The gas has a spatial distribution that can differ from that of the dust, particularly that of large grains observed in the sub-mm.
Furthermore, we find that the gas distribution and mass in transition disks can be very different from primordial disks}. 

Our modeling results predict that gas and (some) dust emission should be detected inside the cavity with
high-sensitivity and high-spatial resolution sub-mm observations.
ALMA measurements of continuum emission inside 30 AU would be a great help in determining 
the dust size distribution and dust mass inside the cavity, quantities that at present are free parameters in the model.
ALMA gas observations at high spatial resolution, such as the CO 6-5 line, 
used in conjunction with other gas tracers will enable the gas mass inside the cavity, 
and the gas mass and surface density in the outer disk to be better constrained.

{Future detailed dynamical disk models that include the dynamics of the gas, 
the interaction between the gas and the dust,
the photochemistry of the disk, and carbon destruction would be a big help for establishing
whether the carbon innermost disk suggested by our modeling is viable
or whether including refractory grains different from carbon is necessary.} 

In this paper we showed that the simultaneous modeling of gas and dust observations is required to 
address the problem of the dust and gas structure of protoplanetary disks.

\begin{acknowledgements}
A. Carmona and C. Pinte acknowledge funding from the
European Commission's 7$^\mathrm{th}$ Framework Program (EC-FP7)
(contract PERG06-GA-2009-256513) and from
the Agence Nationale pour la Recherche (ANR) of France under contract
ANR-2010-JCJC-0504-01.
Calculations were performed at the Service Commun de Calcul Intensif de
l'Observatoire de Grenoble (SCCI) on the Fostino supercomputer
funded by the ANR (contracts ANR-07-BLAN-0221, ANR-2010-JCJC-0504-01 and
ANR-2010-JCJC-0501-01) and  the
EC-FP7 contract PERG06-GA-2009-256513. 
The research leading to these results has received funding from the EC-FP7 under grant agreement no 284405.
FM acknowledges support from the Millennium Science Initiative (Chilean Ministry of Economy), through grant ÒNucleus P10-022-FÓ.
PIONIER is funded by the Universit\'e Joseph Fourier (UJF), the Institut de Plan\'etologie et d'Astrophysique de Grenoble (IPAG), the Agence Nationale pour la Recherche (ANR-06-BLAN-0421 and ANR-10-BLAN-0505), and the Institut National des Science de l'Univers (INSU PNP and PNPS). The integrated optics beam combiner is the result of a collaboration between IPAG and CEA-LETI based on CNES R\&T funding.
Part of this work was funded by the Agence National pour la 
Recherche of France through the Chaire d'Excellence grant ANR (CHEX2011 SEED).
This research has made use of the SIMBAD database,
operated at the CDS, Strasbourg, France.
This work is based (in part) on archival data obtained with the Spitzer Space Telescope, which is operated by the Jet Propulsion Laboratory, California Institute of Technology under a contract with NASA. 
We thank Alex Brown for providing the Chandra X-ray luminosity.
\end{acknowledgements}

% for the bibliography, at the end
\bibliographystyle{aa} % style aa.bst
\bibliography{Carmona_HD135344B} % your references Yourfile.bib
%\newpage

\begin{appendix}

\section{}

\subsection{Photometry}
In this appendix, we summarize the different sources of the photometry used to construct the SED.
%\vspace{-0.5cm}
\begin{table*}[ht]
\begin{center}
\caption{Photometry}
\label{table_photometry}
\begin{tabular}{llll|llll}
\hline
$\lambda$ & $\lambda$~F$_{\lambda}$ & Error & References &  $\lambda$ & $\lambda$~F$_{\lambda}$ & Error & References \\
$[\mu$m] & [W m$^{-2}$] & [W m$^{-2}$]  & &  $[\mu$m] & [W m$^{-2}$] & [W m$^{-2}$] \\
\hline
0.105 & 2.8$\times 10^{-15}$ & 1.6$\times 10^{-15}$ & FUSE  & 10.70 & 3.047$\times 10^{-13}$ & 3.1$\times 10^{-16}$ & Spitzer\\
0.110 & 4.6$\times 10^{-15}$ & 1.7$\times 10^{-15}$ & FUSE  & 12.25 & 2.038$\times 10^{-13}$ & 2.8$\times 10^{-16}$ & " \\
0.115 & 5.1$\times 10^{-15}$ & 1.9$\times 10^{-15}$ & FUSE  & 14.03 & 1.514$\times 10^{-13}$ & 1.9$\times 10^{-16}$ & " \\
0.118 & 7.5$\times 10^{-15}$ & 2.7$\times 10^{-15}$ & FUSE  & 16.34 & 2.047$\times 10^{-13}$ & 2.3$\times 10^{-16}$ & " \\
0.119 & 3.7$\times 10^{-15}$ & 3.7$\times 10^{-16}$ & COS   & 19.39 & 3.644$\times 10^{-13}$ & 3.6$\times 10^{-16}$ & " \\
0.123 & 4.8$\times 10^{-15}$ & 3.7$\times 10^{-16}$ & COS   & 22.15 & 4.955$\times 10^{-13}$ & 4.2$\times 10^{-16}$ & " \\
0.129 & 3.6$\times 10^{-15}$ & 3.2$\times 10^{-16}$ & COS 	& 25 & 8.0$\times 10^{-13}$ & 1.1$\times 10^{-13}$ & IRAS $\,^{b}$\\
0.137 & 2.7$\times 10^{-15}$ & 3.2$\times 10^{-16}$ & COS  & 25.39 & 6.280$\times 10^{-13}$ & 5.8$\times 10^{-16}$ & Spitzer\\
0.142 & 2.2$\times 10^{-15}$ & 1.3$\times 10^{-15}$ & COS  & 29.73 & 8.155$\times 10^{-13}$ & 8.8$\times 10^{-16}$ & " \\
0.152 & 3.4$\times 10^{-15}$ & 2.1$\times 10^{-15}$ & COS  & 47 & 1.5$\times 10^{-12}$ & 2.2$\times 10^{-13}$ & \citet{Harvey1996}\\
0.163 & 7.6$\times 10^{-15}$ & 6.6$\times 10^{-16}$ & COS  & 60 & 1.3$\times 10^{-12}$ & 1.8$\times 10^{-13}$ & IRAS$\,^{b}$\\
0.169 & 9.2$\times 10^{-15}$ & 8.2$\times 10^{-16}$ & COS  & 60 & 1.4$\times 10^{-12}$ & 2.0$\times 10^{-13}$ & ISOPHOT\\
0.174 & 1.4$\times 10^{-14}$ & 1.1$\times 10^{-15}$ & COS  & 63.18 & 1.20$\times 10^{-12}$ & 4.7$\times 10^{-15}$ & GASPS \\
0.37 & 3.4$\times 10^{-12}$ & 1.6$\times 10^{-13}$ &  \citet[][]{CoulsonWalter1995} & 72.85 & 9.84$\times 10^{-13}$ & 4.1$\times 10^{-15}$ & "\\
0.44 & 6.7$\times 10^{-12}$ & 1.4$\times 10^{-13}$ & " & 79.36 & 8.96$\times 10^{-13}$ & 3.8$\times 10^{-15}$ & "\\
0.55 & 7.4$\times 10^{-12}$ & 1.1$\times 10^{-13}$ & " & 80 & 9.8$\times 10^{-13}$ & 1.5$\times 10^{-13}$ & ISOPHOT\\
0.65 & 7.25$\times 10^{-12}$ & 9.2$\times 10^{-14}$ & " & 89.99 & 7.77$\times 10^{-13}$ & 3.3$\times 10^{-15}$ & GASPS \\
0.77 & 5.3$\times 10^{-12}$ & 1.2$\times 10^{-13}$ & \citet[][]{Grady2009} & 90.16 & 7.68$\times 10^{-13}$ & 3.3$\times 10^{-15}$ & "\\
0.8 & 6.18$\times 10^{-12}$ & 7.5$\times 10^{-14}$ & \citet[][]{CoulsonWalter1995} & 100 & 6.9$\times 10^{-13}$ & 1.2$\times 10^{-13}$ & ISOPHOT \\
1.2 & 4.5$\times 10^{-12}$ & 1.0$\times 10^{-13}$ & " & 100 & 7.7$\times 10^{-13}$ & 7.7$\times 10^{-14}$ & IRAS $\,^{b}$\\
1.22 & 4.80$\times 10^{-12}$ & 9.8$\times 10^{-14}$ & 2MASS$\,^{a}$ & 144.78 & 3.97$\times 10^{-13}$ & 2.1$\times 10^{-15}$ & GASPS\\
1.6 & 3.96$\times 10^{-12}$ & 9.4$\times 10^{-14}$ & \citet[][]{CoulsonWalter1995} & 145.53 & 3.96$\times 10^{-13}$ & 2.1$\times 10^{-15}$ & "\\
1.63 & 4.4$\times 10^{-12}$ & 1.3$\times 10^{-13}$ & 2MASS $\,^{a}$ & 157.75 & 3.78$\times 10^{-13}$ & 1.9$\times 10^{-15}$ & "\\
1.7 & 4.1$\times 10^{-12}$ & 1.8$\times 10^{-13}$ &  \citet[][]{Grady2009} &179.52 & 2.85$\times 10^{-13}$ & 1.7$\times 10^{-15}$ & " \\
2.19 & 4.19$\times 10^{-12}$ & 5.5$\times 10^{-14}$ & 2MASS$\,^{a}$ & 180.42 & 2.83$\times 10^{-13}$ & 1.7$\times 10^{-15}$ & " \\
2.2 & 3.5$\times 10^{-12}$ & 1.4$\times 10^{-13}$ & \citet[][]{CoulsonWalter1995} & 200 & 1.4$\times 10^{-13}$ & 1.0$\times 10^{-14}$ & ISOPHOT\\
3.8 & 2.10$\times 10^{-12}$ & 7.9$\times 10^{-14}$ & \citet[][]{CoulsonWalter1995} & 350 & 5.0$\times 10^{-14}$ & 1.5$\times 10^{-14}$ & \citet[][]{CoulsonWalter1995}\\
4.75 & 1.48$\times 10^{-12}$ & 6.3$\times 10^{-14}$ & " & 450 & 2.5$\times 10^{-14}$ & 1.0$\times 10^{-14}$ & "\\
5.60 & 1.681$\times 10^{-12}$ & 1.2$\times 10^{-15}$ & Spitzer & 800 & 2.4$\times 10^{-15}$& 1.5$\times 10^{-16}$ & "\\
6.80 & 1.083$\times 10^{-12}$ & 8.8$\times 10^{-16}$ & " & 850 & 1.73$\times 10^{-15}$ & 3.5$\times 10^{-17}$ &\citet{Sandell2011}\\
7.42 & 8.949$\times 10^{-13}$ & 8.2$\times 10^{-16}$ & " & 1100 & 7.1$\times 10^{-16}$ & 1.1$\times 10^{-16}$ & \citet[][]{CoulsonWalter1995}\\
8.45 & 6.565$\times 10^{-13}$ & 7.1$\times 10^{-16}$ & " & 1300 & 3.3$\times 10^{-16}$ & 9.7$\times 10^{-17}$ & \citet{Sylvester1996}\\
9.66 & 4.965$\times 10^{-13}$ & 5.2$\times 10^{-16}$ & "\\
\hline
\end{tabular}
\tablefoot{ 
%\tablefoottext{a}{Error assumed to be 10\%},
 \tablefoottext{a}{\citet[][]{Skrutskie2006}}
 \tablefoottext{b}{ \citet{Walker_Wolstencroft1988}}
}
\end{center}
\end{table*}
%\clearpage
\subsection{Stellar parameters}
\label{star_parameters}
The stellar properties of HD~135344B have been studied by previous authors
\citep[e.g.,][]{Dunkin1997,Mueller2011,Andrews2011}.
While authors agree on a spectral type F4, an effective temperature  
6\,750$\pm$150 K  (6\,660 K, Dunkin et al. 1997; 6\,590 K, Andrews et al. 2011;  
6\,810$\pm$80 K, M\"ueller et al. 2011), and a mass of 1.6 to 1.7 $M_\odot$,
there is a discrepancy in estimations of the radius of the star. 
\citet{Mueller2011} suggest a radii of 1.4$\pm$0.25 $R_\odot$, 
while \citet{Andrews2011} suggest a radii of 2.15 $R_\odot$.

To have an independent assessment of the star properties,
we downloaded reduced archival high-resolution (R=75\,000) ESO/VLT-UVES spectra of HD~135344B in the 4\,800 to 5\,500 \AA~
and 5\,830 $-$ 6\,800 \AA~
ranges and used the interactive spectra visualization tool described in \citet{Carmona2010}
to compare the UVES spectrum to BLUERED \citep{Bertone2008} high-resolution synthetic spectra,
in regions not contaminated by emission lines or telluric absorption.
We found that the optical spectra of HD~135344B is compatible with spectral templates 
with $T_{\rm eff}$ ranging from 6\,500 to 7\,000 K for log$~g$ ranging between 4.0 and 5.0 (see Fig.~\ref{Teff}).
Naturally,  log$~g=$ 5.0 is not realistic as log$~g=4.33$ for an F-type star on the main sequence.
The minimum in the $\chi^2$ statistic suggests a $T_{\rm eff}$ around 6750 K.

To have a second constraint on the spectral type and A$_{\rm V}$,
we employed the HST/COS\footnote{Cosmic Origins Spectrograph} 
spectrum of HD~135344B.
In Fig.~\ref{HST-COS}, we display coadded COS spectrum of HD~135344B, 
smoothed by a running nine-point boxcar filter, and
dereddened by a \citet[][]{Cardelli1989} R=3.1 extinction curve with E(B-V)=0.129 (Av=0.4).
We overplot F3V (HR 9028, IUE SWP 14002, red) and F4V (HD 27901, SWP 45935, blue) 
spectral type comparison stars scaled by the difference in V magnitudes between them and HD~135344B.
The spectral template F4V provides the best fit to the observed COS spectrum.
The F4V spectrum corresponds to a $T_{\rm eff}$ of 6620 K using 
\citet{SchmidtKaler1982}.

%\citep{SchmidtKaler1982}. 
\begin{figure}[h]
\begin{center}
\includegraphics[width=0.5\textwidth]{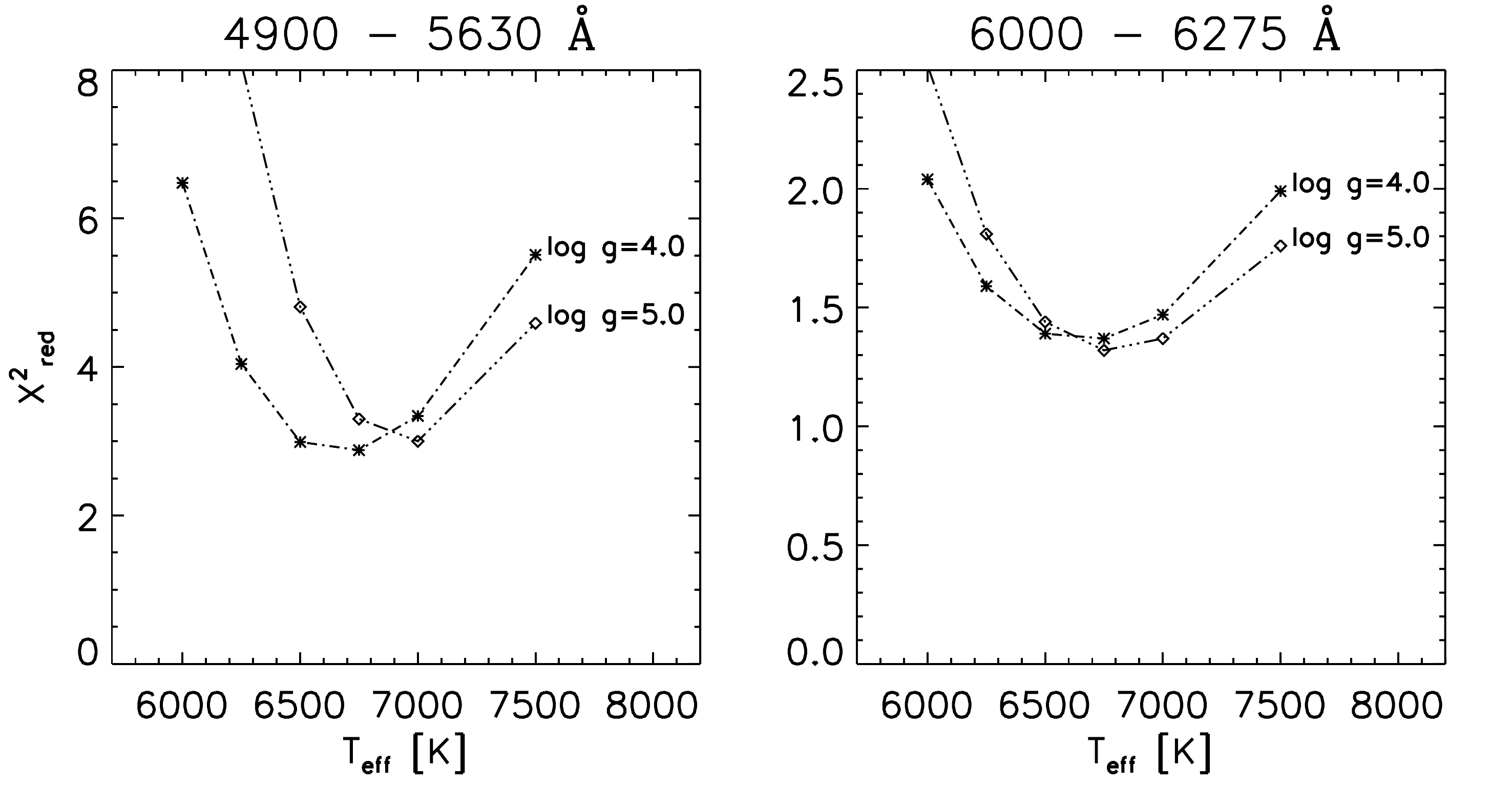}
\caption{$\chi^2_{\rm red}$ as a function of the T$_{\rm eff}$ of the observed VLT/UVES high-resolution (R=75000) spectrum and 
a rotational broadened BLUERED  \citep{Bertone2008} high-resolution synthetic spectra. $\upsilon$ sin$\,i$ was set to minimize 
$\chi^2_{\rm red}$ for each T$_{\rm eff}$ and log $g$.} 
\label{Teff}
\end{center}
\end{figure}

\begin{figure}[t]
\begin{center}
\includegraphics[width=0.48\textwidth]{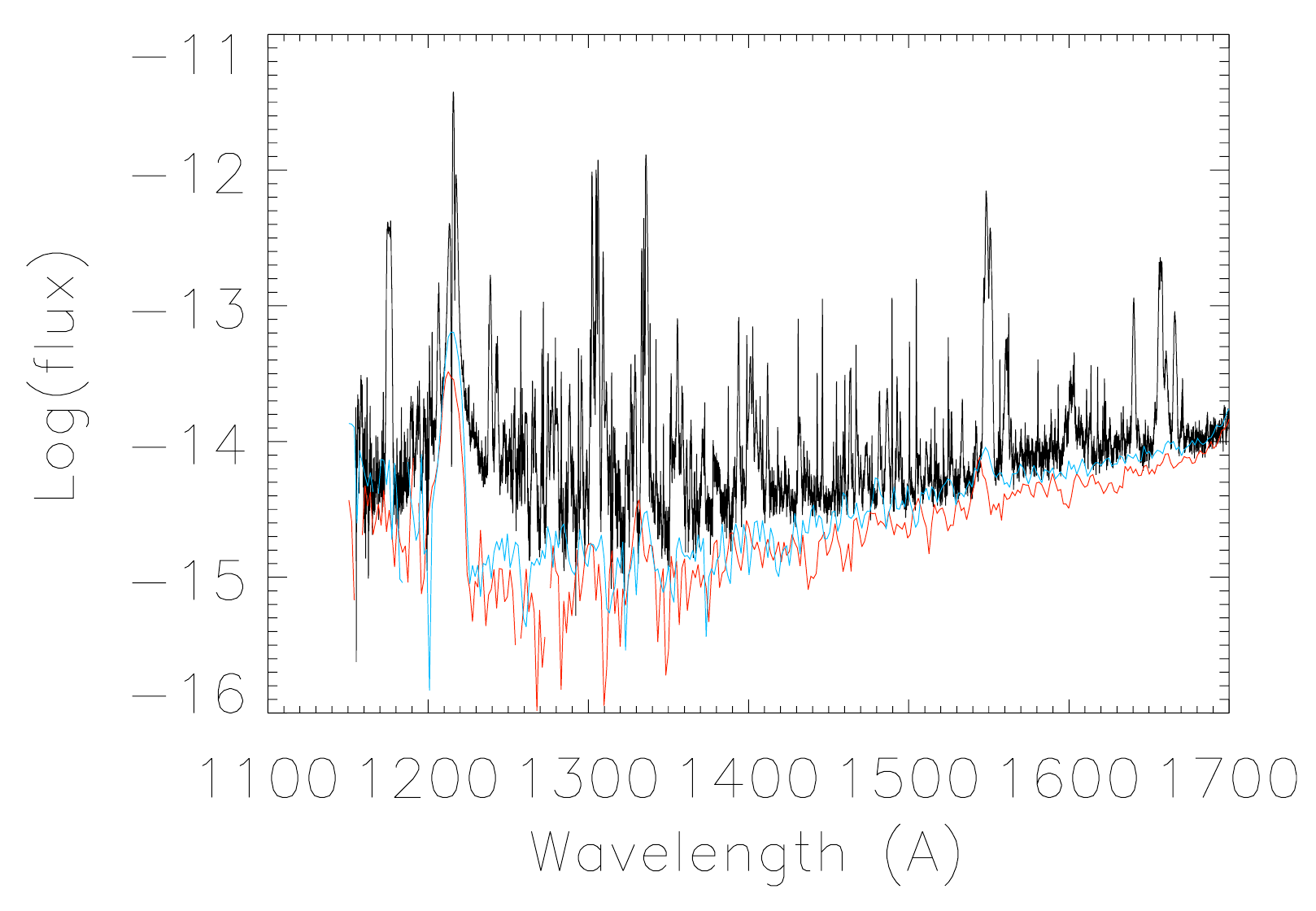}
\caption{
Coadded HST-COS  spectrum of HD~135344B smoothed and dereddened (see details in the text).
F3V (HR 9028, IUE SWP 14002, red) and F4V (HD 27901, SWP 45935, blue) spectral type comparison
stars are shown, scaled by the difference in V magnitudes between them and HD~135344B. 
HD~135344B shows a distinct FUV excess for wavelengths shortward of 1600~\AA, demonstrating
the accretion luminosity. The structure seen in the COS data is not noise, but fluorescent H$_2$ emission
\citep[][]{France2012}.  The rise in flux for the comparison stars shortward of H~I Lyman $\alpha$ is 
an artifact.  No correction in all of the spectra has been made for geocoronal Lyman $\alpha$. 
}
\label{HST-COS}
\end{center}
\end{figure}

To obtain an estimate of the luminosity and to constrain the value of the radius, 
we used a distance of 140 pc \citep{vanBoekel2005}, 
the $B$ and $V$ photometry (9.2 and 8.7, SIMBAD database),
and the $T_{\rm eff}$, $M_{\rm V}$, $BC$ values of \citet{SchmidtKaler1982}
for stars of luminosity class V (i.e., smallest possible radii). 
We found that for a $T_{\rm eff}$ equal to 6\,440 K (F5) and 6\,890 K (F2),
$R$= 2.09 and 2.07 R$_\odot$ respectively.
Interpolating for $T_{\rm eff}$ 6\,620~K, we obtained 2.08 R$_\odot$.
These values are closer to the radius estimate of 2.15 R$_\odot$ suggested by \citet[][]{Andrews2011}
than the 1.4 R$_\odot$ claimed by  \citet[][]{Mueller2011}.

For our models, 
we used a star with $T_{\rm eff}$=6\,620K (F4V), a stellar radius of 2.1 R$_\odot$, a mass of 1.65 M$_\odot$, and $A_{\rm v} = 0.4$.

\newpage
\subsection{CO 4.7 $\mu$m emission}
%\vspace{-1cm}
\begin{figure}[h]
\begin{center}
\includegraphics[width=0.4\textwidth]{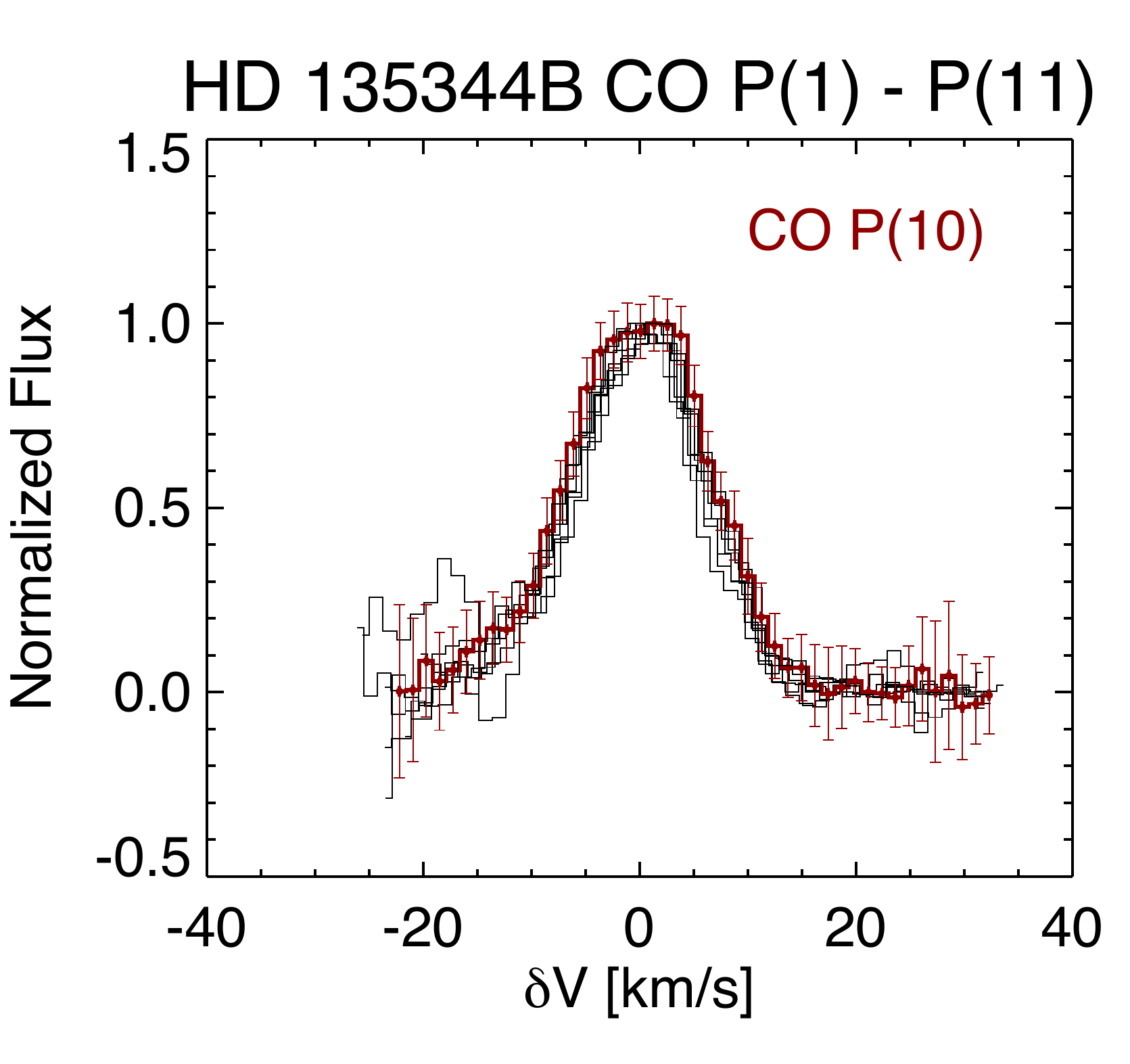}
\label{CO_lines} 
\caption{The CO P(1) to P(11) line profiles are available within the CRIRES spectrum.
The lines have been continuum subtracted and normalized by the peak flux.
The CO P(10) line is in red.}
\end{center}
\end{figure}

\subsection{Model 5: Herschel CO and water lines}
%\vspace{-1cm}
\begin{table}[h]
\caption{Predicted and observed CO rotational emission line fluxes by Model 5
at Herschel, sub-mm, and mm wavelengths}
\label{Table_CO_line_fluxes}
\begin{center}
%\script size
\begin{tabular}{ccccc}
\hline
Transition & $\lambda$ & Flux  & Observed  & Ref.\\
 & [$\mu$m] & [$10^{-19}$ W m$^{-2}$] & [$10^{-19}$ W m$^{-2}$] \\
 %\end{tabular}
% \begin{tabular}{lrrc}
\hline
$J = $ 36 $-$ 35 & 72.84 & 0.4 & $<246$ & 1\\
$J = $ 29 $-$ 28 & 90.16 &  1.7 & $<191$ & 1 \\
$J = $ 23 $-$ 22 & 113.46 & 3.9 & $<128$ & 1\\
$J = $ 18 $-$ 17 & 144.78 & 5.5 & $< 69$ & 1\\
$J = $ 15 $-$ 14 & 173.63 & 9.1 & $< 155$ & 1\\
$J = $ 6 $-$ 5 & 433.55 & 8.3 & ...\\
$J = $ 5 $-$ 4 & 520.23 & 5.6 & ...\\
$J = $ 4 $-$ 3 & 650.25 & 3.3 & ...\\
$J = $ 3 $-$ 2 & 866.96  & 1.5 & 1.2 & 2\\
$J = $ 2 $-$ 1 & 1300.40 & 0.5 & 0.8 & 3\\
$J = $ 1 $-$ 0 & 2600.75 & 0.06   & ...\\
\hline
\end{tabular}
\end{center}
{{\bf References:} (1) \citet{Meeus2013}; 
(2) \citet{Dent2005};
(3) \citet{Thi2001} 
}\\
\end{table} 

\begin{table}[h]
\caption{Predicted rotational emission line fluxes by Model 5
in selected H$_2$O water lines at Herschel wavelengths.}
\label{Table_H2O_line_fluxes}
\begin{center}
%\vspace{-1cm}
\begin{tabular}{ccccc}
\hline
Transition & $\lambda$ & Flux  & Observed  & Ref.\\
 & [$\mu$m] & [$10^{-19}$ W m$^{-2}$] & [$10^{-19}$ W m$^{-2}$] \\
 %\end{tabular}
% \begin{tabular}{lrrc}
\hline
o-H$_2$O $8_{18}-7_{07}$& 63.32 & 3.7 & $<120$ & 1\\
o-H$_2$O $7_{16}-6_{25}$& 66.09 & 2.8 & $<120$ & 1\\
o-H$_2$O $3_{30}-2_{21}$& 66.44 & 23.7 & $<120$ & 1\\
o-H$_2$O $7_{07}-6_{16}$& 71.96 & 7.8 & $<120$ & 1\\
o-H$_2$O $3_{21}-2_{12}$& 75.39 & 26.1 & $<120$ & 1\\
o-H$_2$O $4_{23}-3_{12}$& 78.74 & 19.6 & $<120$ & 1\\
o-H$_2$O $6_{16}-5_{05}$& 82.03 & 12.2 & $<120$ & 1\\
o-H$_2$O $2_{21}-1_{10}$& 108.07 & 18.8 & $<120$ & 1\\
o-H$_2$O $3_{03}-2_{12}$& 174.62 & 12.5 & $<120$ & 1\\
o-H$_2$O $2_{11}-1_{01}$& 179.53 & 16.6 & $<120$ & 1\\[2mm]
p-H$_2$O $7_{26}-6_{15}$& 59.99 & 0.8 & $<120$ & 1\\
p-H$_2$O $6_{15}-5_{24}$& 78.93 & 1.0 & $<120$ & 1\\
p-H$_2$O $6_{06}-5_{15}$& 83.29 & 3.7 & $<120$ & 1\\
p-H$_2$O $4_{13}-4_{04}$& 187.11 & 1.0 & $<120$ & 1\\
\hline
\end{tabular}
\end{center}
{{\bf References:} (1) Upper limit at 145 $\mu$m is assumed \citep{Fedele2013};}\\
\end{table}

\subsection{Details of Model 5}
In this Appendix, we show additional plots of the structure and emitting regions of
several gas tracers in Model 5.
\begin{figure*}[t]
\begin{center}
\begin{tabular}{cc}
{\bf \sffamily \large ~~~~~Gas temperature} & {\bf \sffamily \large ~~~~~~~Dust temperature} \\[2mm]
\includegraphics[width=0.4\textwidth]{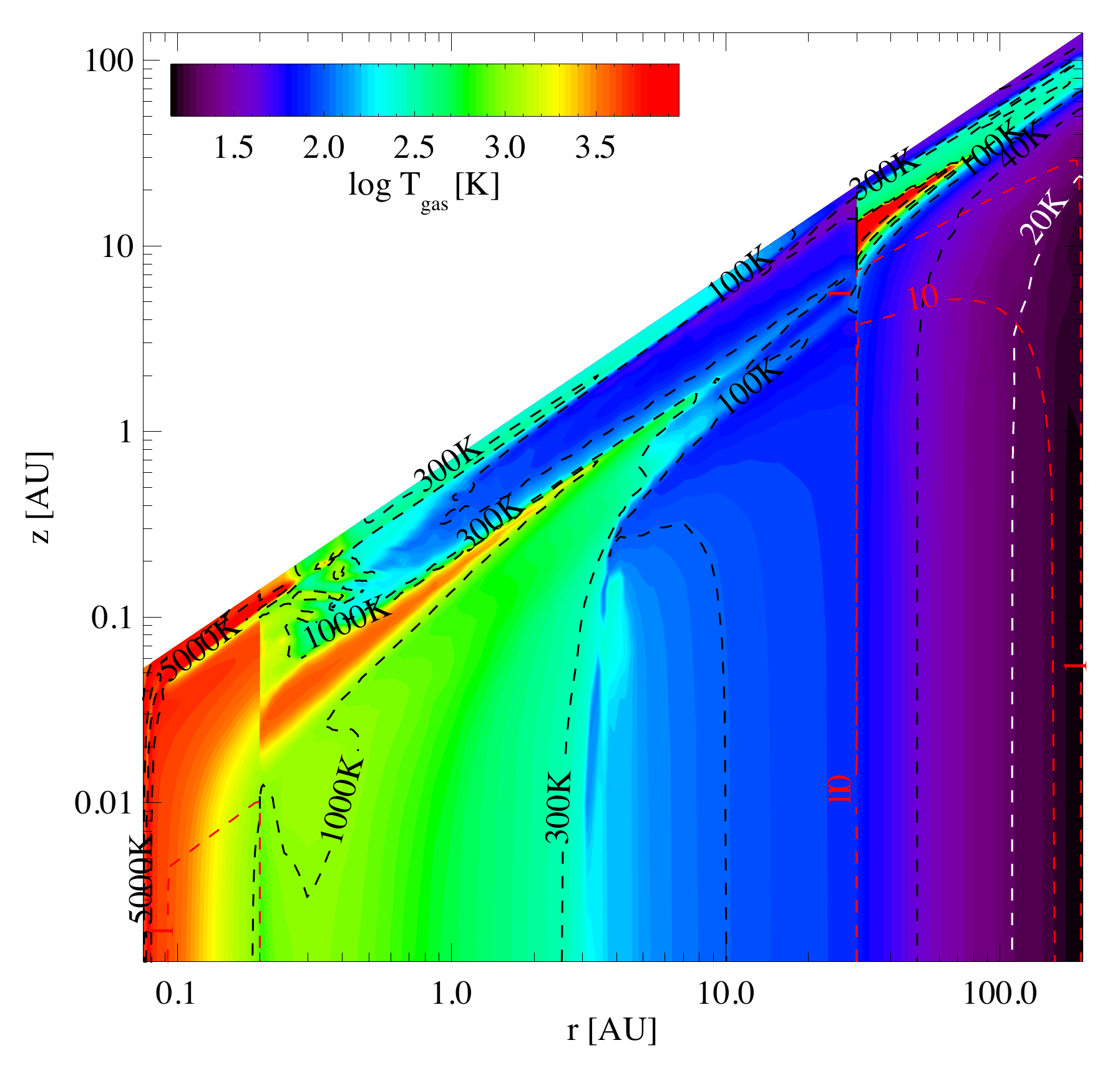} &
\includegraphics[width=0.39\textwidth]{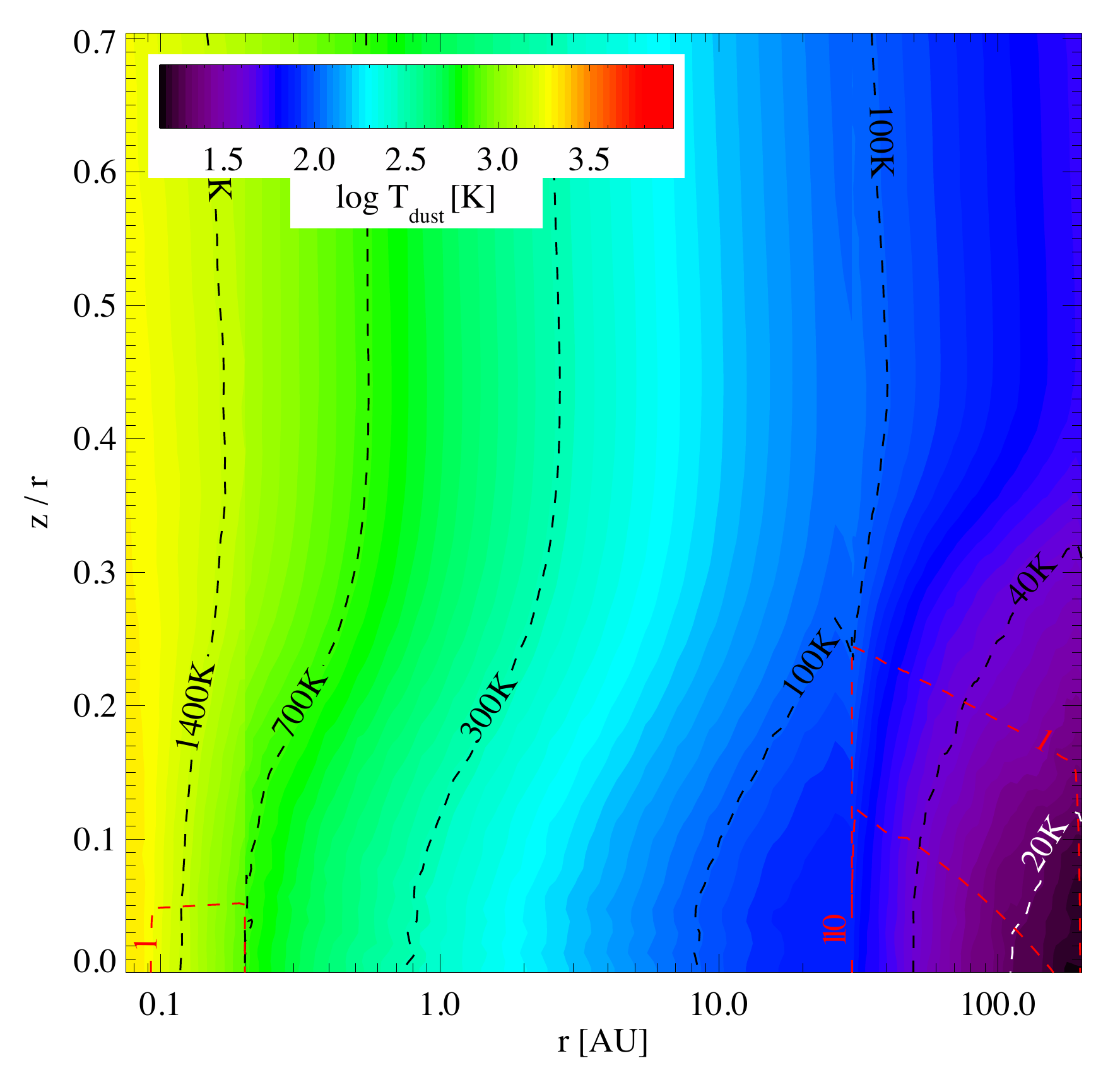}  \\[1mm]
{\bf \sffamily \large ~~~~~Hydrogen number density} \\[2mm]
\includegraphics[width=0.39\textwidth]{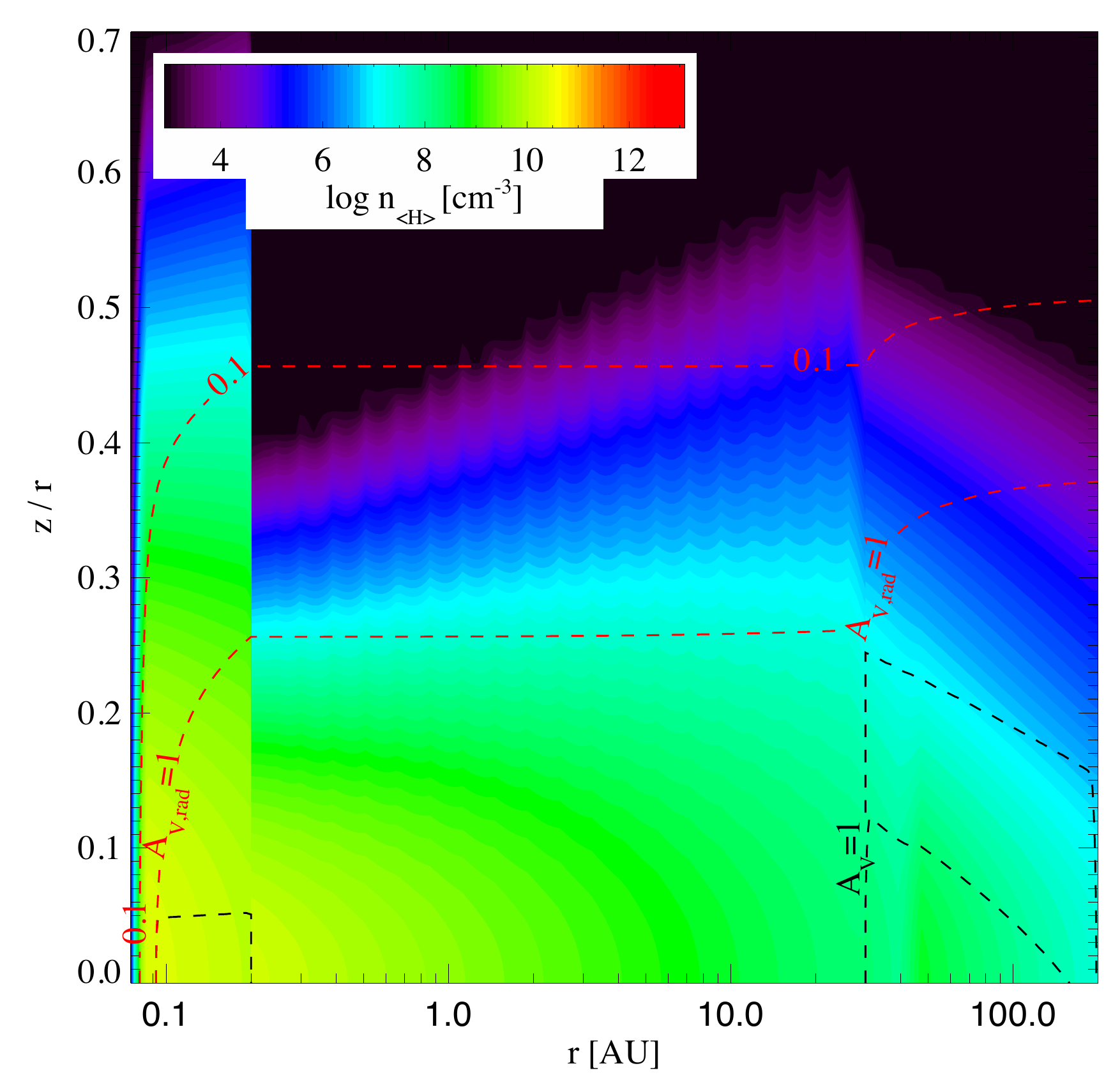}&  
\includegraphics[width=0.4\textwidth]{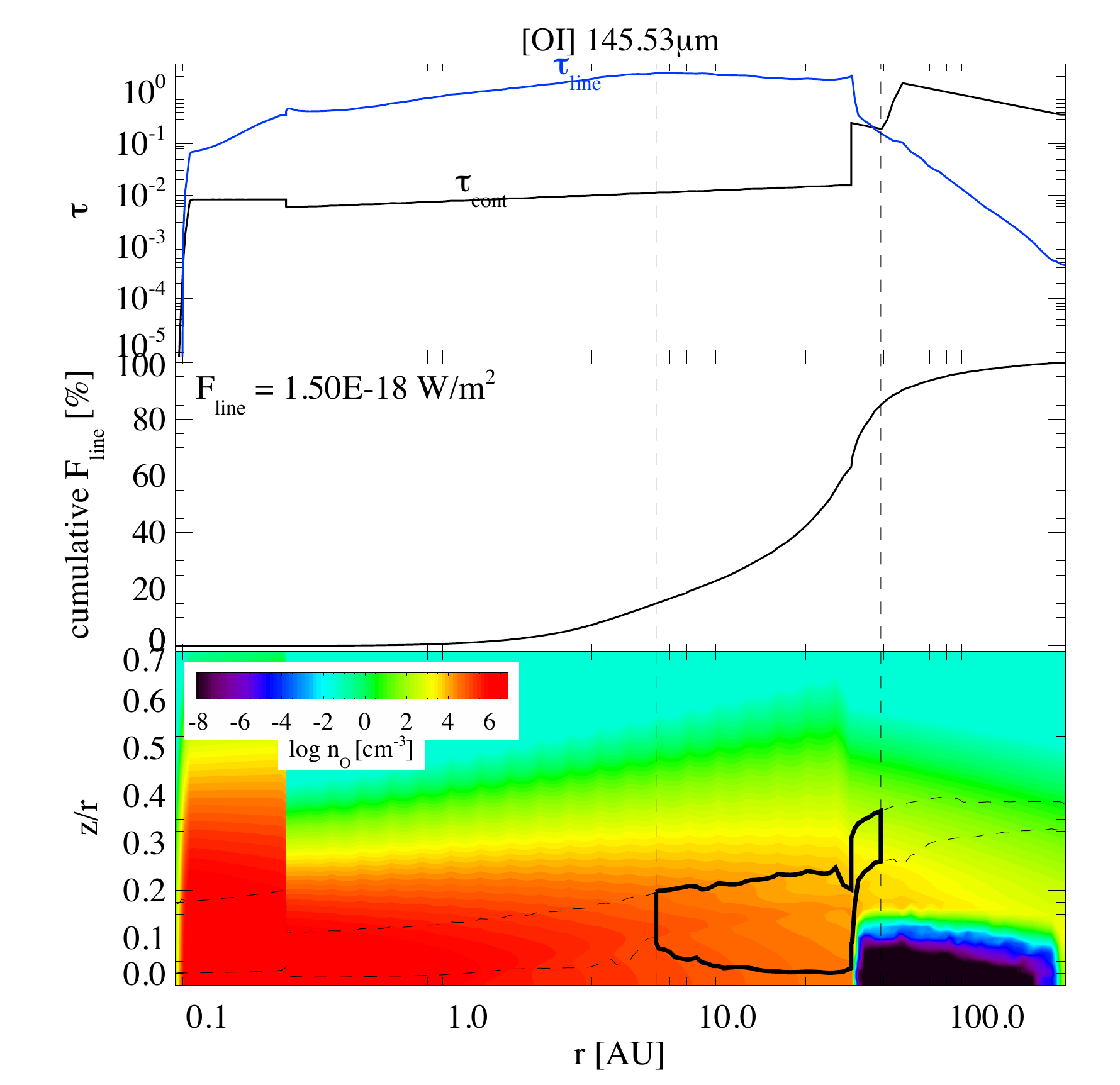}\\[1mm]
\includegraphics[width=0.4\textwidth]{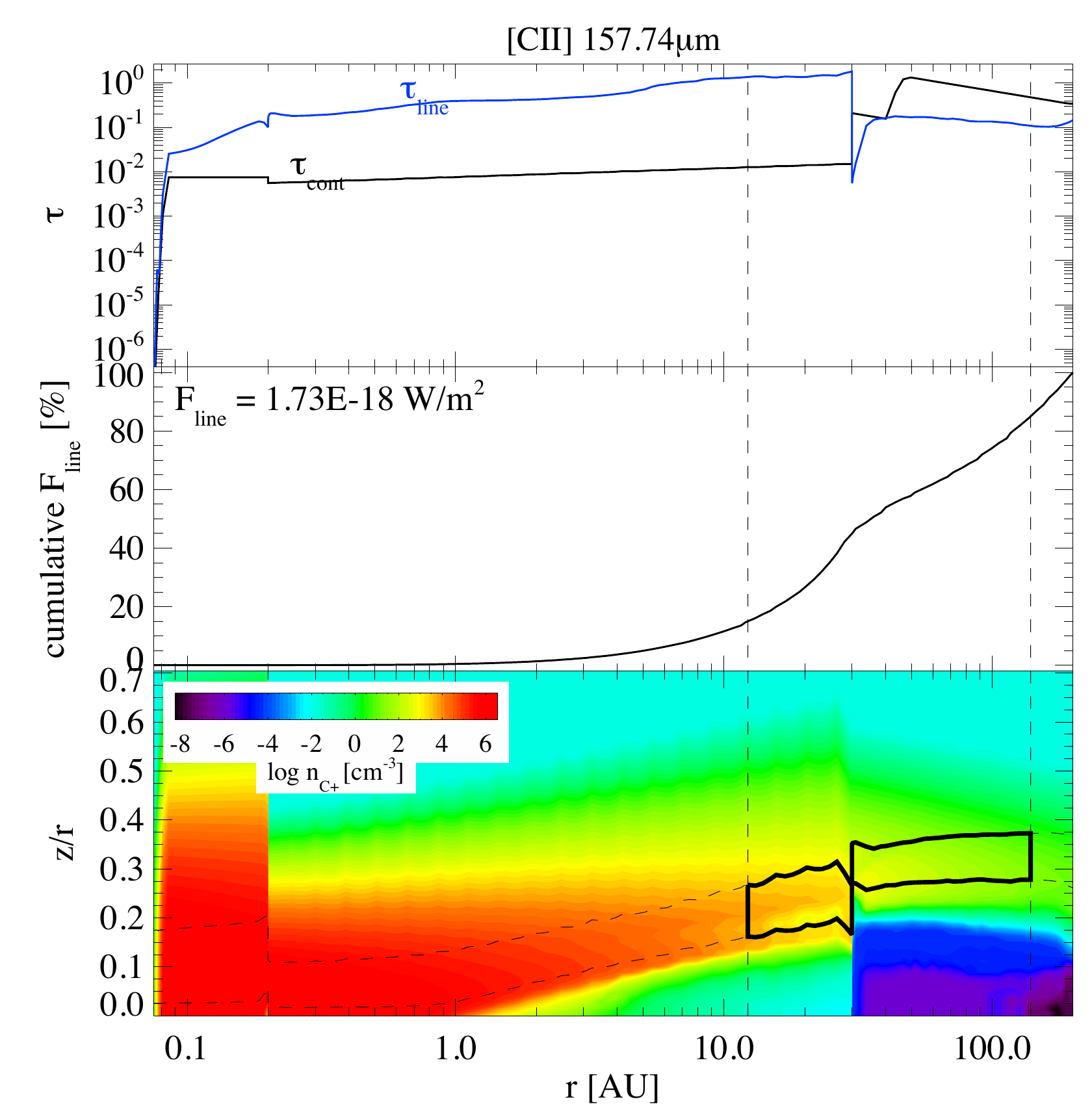}&  
\includegraphics[width=0.42\textwidth]{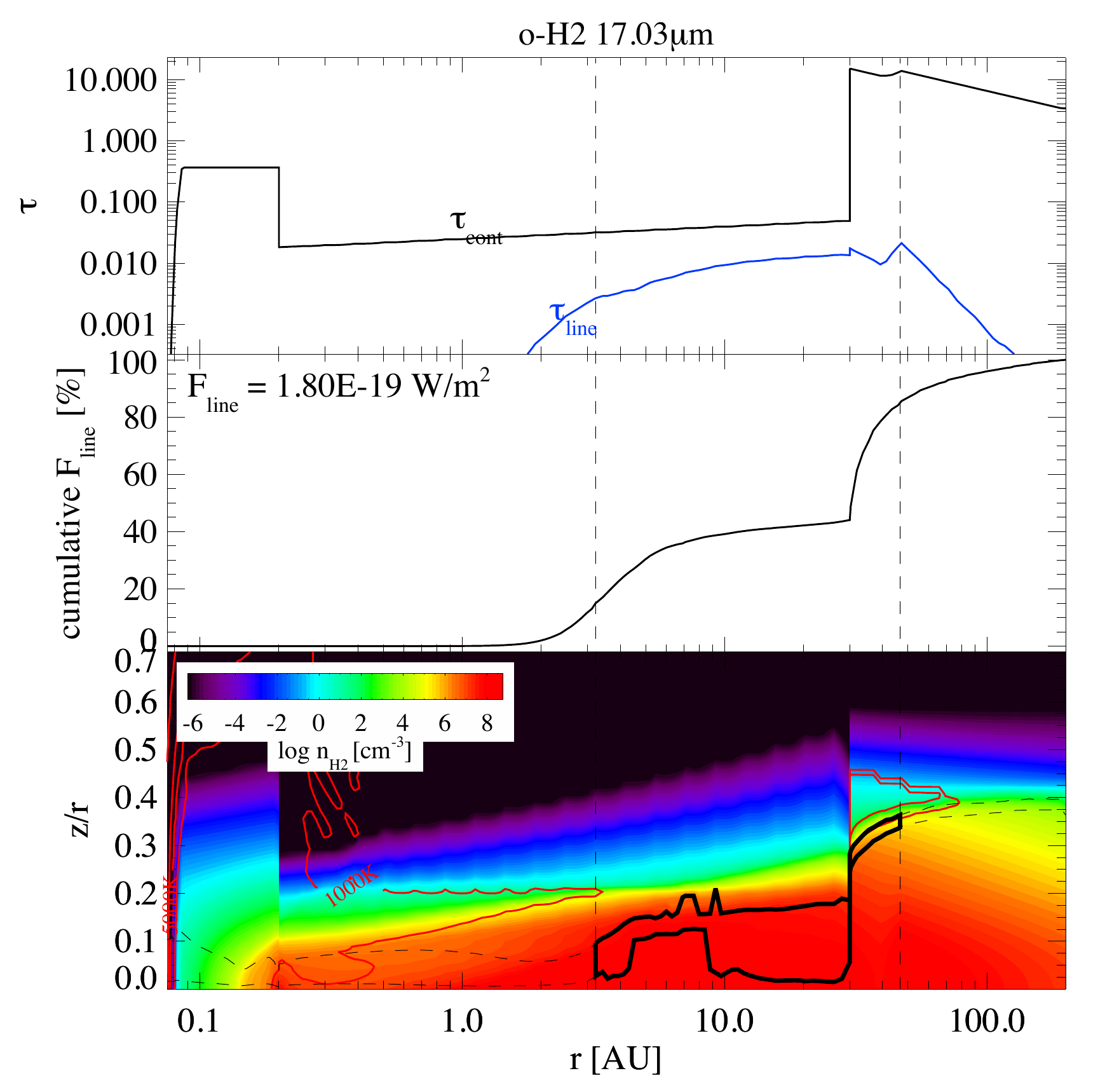}\\[1mm]
\end{tabular}
\end{center}
\caption{ Model 5. 
{\it Upper panels:}  ({\it left)} Gas temperature, ({\it right}) Dust temperature.
{\it Central panels:}  ({\it left)} Hydrogen number density,  ({\it right)} optical depth of the line ($\tau_{\rm line}$), of the continuum ($\tau_{\rm cont}$), cumulative vertical flux, 
and number density as a function of the radius for the [\ion{O}{i}] line at 145 $\mu$m. 
The box in thick black lines represents the region in the disk that emits 70\% of the line radially and 70\% of the line vertically,
thus approximately $\sim$ 50\% of the line flux.  
{\it Lower Panels:} Similar plots for the [\ion{C}{ii}] line at 157 $\mu$m ({\it left)} and the ortho H$_2$ 0-0 S(1) at 17 $\mu$m {\it (right)}.
}
\label{Model5_appendix} 
\end{figure*}

\end{appendix}

\end{document}